\documentclass[onecolumn]{emulateapj}
\usepackage{lscape}
\newcommand{\mic}{\hbox{$\mu$m}}             % \mic
\newcommand{\kms}{\mbox{km s$^{-1}$}}        % \km/s
\newcommand{\thC}{\mbox{$\theta^1$Ori~C}}    % \theta 1 Ori C
\received{31 March 2004}
\accepted{13 December 2004}

\begin{document}

\title{The Orion Nebula in the mid-IR}

\author{M. Robberto\altaffilmark{1}, S. V. W. Beckwith, and N. Panagia\altaffilmark{1}}
\affil{Space Telescope Science Institute, 4300 San Martin Drive,
    Baltimore, MD 21218}

\author{S. G. Patel\altaffilmark{2}}
\affil{University of California, Santa Cruz, 1156 High Street, Santa Cruz, CA 95064}

\author{T. M. Herbst, S. Ligori}
\affil{Max-Planck-Institut f\"ur Astronomie, K\"onigstuhl 17,
    D-69117 Heidelberg, Germany}

\author{A. Custo\altaffilmark{3}, P. Boccacci, and M. Bertero}
\affil{INFM and DISI, Universit\'a di Genova, Via Dodecaneso 35,
     I-16146 Genova, Italy}

\altaffiltext{1}{Affiliated with the Space Telescope Division of the European Space Agency, ESTEC,
Noordwijk, the Netherlands}
\altaffiltext{2}{STScI Summer Student program}
\altaffiltext{3}{Present address: Massachusetts Institute of Technology, Computer Science and Artificial Intelligence Laboratory (CSAIL), 200 Technology Sq. Cambridge, MA}

\begin{abstract} 
We present two wide-field ($\approx 5\arcmin\times 3.5\arcmin$), diffraction limited ($\lambda/D\simeq0.5\arcsec$ at 10~\micron), broad-band 10~\micron\ and 20~\mic\ images of the Orion Nebula, plus  six 7--13~\micron\ narrow-band ($\lambda/\Delta\lambda\simeq1$) images of the BN/KL complex taken at the 3.8m UKIRT\footnote{The United Kingdom Infrared Telescope is operated by the Joint Astronomy Centre on behalf of the U.K. Particle Physics and Astronomy Research Council.} telescope with the MPIA MAX camera. The wide field images, centered on the Trapezium and BN/KL regions, are mosaics of $35\arcsec\times35\arcsec$ frames obtained with
standard chopping and nodding technique and reconstructed using a new restoration method developed for this project. They show the filamentary structure of the dust emission from the walls of the HII region and reveal a new remarkable group of arc-like structures $\approx 1\arcmin$ to the South of the Trapezium. The morphology of the Ney-Allen nebula, produced by wind-wind interaction in the
vicinity of the Trapezium stars, suggests a complex kinematical structure at the center of the Cluster. We find indications that one of the most massive members of the cluster, the B0.5V star $\theta^1$Ori-D, is surrounded by a photoevaporated circumstellar disk. Among the four historic Trapezium OB stars, this is the only one without a binary companion, suggesting that stellar multiplicity and the presence of massive circumstellar disks may be mutually exclusive. In what concerns the BN/KL complex, we find evidence for extended optically thin silicate emission on top of the deep 10~\micron\ absorption feature. Assuming a simple two component model, we map with $\simeq 0\farcs5$ spatial resolution the foreground optical depth, color temperature and mid-IR luminosity of the embedded sources. We resolve a conspicuous point source at the location of the IRc2-A knot, approximately $0\farcs5$ north of the deeply embedded HII region ``I''. We analyze the spectral profile of the 10~\micron\ silicate absorption feature and find indication for grain crystallization in the harsh nebular environment. In the OMC-1 South region, we detect several point sources and discuss their association with the mass loss phenomenology observed at optical and millimeter wavelengths. Finally, we list the position and photometry of 177 point sources,  the large majority of which detected for the first time in the mid-IR. Twenty two of them lack a counterpart at shorter wavelengths, and are, therefore, candidates for deeply embedded protostars. The comparison of photometric data obtained at two different epochs reveals that source variability at 10{~\mic} is present up to $\approx 1$~mag level on a time-scale $\sim 2$~yr. With the possible exception of a pair of OB stars, all point sources detected at shorter wavelengths display 10~\mic\ emission well above the photospheric level, that we attribute to disks circumstellar emission. The model of Robberto, Beckwith, \& Panagia (2002, ApJ. 578, 897) provides the simplest explanation for the observed mid-IR excess. 
\end{abstract}


\section{Introduction}

The Great Orion Nebula (M42=NGC1976), with its associated cluster of Pre-Main-Sequence stars, represents the nearest ($d\simeq 450~$pc) and best studied episode of massive star formation \citep{LL03}. The stellar cluster,
distributed on a volume of $\sim 3$~pc radius \citep{LAH+Hart98}, contains approximately 2500 stars $\simeq 10^6$~yr
old spanning the mass range 45--0.02~M$_\odot $\citep{LAH97,HC00}. There is clear evidence of mass
segregation, with the most massive OB stars (the Trapezium system) concentrated in the central $\approx 0.15$~pc, where the
stellar density approaches $2\times10^4$~stars~pc$^{-3}$. Ultraviolet radiation from the Trapezium stars, mostly
the O6.5V star \thC, has produced the Great Orion Nebula (O'Dell 2001a,b), a blister cavity excavated on the surface of
the parental molecular cloud OMC-1. The same UV radiation also photo-ablates the accreting circumstellar disks of the nearest stars, producing the characteristic photoionized ``proplyds'' \citep{LV79, OD+93}. Proplyds provide the most compelling evidence that the star formation process can be dramatically affected by the mutual interaction between cluster members.

The Orion Nebula cluster, especially the core region (Trapezium cluster), has been investigated by several authors using data collected at visible \citep{Prosser+95,LAH97,Robberto+04} and near-IR wavelengths \citep{MJM+Stauff94,HC00,LR00,Luhman+00,Lada+00}. 
Since circumstellar disks emit in the IR the energy they ``passively'' absorb from their central stars, or ``actively'' produce by the viscous dissipation processes associated to mass accretion, there is a need for similar observations at longer wavelengths, in particular in the mid-IR ($\lambda\simeq5-20~\micron$), where ground based telescopes can reach sub-arcsecond spatial resolution. Mid-IR observations have also the potential of discovering objects more deeply embedded within the molecular cloud, otherwise invisible at shorter wavelengths (with the possible exception of hard X-rays) and eventually identify  sites where protostellar contraction is still ongoing. Moreover, mid-IR data may also provide complementary information on the structure of the HII region, through the diffuse emission of the dust grains and PAH molecules mixed with the ionized gas, or located in the surrounding photodissociation region.

The first 3.5~\micron\ image of the Trapezium cluster taken with a panoramic detector was presented, but only briefly discussed,
by \citet{MJM+96}. A more complete study of the Trapezium cluster at this wavelength has been published by \citet{Lada+00}. 
They found that a fraction 80\%--85\% of stars are surrounded by circumstellar disks, in agreement with earlier studies
made at near-IR wavelengths, and identify 78 sources with extremely red colors that they classify as candidate prostostars.

At wavelengths $\lambda>3.5$~\micron, the region remains largely unexplored. Maps made with single element bolometers in the early days of IR astronomy revealed two prominent sources, the BN/KL complex \citep{BN67,KL67} and the Ney-Allen nebula in the vicinity of the Trapezium \citep{NA69}. Both fields have been widely investigated with increasing sensitivity and spatial and spectral resolution. Images at 10 and 20~\micron\  made with panoramic detectors have been published by \citet{Gezari+98, Greenhill+04, SMB04} for BN/KL and by \citet{Hayward+94} for the Ney-Allen nebula. \citet{Smith+04} have recently presented an image of the OMC-1S region.

We have undertaken a large observational program on the Orion Nebula using the mid-IR camera MAX on the UKIRT telescope. Our research program consists of two parts: first, to observe at $10~\micron$ with relatively long integration times the most prominent
proplyds, in order to detect and resolve their structure in the mid-IR. Second, to map the entire nebular core, i.e. an area of approximately $5\arcmin \times 5\arcmin$ covering both the Ney-Allen and BN/KL regions, with shorter integration times. In this paper we present the results relative to this second study, complemented by narrow-band observations of the BN/KL complex. In Section~2 we present our observations and our data reduction strategy, based on a reconstruction method originally developed for this project. In Section~3, we present our main results, the 10 and 20~\mic~ wide field images and the photometric catalog of 177 point sources detected at 10~\mic. In Section~4, we discuss the overall structure of the nebula (Section~4.1), the Ney-Allen region (Section~4.2), the BN-KL complex (Section~4.3), the OMC-1 South region (Section 4.4), and finally the main properties of our sample of point sources (Section 4.5). In Section 5, we summarize our findings.

\section{Observations and data reduction}
\subsection{Data acquisition}
The wide field maps presented in this paper were obtained combining observations made in November 1998 and December 2000 at the 3.8m United Kingdom Infrared Telescope (UKIRT) on Mauna Kea, Hawaii. We used MAX\footnote{the instrument has been decommissioned after December 2000. The 20~\micron\ mosaic represents the very last set of images taken by MAX.} 
\citep{RobbertoHerbst98}, a mid-IR imager based on a Rockwell $128\times 128$ Si:As Blocked Impurity Band detector optimized for high background observations. MAX's optics delivered a plate scale of $0\farcs26$, corresponding to approximately $35\arcsec\times 35\arcsec$ field of view. Images have been obtained through broad N-band and Q-band filters. M.~Cohen provided detailed modeling of our photometric system, including typical Mauna Kea transmission, optics throughput, and detector response. For an A0V star, the N-band filter has an effective wavelength $\lambda_{10} = 10.1$~\micron\ and $\Delta\lambda = 3.13$~\micron, whereas the Q-band filter has $\lambda_{20} = 19.7$~\micron, and $\Delta\lambda = 2.44$~\micron. The narrow-band images were obtained on December 18, 2000, through the standard set of silicate filters centered at 7.7, 8.7, 9.8, 10.2, 11.2 and 12.4~\micron~ \citep[see][for the detailed filter parameters]{RobbertoHerbst98}.

UKIRT is equipped with tip-tilt adaptive correction at the secondary mirror \citep{Hawardeen+99}.  Exploiting this capability, MAX routinely reaches diffraction limited performances at $\lambda \ge 8$~\micron\ with $\lambda/D\simeq 0\farcs5$. The guide system, however, only allows for chopping throws $\leqq 30\arcsec$ (a maximum of $\simeq 20\arcsec $was in fact suggested for the $\simeq 5$~Hz chopping frequency we used). Given the extent of the Orion nebula and the MAX sensitivity, we could not chop onto empty sky areas for background subtraction. We experimented with the use of sky frames taken several arcminutes away, nodding the telescope at a frequency $\simeq 0.1$~Hz, constrained by the telescope pointing and tip-tilt reference star acquisition, but this method 
produced significant losses of sensitivity and observing efficiency. We therefore opted for the conventional chopping and nodding technique, adopting an observing strategy optimized for image post-processing, namely the reconstruction method envisioned in 
\citet[][and references therein]{Bertero+00}. In Table~\ref{Tab_parameters}, we list the main observing parameters:  night of the observation, filter, integration time per single frame, chopping frequency, total exposure time  on the main beam per each nodding cycle and chopping throw amplitude and direction. 

We scanned the survey region both in declination and right ascension using different chopping and nodding amplitudes. At each step, the telescope was offset (nodding) by an amount equal to the chopping amplitude, i.e. the new pointing position of the telescope on the main beam was coincident with the position imaged in the previous offset beam.  Directions and amplitudes of chopping and nodding were carefully matched and aligned with the orientation of the array columns, or rows. Due to a $\sim1^\circ$ rotation of the detector with respect to the equatorial coordinates, the chopping/nodding direction were not exactly coincident with the NS or EW axis. Therefore, the combined rows or columns have a slight drift with respect to the scan direction.  The setup process and observing strategy is detailed further in \citet{Bertero+00}.

Scans were typically $\sim 4$~arcmin long, with a small side overlay ($\sim 10$~pixels) to maximize field coverage. Since in the two runs the camera was mounted on different focal positions of the telescope, images are rotated by $90^\circ$ between the two runs.
The rotation of the optical path allows us to easily individuate ghost images produced by the brightest sources, BN in particular.

Observations were generally performed up to an airmass $z=2$ under excellent weather conditions. Absolute flux calibration in the N~band has been obtained by observing at various airmasses the following standard stars:
HR~337 ([N]=-2.06), HR~1040 ([N]=2.55), HR~1713 ([N]=0.06), HR~2061 ([N]=-5.13), HR~4069 ([N]=-0.94), HR~2943 ([N]=-0.76), HR~4554 ([N]=2.32) and HR~5340 ([N]=-3.17). Our absolute flux calibration assumes for $\alpha$~Lyrae [N]=0  with a flux $F=1.116\pm0.015\times10^{-16}$~Wcm$^{-2}\mu$m$^{-1}$ at $\lambda_{10}$, corresponding to 37.9~Jy. Calibration at 20\micron\ was based on HR~1760 ([Q]=-0.30), HR~2061 ([Q]=5.7), and HR~4069 ([Q]=-1.00).  The zero magnitude flux in this case is $F=7.887\pm0.011\times10^{-18}$~Wcm$^{-2}\mu$m$^{-1}$ at $\lambda_{20}$, corresponding to 10.2~Jy. The narrow-band images of BN were calibrated against HR~1040, HR~1047, HR~1713, HR~4069, HR~2943 and HR~5340, with magnitudes and zero points given in Table~\ref{Tab_SILICATES}. It is clear that the depth of our images is not uniform, since the total integration time in each scan is a function of the adopted pattern (amplitude of the chopping throws and offset between adjacent rows/columns), airmass, repeated or discarded frames, etc. Moreover, in the final images, part of the field has been observed on more than one night. In general, we estimate that the average integration time in the N-band is of  order 200 seconds. Concerning the narrow-band images, the total integration time is 103.5~s at all wavelengths except the 8.7~\micron\ band, where is 102.5~s.

As a final remark, we note that we were able to obtain several scans of the region only at 10~\mic. The 20~\mic\ data were all collected in one night, with only one chopping and nodding amplitude.  The final mosaic is therefore shallower and potentially more affected by reconstruction artifacts.

\subsection{Data reduction}
The raw images were first median filtered to remove a small number ($\approx 15$) of dead pixels and occasional extra-noise in the
frames containing the saturated BN peak. Cleaned frames were then normalized to the same integration time and combined to form chopped and nodded pairs. The final combination in the wide area mosaic required a special effort, described in the following section.

\subsubsection{Mosaics}
\begin{figure}[h]%1
\plotone{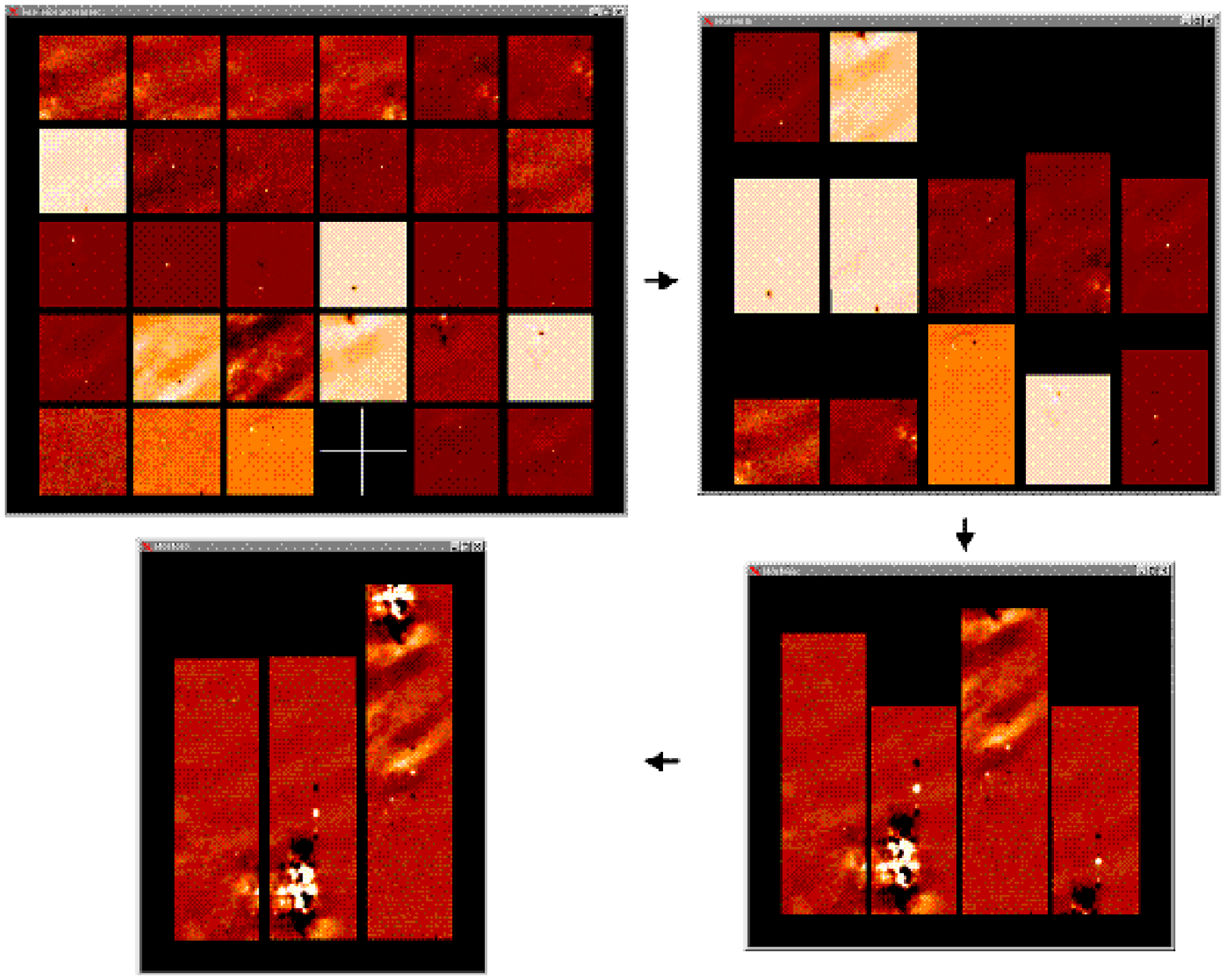}
\caption{Example of the construction of the 10~\micron\ mosaic. Clockwise from top-left, individual chopped and nodded frames 
are progressively assembled into a final strip. The restoration method works on each indivial column (or row). The column, approximately $35\arcmin$ wide, contains the BN/KL complex and the OMC-1S region. North is at the top and west is on the right.}
\label{Fig1}
\end{figure}

All chopped and nodded pairs taken on the same night were combined into rows or columns and then into a single mosaic. This operation was performed using MOSAIC, an IDL package designed for mid-IR data and publicly available on the Internet \citep{Varosi+Gezari93}. The images were combined by integer pixels, rounding the coordinates of the FITS header (accurate to $\simeq 0\farcs1$) to the nearest value. Figure~\ref{Fig1} illustrates the mosaicing process using images directly grabbed from the MOSAIC output windows. The mosaics contain for each source two negative counterparts at the position of the two offset beams. Hence, we needed to reconstruct a true image.

The restoration method we have used was specifically developed for this project. The theoretical basis have been illustrated in a number of papers \citep{Bertero+98,Bertero+99,Bertero+00}, thus we provide here only a brief outline. Since the problem is underconstrained, it admits infinit solutions. In particular, the set of solutions whose chopped and nodded images are identically zero constitutes the null space, infinite dimensional, of the chopping and nodding operator. Any linear combination of these functions can be added to a reconstructed image without affecting the observed, chopped and nodded, image. In the null space it is possible to define a metric and apply minimization processes. Our strategy is to search for the minimal positive solution, assuming the celestial signal cannot be negative. The implementation of the algorithm is based on the Landweber method \citep{Bertero+Boccacci98}. It is iterative and therefore produces an approximate solution at each step. The number of iterations needs to be optimized to control noise propagation, and our criterion has been described in \citet{Bertero+00}. When applied to real images, it produces excellent results.
In some situations, however, it may generate artifacts, i.e. artificial ghost images of the stars (appearing as dark images over a bright background or as bright images over a dark background) or discontinuities in the background level. The location and periodicity of 
the artifacts are related to the values of image size and chopping  amplitude through the mathematical structure of the problem \citep{Bertero+99}. For our 10~\mic\ mosaic, they have been minimized by replacing at their positions  the information provided by the reconstructed images obtained with different chopping throws and orientation. Another problem is that the  inversion method we have used provides less than ideal results when the signal in the image to be inverted is negative. In this case, the number of iterations needed to recover the signal may be larger than the number of iterations corresponding to optimal noise reduction. Also in this case, these problems have been greatly mitigated at 10~\mic\ by using images taken with different orientation and chopping amplitudes. 

\subsubsection{Narrow-band images}
The narrow-band images were registered using the peak of the BN object, then combined and trimmed to the same size of $101\times117$~pixels, corresponding to $26\farcs8\times31\farcs0$. The astrometric solution was derived assuming for the BN peak the following 
coordinates: RA(2000.0)=$05^h35^m14\fs10$ and DEC(2000.0)=$-05\arcdeg22\arcmin22\farcs7$, a nominal pixel scale of $0\farcs26$
constant across the field, a rotation angle of $-0\fdg8$ east of north, as measured on the same night for the wide field area project. 
The BN center was set at the pixel position $64.2,82.2$, and measured using a centroid procedure with 5 pixel ($1\farcs3$) radius.

\subsubsection{Photometry}
Point sources have been identified by visual inspection of the original (chopped and nodded) images, and then confirmed on the final reconstructed image. A few sources, indicated in Table~\ref{Tab_stars10mic}, appear very faint only in the final mosaic and don't show any counterparts at shorter wavelengths. It is possible that some are artifacts surviving the image cleaning  processes described in the previous section, and an independent confirmation is needed to make sure they are real. On the other hand, the variations of exposure time and airmass across the image may have rendered invisible sources otherwise within reach with our instrument in more ideal conditions.
 
Aperture photometry was obtained on individual chop/nod pairs using circular  apertures with radii between 2 and 5~pixels, and applying an aperture correction based on the average encircled energy of the standard stars observed during the same night. Typically, all standards were used, with the exception of HR~2061 ($\alpha$~Ori), known to be extended in the mid-IR \citep{Rinehard+98} and easily resolved by MAX. An airmass correction was applied to each frame. In several cases, the non-uniformity of the background emission dominates the final photometric accuracy. For point sources, an error in the sky estimate affects in a predictable way the results of multiaperture photometry with curve of growth correction: an overestimate of the sky will cause the source to become fainter with the radius, and vice versa. We used this property to tune the local sky value to obtain photometric results as much as possible independent of the aperture. Sometimes, however, the variations of the sky brightness underneath the source or the extended PSF profile produced unstable results. In these cases, we adopted a different strategy. Using the IDL procedure MIN\_CURVE\_SURF, we estimated the brightness distribution of the sky below the source  using the minimum curvature surface fitting a sky annulus surrounding the star. The assumption here is that in the presence of strongly variable background a higher order interpolation provides a more reliable estimate of the sky distribution under the source. The subtraction of this {\sl sky} frame from the original one provides stellar images that sit on a relatively flat background, and therefore, can be measured with greater accuracy. Results obtained with this method turned out to depend on the choice of the inner and outer radii of the annulus, typically 5 and 8~pixels. Our final results were then obtained by averaging the results provided by a few different, but reasonable on the basis of visual inspection, combinations of inner and outer radius. Their scatter has also been added in quadrature to the total uncertainty. In general, photometric measurements obtained with different methods coincide within their errors, especially for the smaller apertures ($1\arcsec-1.5\arcsec$) and within the same run. The uncertainties we quote combine quadratically the uncertainties of the zero points (0.12~mag at 7.7~\micron, 0.06~mag at 8.7~\micron, 0.07~mag at 9.6~\micron, 0.06~mag at 10.2~\micron, 0.05~mag at 11.6~\micron, 0.05~mag at 12.4~\micron, 0.06~mag in the broad N-band, and 0.11~mag in the Q-band), the source photon statistics and sky uniformity, and the scatter between the best estimates obtained with different methods, or between different choices of radius/annulus, as explained. 

All sources that had photometric measurements different between the two runs by an amount comparable to, or larger than, the quoted 1-$\sigma$ error were double checked. A direct comparison with nearby sources provides in several cases strong evidence for source variability. For sources isolated, or superimposed on a large and complex background (observed with different chopping  throws/orientation in the two runs), it is not possible to safely discriminate between variability and unaccounted photometric errors.
Whereas the main photometric catalog has been obtained analyzing the single individual images before mosaicing and reconstruction, 
we have also examined the final reconstructed mosaics, both as a sanity check and to find possible evidence of faint sources that may appear only after all frames have been stacked together. 

Concerning the overall sensitivity, the final signal-to-noise ratios of our reconstructed mosaics depend on a number of factors such as the integration time of each exposure, the number of overlaying exposures contributing at each point, airmass, background brightness, chopping direction and noise propagation in the reconstruction algorithm. Therefore, the sensitivity and completeness limits vary across the field in a complex way. To provide an indicative value of our sensitivity limit, we observe that our catalog of 10 micron sources contains objects down to $[N]\simeq8$~mag (see Section~4.5). Since all  stars brighter that $[N]\simeq6.0$~mag are well detected in all partial mosaics, we can assume 6.0~mag as our completeness limit at 10~\micron, at least in the central $2\arcmin\times 2\arcmin$ of our field. However, according to this criteria, this value raises to $[N]\simeq5$ in the immediate vicinity of the Trapezium, where the nebular emission becomes stronger. This because six sources in the range $[N] = 5.0 - 6.0$ have been detected only with favorable chopping throw and measured only at one epoch. For the narrow-band images, a recursive standard deviation algorithm with sigma clipping gives,  over an empty sky area at the SE corner of our images, the following values, scaled over 1~arcsecond area:  13.1~mJy at 7.7~\micron, 4.3~mJy at 8.7~\micron, 11.6~mJy at 9.6~\micron, 6.8~mJy at 10.2~\micron, 7.5~mJy at 11.6~\micron~ and 9.7~mJy at 12.4~\micron.

\section{Results}
\begin{figure}%2
\plotone{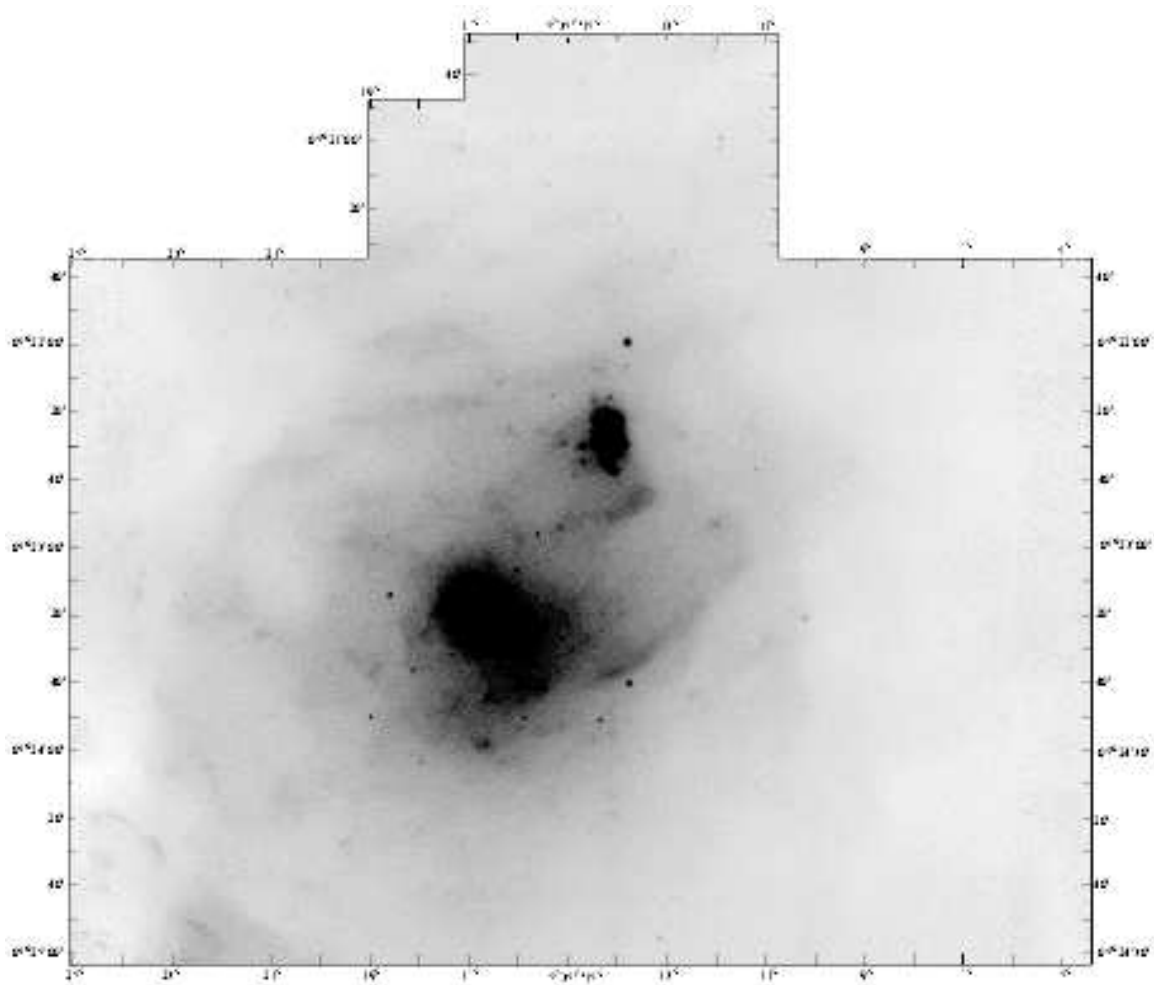}
\caption{Final 10~\micron\ mosaic of the Orion nebula. The two saturated sources at the center of the field are 1)~to the southeast, the Ney-Allen nebula, in correspondence of the Trapezium stars, and 2)~to the northwest, the BN/KL complex.}
\label{Fig2}
\end{figure}

\begin{figure}%3
\plotone{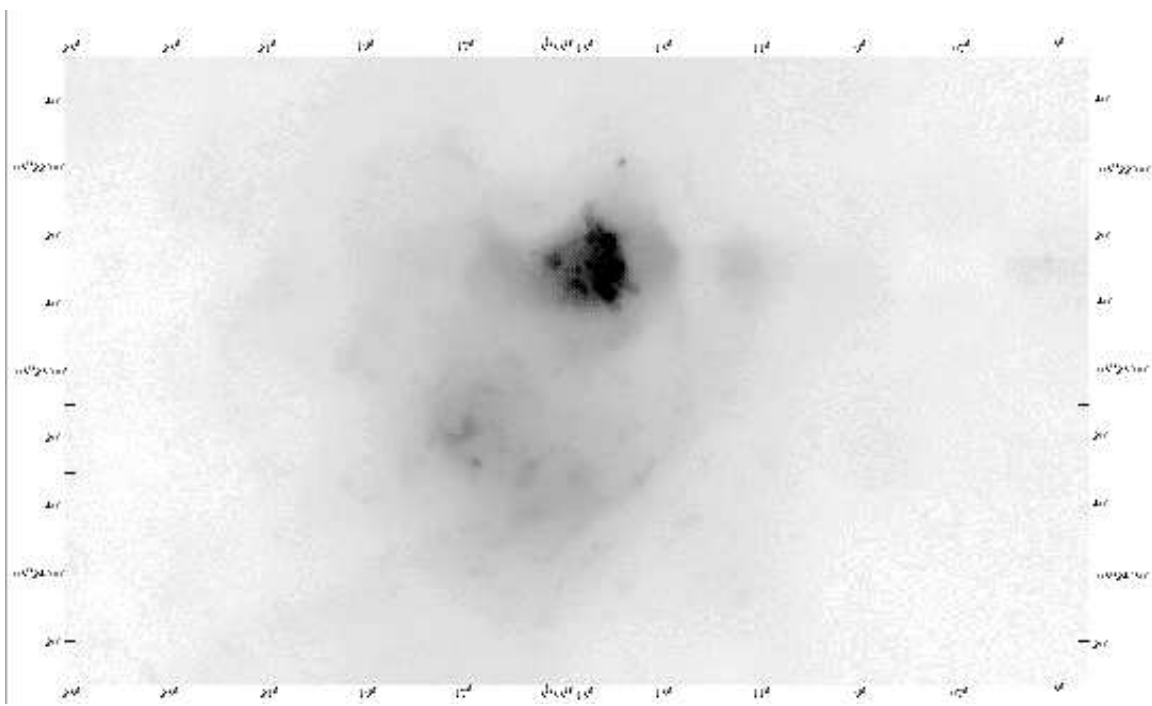}
\caption{Same as Figure~1, at 20~\micron. The only prominent source at this wavelength is the BN/KL complex.}
\label{Fig3}
\end{figure}

The final reconstructed images are shown in Figure~\ref{Fig2} (10~\micron) and Figure~\ref{Fig3} (20~\micron). The size of the final 10~\micron\ mosaic is  $1144 \times 1072$ pixels, corresponding to approximately $5.0\times 4.7$~arcmin. However, as we were able to complete only the central columns to their originally planned extent, the rectangular area fully imaged is smaller, i.e. $1144 \times 800$~pixels = $5.0\times 3.5$~arcmin. The 20~\micron~ image has a size of $1144\times 702$ pixels, corresponding to approximately $5.0\times 3.0$~arcmin.

The morphology at 10~\micron\ is clearly dominated by the BN/KL complex and by the Ney-Allen nebula, the two bright sources visible at the center of the field (BN/KL to the north,  Ney-Allen to the south). Extended emission permeates the entire region, with a general decrease of surface brightness with increasing angular distance from the center. This intense, diffuse emission is barely visible in the original, non-reconstructed images, since the chopping and nodding process filters low frequency spatial components. At 20~\micron,  the BN/KL complex dominates. Several spatially resolved knots of diffuse emission are also visible at both wavelengths. The Bright Bar, a photo-dissociation region viewed edge-on approximately $2\arcmin$ to the southwest of the Trapezium, is visible at the lower-left corner of our 10\mic\ image, in the vicinity of $\theta^2$Ori~A (source MAX-172). Several other filaments are visible at 10~\micron\ in the nebular background. Their length range from $\sim10\arcsec$ to $\sim 1\arcmin$ and their surface brightness is comparable, with peaks in the range $\sim 0.1-0.2$~Jy~arcsec$^{-2}$ at 10~\mic. In the central region the peak brightness of the clumps and filaments
rises to $\sim 0.7$~Jy~arcsec$^{-2}$, in particular to the south-west and to the north-west of the Ney-Allen nebula and to the south of the BN/KL, in the vicinity of the Orion-S region (see Section~4.4).

The most remarkable new feature is represented by three, possibly four, large arcs of mid-IR emission  located approximately 1~arcmin to the south of the Trapezium/Ney-Allen nebula, at R.A.(2000)=$5^h35^m15^s$, DEC.(2000)=$-5^h24\arcmin10\arcsec$ (Figure~\ref{Fig4,5}a,b). They point toward the Trapezium stars (see also Figure~\ref{Fig12} for a sketch of their location inside the nebula) and appear brighter on the side facing the Trapezium. These structures are also visible at shorter wavelengths, in particular in the HST/WFPC2 image of \citet{CRO+Wong96}, and most probably represent surfaces illuminated by grazing UV flux from the central stars. 

\subsection{The Trapezium region}
Our 10 and 20~\mic\ images of the Trapezium region (Figure~\ref{Fig6,7}a,b) show, with improved sensitivity and spatial resolution, the
general morphology first observed at 10~\micron\ by \citet{McCaughrean+Gezari91}, and studied more in detail by \citet{Hayward+94}. The extended mid-IR emission discovered by \citet{NA69} is resolved into a number of diffuse arc-like structures pointing toward \thC. The brightest one, corresponding to the brightness peak of the Ney-Allen nebula, is centered on the B0V star $\theta^1$Ori~D, undetected in our images.  In the rest of this paper, we shall use the name Ney-Allen nebula to indicate this feature only. Another five arcs 
encircle the sources LV1-LV5 of \citet{LV79}, the first photoionized disks observed in the vicinity of \thC. All arcs are spatially resolved in our images, and display different geometries and surface brightnesses. A most remarkable object lies in the immediate vicinity of \thC. This is the circular knot at the center of Figure~\ref{Fig6,7}a first identified as SC3 by \citet{Hayward+94}. \thC{} itself is also visible at 10~\micron{}, but not at 20~\micron, as the fainter source $\simeq 1\farcs8$ to the east of SC3. All together, the arcs appear to trace the edge of a cavity centered on \thC{}, filled with diffuse mid-IR emission, and with surface brightness increasing to the north-east.
The morphology at 20~\mic{} is similar, but we do not detect the peak that \citet{Hayward94} found 20\arcsec\ to the SW of the Ney-Allen nebula at wavelengths longer than 20~\mic.

\subsection{The BN-KL complex}
\subsubsection{Morphology}
In Figure~\ref{Fig8,9} we show the 10 and 20~\micron\ enlargements of the BN-KL complex from the data taken on December 18, 2000, whereas in Figure~\ref{Fig10} the same region is shown in the six narrow-band silicate filters. All images are absolutely calibrated and presented in logarithmic scale to enhance the faintest emission across the field.  A few artifacts are present, generally related to the extraordinary brightness of the BN peak: in the broad N-band image (Figure~\ref{Fig8,9}a) there are two ghost images of BN, approximately at (4.5\arcsec, $-$0.5\arcsec) and (0.5\arcsec, 6.0\arcsec), caused by inner reflections in the camera optics (see also Figure~\ref{Fig11,12}a). Bandwidth limitations of the readout electronics create in all narrow-band images an anomalous ``spike,'' approximately 4~pixels wide, protruding to the West of BN. Some extra noise is also visible in the highly stretched N and Q-band images as a horizontal periodic pattern and in the vertical direction, in the rows and columns corresponding to the BN peak. In general, however, the images have excellent quality and are diffraction limited at all wavelengths ($\lambda/D$ on UKIRT ranges from $0\farcs41$ at 7.7~\micron\ to $0\farcs67$ at 12.4~\micron, and $1\farcs08$ at 19.9~\micron). Due to the high surface brightness of the region, it is not easy to see the spectacular diffraction pattern of BN, with the exception of the first ring shown saturated in all images except the Q-band. Several diffraction rings appear in the differential images that we will present later in Section~4.3. To preserve the highest resolution, considering also that we are mostly dealing with the surface brightness of extended objects, we have not rebinned or convolved our images to our lowest resolution. In Figure~\ref{Fig13.} we show a color composite image of the BN/KL complex obtained combining our 7.7~\micron\ (blue), 12.4~\micron\ (green) and 19.9~\micron\ (red) images.
 
Early mid-IR studies \citep{BN67,Rieke+73} resolved this region into 8 sources (IRC1-8, BN corresponding to IRc1). Another 10  compact sources were added  by \citet{Gezari+98} to the IRc list, and recently \citet{SMB04} added five more. Our images have resolution and sensitivity comparable to those of \citet{Greenhill+04} and  \citet{SMB04} but cover the entire field in 8~bands. Our narrow-band 12.4~\micron~ image (Figure~\ref{Fig10}f), in particular, is comparable to Figure~1 of \citet{SMB04} and Figure~2 of \citet{Greenhill+04}.  We detect new features, especially in the unexplored outer regions of the complex.  Like the large majority of compact sources already cataloged, all new features appear extended and resolved into clumps, ridges, and arc-like structures down to our $\sim0\farcs5$ resolution limit. Instead of adding new entries to the already long list of IRc sources, we indicate the new sources using their offset ($\Delta$RA\arcsec, $\Delta$DEC\arcsec)  relative to BN. For the 30 IRc sources and sub-components already classified, we will still use the list of \citet{SMB04}. The position of the most relevant sources is indicated in Figure~\ref{Fig11,12}.

BN, the main luminosity peak longward of 2~\mic, remains unresolved at all wavelengths. Gaussian fits to the image profile at 10~\mic\  show a $\simeq3.2$~pixel full-width at half-maximum (FWHM) both in right ascension and declination, corresponding to $FWHM\simeq0\farcs87$. For an IR source with a spectrum increasing with wavelength,  like BN, the diffraction limit in broad-band filters is approximately
set by the longest wavelength passing through the filter, $\lambda_{max}\simeq13\mu$m in our case. For an effective telescope diameter $D=3.7$~m, it is $\lambda/D = 0\farcs72$. The pixel sampling and the mosaicing process, performed by integer pixels, add another $\simeq 1/2$ pixel of image blur, bringing the FWHM right to the measured value. The BN peak sits on top of an extended clumpy plateau. \cite{SMB04} have already identified the remarkable SW arc, and we just add to the list a compact source immediately to the south at ($-$2\arcsec, 2\arcsec).

Source {\sl n} of \citet{Lonsdale+82}, coincident with a bipolar radio source resolved by \citet[][their source ``L'']{Menten+Reid95}, has been recently analyzed in detail by \citet{Greenhill+04} and \citet{SMB04}. Using LWS on Keck, both team find that source {\sl n} is extended, with a {\sl FWHM} of approximately $(0\farcs75 \times 0\farcs50) \pm 0\farcs10$ and major axis oriented $\simeq 120^\circ$. 
Our pixel scale, approximately 3 times larger than that of LWS, does not sample adequately the core of the diffraction figure at 7.7~\micron. A two dimensional Gaussian fit at this wavelength gives for source~{\sl n} a {\sl FWHM} of $(0\farcs47 \times 0\farcs51)$, but it must be remarked that the Airy function is poorly reproduced by a Gaussian curve. If we use the BN peak as a reference for the PSF core and compare horizontal and vertical cuts across both sources, then source {\sl n} remains unresolved at all wavelengths in our images.
 
Other two sources, IRc12 and IRc7 appear point-like at 7.7~\micron\ but become extended at longer wavelengths. Sources IRc11 and IRc6, on the other hand, show compact cores at 12.4~\micron. The point source, MAX-69 (see Section~3.4), is clearly visible  at (11\arcsec,$-$6.5\arcsec), $\simeq20\arcsec$ south-east of BN. This is a well known double star which remains unresolved in our images
and most probably lies in the foreground of the BN/KL complex. 

Diffuse emission becomes brighter to the south-east with longer wavelengths. At 7.7~\micron, sources IRc2,  IRc4, and IRc5 appear aligned in the northeast-southwest direction and abruptly truncated to the south-east, but at longer wavelengths this sharp edge, especially for IRc4,  becomes less prominent and a broad plateau of emission raises around IRc8 and IRc22. At 19.9~\micron, extended emission fills the entire field and two dark lanes become clearly visible,  one oriented NS along a line roughly passing through positions (8\arcsec, 0\arcsec) and (3\arcsec, $-$18\arcsec), the other, almost perpendicular, immediately to the south of the ``equatorial belt'' composed of IRc14, Irc12, Irc11, Irc2 and Irc7. These dark lanes are coincident with the ammonia emission of \citet{Wilson+00}, which trace the density distribution of the Orion hot core \citep[see also]{Greenhill+04}. Another remarkable feature is the peculiar ``eastern ring" of sources centered at (10\arcsec, $-$4\arcsec) and traced by IRc16, IRc 15, Irc6E, IRc21, IRc2, IRc11, IRc12, IRc14, Irc13, plus other faint emission around (12\arcsec, $-$2\arcsec).  The area inside the ring is mostly faint clumpy emission and contains the majority of the H$_2$O masers detected by \citet{Johnston+89} in OMC-1. The similar shape of the prominent IRc3 and IRc4 sources is also noticeable; both are arc-like and point to the IRc2 region. The tails of their arcs are part of another remarkable new feature, a group of equally spaced knots tracing an arc-like ``necklace'' between IR5 and IRc3. The knots have all comparable brightness, are elongated and radially oriented and point to the IRc2 central region. 

Finally, IRc2 is well resolved at 10\micron\ (Figure~\ref{Fig8,9}a), and shows a morphology that matches the structure of sources 
IRc2-A to -D resolved by \citet{Chelli+84} and \citet{Dougados+93} at shorter wavelengths. Its morphology has been discussed in detail by \cite{Greenhill+04} and \cite{SMB04}. We will return on this object in Section~4.3.

\subsubsection{Photometry}
In Table~\ref{Tab:BN_sil}, we present the photometric data at 35 positions within the BN/KL complex. The first 30 are the sources listed in Table~1 of \cite{Greenhill+04}, the last 5 are the three brightest knots to the southwest of IRc4 (Knot~1-3), the compact source south of the SW~Arc (BN~SW) and the star MAX-69. Photometry has been obtained using a circular aperture of $1\farcs128$ diameter, corresponding to 2~square arcsec area, but the fluxes are normalized to Jy~arcsec$^{-2}$. The only exception is the BN peak, for which we give the total flux within a synthetic aperture of $5\farcs2$.  All values are uncorrected for the background emission, and the zero point errors listed in Section~2.2.3 represent the main source of uncertainty. The corresponding spectra are shown in Figure~\ref{Fig14}. The silicate absorption feature is prominent at all positions except for the very last object MAX-69, which is probably a foreground member of the Trapezium cluster.  A secondary maximum at the center of the absorption feature is also generally present, with the exception of the main BN peak and its immediate surroundings. This emission feature sits at the center of telluric ozone absorption band and, especially at the lowest signal levels, may be hard to detect in less than excellent weather conditions. Having carefully checked our absolute calibration, and  verified that our photometry for the main BN peak agrees with the previously published spectrophotometric data (see Section~4.3), we believe this emission feature is real. 

\subsection{OMC-1 South}
The OMC-1S region, located $\simeq100\arcsec$ south of BN/KL complex, is characterized by intense outflow activity probably generated  by deeply embedded sources \citep{ODeARAA01}. In Figure~\ref{Fig11}, we show the position of the point sources found at shorter wavelengths together with the contours of the high-velocity outflow of \citet{Rodriguez+99b}. Part of the field is also visible in Figure~\ref{Fig4,5}. The \citet{Lada+00} list of candidate protostellar sources contains five objects in this area, TPSC-1, TPSC-2, TPSC-16, TPSC-48, and TPSC-78. Three of them, namely TPSC-1, TPSC-2, and TPSC-78 are the sources with the reddest $K-L$ color of the entire L-band survey: the first two, undetected in the K-band by \citet{Lada+00}, have lower limits at [K-L]$>6.09$~mag and [K-L]$>4.90$~mag, respectively, whereas TPSC-78 has [K-L]$=4.85$~mag. TPSC-1 is nearly coincident with our source, MAX-61 
that shows significant variability at 10\micron, with $[N]=4.40$~mag in November 1998 and $[N]=3.68$~mag in December 2000 (see Section~4.5.1).  Object TPSC-2 is MAX-64, and with $[N]\simeq5.3$~mag it also shows a conspicuous $L-N\simeq 3.8$~mag color index.

\subsection{Point sources}
In Figure~\ref{Fig12}, we map the location of the 177 point sources detected at 10 and 20~\micron. In Table~\ref{Tab_stars10mic}, we 
present their cross-identification, location and 10~\mic~ photometry. In particular, we list their sequential running number (column a) and their entry number in the catalogs of \citet{LAH97}, \citet{HC00}, and \citet{Muench+02} (column b-d), all at shorter wavelengths; coordinates (columns e,f) are derived from our final mosaic with typical accuracy $\simeq0\farcs5$, accurate enough to unambiguously identify each source. In a few cases, residual pointing errors at the edges of the mosaic cause larger discrepancies, listed in the last column~(l). The magnitudes are given for the two epochs (columns g,h). For a few sources, we could derive photometry only in the final mosaic (column~i). For 22 sources, there is no counterpart identified at shorter wavelengths. Those without counterpart can therefore be regarded as deeply embedded objects. A few of them, however, are close to our sensitivity limits and appear only on the final mosaic.
As explained is Section~2.2.1, we cannot rule out in these cases the possibility that they are artifacts of our data reconstruction algorithm. Further observations are needed to confirm them as real sources.

\section{Discussion}
\subsection{Diffuse emission}
With the exception of the bright BN/KL complex and the Ney-Allen nebula, the morphology of the Orion Nebula in the mid-IR is more similar to the radio continuum and dereddened line images obtained by \citet{ODell+Yusef-Zadeh00} than to the optical or near-IR images. This because the foreground ``veil'' of neutral material immediately in front of the nebula, which affects the morphology of the nebula in the visible, becomes optically thin at mid-IR (and radio) wavelengths, whereas the stellar emission, prominent at shorter wavelengths, becomes fainter. For this reason, the Dark Bay, a region of high optical depth ($A_V\ge3$) approximately parabolic
in shape, that is protruding from the east of the nebula up to $\approx 30\arcsec$ east of the Trapezium, is essentially invisible in our images, confirming its nature as an enhancement of dust column density within the foreground neutral lid (O'Dell \& Yusef-Zadeh 2000). 

The filamentary structures seen in emission at 10~\mic~ appear to be associated to the extinction-corrected color images of \citet{ODell+Yusef-Zadeh00}. In particular, the ``East Bright Bar'' and ``E-W Bright Bar'' shown in Figure~5 of \citet{ODell+Yusef-Zadeh00}
have 10~\mic\ counterparts. On the other hand, their ``North Bright Bar,'' which is close to the top edge of our 10~\mic\ field, remains undetected, as well as the comet-like structure approximately $1\arcmin$ to the east of the Trapezium, but they are clearly visible on the L~band images of both \citet{Lada+00} and \citet{MJM+96}. Most of the 10~\micron{} filaments appear aligned along arcs roughly centered on the Trapezium. This result suggests that the filaments trace the edges of the main HII cavity rather than, e.g., shocks produced by collimated outflows. The morphological similarity between the filaments and the bars seen in the radio and in extinction corrected line images indicates that they are also associated with the folds in the ionization front seen edge-on. Most probably, the 10\mic\ emission is dominated by PAH emission in the photodissociation region. \Citet{Sloan+96} have shown that the 11.3\mic\ PAH feature, produced by the aromatic C--H bonds when a grain is transiently heated through the absorption of a UV photons, dominates the mid-IR spectrum of the Orion Bar. Similarly, the great arcs visible at 10~\mic\ approximately 45\arcsec\ south of the Trapezium (Figure~\ref{Fig4,5}a,b), reminiscent of the pillars seen in M16, may result from the advance of the ionization front between the dense and clumpy medium of the surrounding molecular cloud.

\subsection{The Trapezium region}
The complex arc-like morphology of the mid-IR emission in the vicinity of \thC\ (Figure~\ref{Fig6,7}a,b) has clear optical counterparts in recombination lines, first detected by \citet{McCullough+95}. Using [OIII] and H$\alpha$ line images taken with the {\sl HST}, \citet{Bally+98} found that similar arcs are present around many proplyds within 30\arcsec{} from \thC{}. \citet{Bally+00} also found arcs in the outskirts of the Orion Nebula, especially in the LL~Ori region (which remains outside of our imaged field). \citet{Hayward+94} first suggested the arcs are the bow shocks produced by the interaction of the fast ($\simeq500~\kms$) wind from \thC{} with the slow ($\simeq10-50~\kms$) photoevaporated flows emerging from the ionization front at the proplyd surfaces. This interpretation is supported by proper motion \citep{Bally+00} and spectroscopic \citep{Henney+02} measurements, showing that these are stationary structures.
The wind-wind interaction originating the arcs has been studied analytically by several authors \citep[e.g.][and reference therein]{Canto+96}, and numerically by \citet{Garcia-Arredondo+01}. There are in fact two basic configurations: the interaction of two spherical winds, of interest in our case, and the interaction of a spherical wind with a plane-parallel flow. The latter  is probably the case in the outskirts of the Orion nebula, where the wind-wind interaction is occurring between T-Tauri winds and the plasma flowing from the main ionization front \citep{Bally+00}.

The analytical analysis of \citet{Canto+96}, which leads to an exact solution of the wind-wind interaction  within the ``thin shell''  approximation, provides an approximate solution that can be easily compared with observations. The ratio of the wind momenta,
\begin{equation}
\beta={\dot{M}_1v_1 \over \dot{M}_2v_2},
\end{equation}
where $\dot{M}_{1,2}$ and $v_{1,2}$ are respectively the mass loss rates and terminal velocity of the two winds, governs both the location of the stagnation point and the shape of the arcs. The ratio between the distance from the stagnation point to the source inside the arc, $R_0$, and the separation of the two outflow sources, $D$, is given by
\begin{equation}
{R_0\over D}={\beta^{1/2} \over {1+\beta^{1/2}}}.
\end{equation}
The shape of the bow shock is characterized by the asymptotic angle $\theta_\infty$ of the tail, $\theta_\infty$, given by
\begin{equation}
\theta_\infty -\tan\theta_\infty={\pi \over 1-\beta}.
\end{equation}
If the bow shock axis is tilted with respect to the line of sight by an angle, $\theta_i$, the projection effect increases both the apparent ${R_0\over D}$ ratio and the apparent asymptotic angle. For large tilt angles, the entire back side of the bow shock becomes visible and the characteristic arc-like geometry disappears. In these cases, the bow shock will look like a diffuse source in the vicinity of the central star.

Whereas the ratio $R_0/D$ can be easily obtained from the data, it is difficult to estimate the asymptotic angle since the surface brightness of the arcs decreases in the tail region. For this reason, we characterize the shape of the bow shock through the apparent distance, $R_{90}$, between the central star and the arc at $90^\circ$ elongation from the direction of \thC. It turns out that different combinations of $\beta$ and $\theta_i$ generate morphologies with ratios $R_0/D$ and $R_{90}/D$ that vary within a relatively narrow range. This variation is shown in Figure~\ref{Fig13}, where we have plotted the combinations obtained for different values of $\beta$ and $\theta_i$. The bow shock morphology, limited to the cases with $\beta=$0.1, 0.04, 0.004, and  $\theta_i=30^\circ$, is also shown to visualize the change of shape along the region.

In Figure~\ref{Fig13}, we also plot the values of $R_0/D$ and $R_{90}/D$ measured for the Ney-Allen nebula and the five LV arcs. In the case of LV1, LV4, and possibly LV5, the parameters are compatible with the thin shell model. But in the other three cases, namely LV2, LV3, and the Ney-Allen nebula, there is a clear disagreement. In correspondence to the observed values of $R_0/D$, the values of $R_{90}/D$ should be much larger, because if the two winds carry a similar momentum ($\beta\simeq1)$ and the stagnation point moves closer to \thC, the bow shock is expected to become more and more open. Instead, the three arcs with stagnation point closer to \thC\, show a roughly circular shape, with $R_0\simeq R_{90}$.

The intrinsic limitations of the algebraic solution of \citet{Canto+96} based on the thin shell approximation cannot explain this anomaly. \citet{Garcia-Arredondo+01} have shown that, despite the simplifying assumption of stationary flow and momentum conservation, the analytic treatment nicely matches their detailed hydrodynamical simulation.  A possible explanation for the disagreement may better be provided by the detailed morphology of the arcs. The LV 3 arc, in particular, is clearly asymmetric with respect to the direction to \thC. It becomes narrower and brighter to the south in the direction of SC3, with a secondary maximum evident to the NE, at $\simeq90^\circ$ from the \thC{} direction. The asymmetry indicates that other dynamical effects, not accounted for by the standard models, are taking place. A relative motion of LV3 relative to \thC{} would break the cylindrical symmetry assumed by simple wind-wind models. A rather modest  proper motion, of the order of a few \kms{}, would be enough to add a significant tangential component to the speed of the photoevaporated flow, which is of the order of the sound speed in a typical ionized gas, $c_s\simeq10$~\kms. Other dynamical effects such as the interaction with the collimated mass outflows produced by the central LV stars may also locally affect the morphology and surface brightness of the arcs, leading to inaccurate estimates of the opening angle.

The bright arcs of the Ney-Allen nebula are peculiar also from another point of view. Let us compare the six brightness profiles along the cuts joining \thC{} to $\theta^1$Ori~D and the LV1-5 sources (Figure~\ref{Fig18}). The LV1, LV3, and LV4 arcs show, on the side facing \thC, brightness distributions at 10 and 20~\micron{} very similar to each other. The LV2 and LV5 arcs, on the other hand, are surrounded by the diffuse luminosity, mostly originated by the Ney-Allen nebula and by the LV4 arc, and their 10 and 20~\mic{} profiles cannot be easily compared. The position of their peaks, however, is the same at both wavelengths, suggesting that also for them the 10 and 20~\micron{} emission may be spatially coincident. The fact that the LV1-5 arcs show a similar morphology at both wavelengths supports our previous hypothesis that the thin shell is a good approximation and that the discrepancies between models and observations can be ascribed to relative motions.  The case of the brightest arc (Ney-Allen nebula), however, is different. Figure~\ref{Fig18}a shows that  the side facing \thC{} does not show a sharp edge at 10 or at 20~\mic. Moreover, the 20\mic{} emission is more extended in the direction of \thC, indicating that a colder dust component becomes brighter at larger distances from the central star. The intensity profiles therefore suggest that (a)~the emission is not concentrated in a thin shell, and (b) the UV flux from \thC\ is not the dominant heating source. Departures from the thin shell approximation have theoretical ground. In particular, \citet{Garcia-Arredondo+01} have shown that if the stellar wind passes through an internal reverse shock, the wind-wind interaction takes place in a sub-sonic regime. In this case, a turbulent mixing layer develops at the interface between the two winds, and filaments and ``fingers'' are produced by the entrainment of the fast wind within the dense subsonic material. Instead of the sharp decay typical of a thin shell, the drop of surface brightness will appear in this case more gradual, and extended.  Among the sources in the vicinity of \thC, this scenario applies naturally, and exclusively, to the Ney-Allen nebula, as it is associated to the B0V star $\theta^1$Ori~D. This hot star originates a substantial stellar wind, and the formation of a internal reverse shock where the gas flows enter a subsonic regime is a natural  outcome in the early adiabatic phase of expansion \citep{Castor+75,Weaver+77}.
 
The main peculiarity of the Ney-Allen nebula, however, is that it contains in an extremely harsh environment  a considerable amount of dust, estimated between $\simeq2\times10^{27}$~g if $T_d=200$~K and $\simeq1\times10^{26}$~g if $T_d=350$~K \citep{Hayward94}. Proplyds in Orion are supported by the photoevaporation of circumstellar disks around low mass pre-main sequence stars \citep{Churchwell+87,Meaburn88,HA98,Johnstone+98,H+CRO99}. In analogy, it is natural to think to a massive circumstellar disk around $\theta^1$Ori~D as the most natural dust reservoir for the Ney-Allen nebula. The presence of a massive circumstellar disk is compatible with the standard view that high-mass stars reach the main-sequence while they are still in their protostellar accretion phase \citep{Stahler+00}.  In this scenario, it is important to notice that $\theta^1$Ori~D is the only star of the Trapezium without a binary or multiple companion, down to a resolution limit $\lambda/D = 57~$mas and 76~mas in the H and K~bands, respectively \citep{Preibisch+99}. 
Conversely, no other Trapezium stars seems to be associated with excess mid-IR emission. This result may therefore indicate that for OB stars the presence of low mass companions is not compatible with substantial circumstellar disk, and vice versa.

Concerning the SC3 source, i.e. the bright extended blob in the immediate vicinity of \thC, we have discussed  in a previous paper \citep{Robberto+02} its apparent morphological anomaly, as it appears undisturbed by the powerful \thC{} wind despite having the shortest projected distance from \thC. On the basis of its high mid-IR brightness, we have proposed that it may represent a disk on the far side of the Nebula, oriented face-on with respect to both  \thC\ and the Earth. High spatial resolution mid-IR observations, possibly with interferometric techniques, should easily resolve the radial distribution of its bright mid-IR emission and allow us to clarify the nature of this most enigmatic object.

\subsection{The BN/KL complex}
\subsubsection{Main physical parameters}
The ubiquity and strength of the silicate absorption feature in the 10~\mic\ region indicates that foreground absorption plays a dominant role in shaping the observed morphology of the BN/KL complex. As a baseline, one can therefore assume a two component model \citep{Aitken+81}, where a background blackbody source is shielded by a foreground absorbing medium too cold to contribute significantly to the mid-IR emission. This assumption allows us to readily obtain a map of the foreground opacity from the line to continuum ratio at 9.8~\micron. In a first order approximation, the continuum can be derived from the interpolation at 9.8~\micron\ of the fluxes measured at 7.7 and 12.4~\micron, whereas for the depth of the silicate feature we interpolate between the fluxes measured at 8.7 and 10.2~\micron~ to avoid the secondary maximum seen in the 9.6~\micron~ filter. The natural logarithm of the absorption line strength is the optical depth and is presented in Figure~\ref{Fig19}. The optical depth in the main central region increases from west to east across the entire field, peaking in the immediate vicinity of the dark lane. It reaches the maximum value at the IRc2 position, $\tau_{9.8}=3.37$ (unless otherwise specified, all values are per pixel, i.e. averages over an area $\simeq0\farcs06$, corresponding to $A_{9.8}=3.66$. This is $\simeq 20\%$ lower than the value reported by \citet{Gezari+98}. The drop of extinction coinciding with the dark lane is due to the absence of flux sufficiently high to allow for a meaningful estimate of the extinction and must therefore be regarded as a selection effect.

The map of the peak optical depth allows us to correct all images for the foreground extinction. There are several possible choices for the spectral profile of the silicate feature. We adopt here the ``Orion Trapezium" emissivity \citep{Forrest+75}, as updated by the spectroscopic observations of \citet{Hanner+95}. We average the \citeauthor{Hanner+95} emissivity over the (cold) filter transmission profiles provided by the manufacturer. We also include the non-silicate extinction, that we parameterize in terms of the visual extinction, $A_V$, using a fit to the extinction data of \citet{RiekeLebofsky85}, $A_\lambda = 0.381\cdot A_V\cdot\lambda^{-1.652}$. Following \citeauthor{RiekeLebofsky85}, we weight the two contributions by assigning to the silicate extinction at the 9.8~\micron\ peak a $90$\% fraction of the total extinction at that wavelength.

The morphology of the BN/KL complex after extinction correction for all 6 silicate filters is shown in  Figure~\ref{Fig20}.  As expected, the changes with respect to the unreddened images (Figure~\ref{Fig10}) are stronger where the silicate absorption feature is more prominent. In particular, at 9.8 and 10.2~\micron, the IRc2 object recovers its rank of second brightest source at all wavelengths after BN peak. In Figure~\ref{Fig21}, we zoom in on the IRc2 region, comparing for all 6~silicate filters the morphology before and after extinction correction.  The most striking result is the appearance of a point source just to the south of the IRc2 peak. It was visible in the raw  7.7~\micron\ images, but now becomes evident also in the dereddened 9.6 and 10.2~\micron\ images. Its position is coincident with knot IRc2-A of \citet{Dougados+93} and lies $\simeq 0\farcs5$ north of radio source I of \citet{Menten+Reid95}, which remains undetected.  The fact that in the dereddened images the  mid-IR peak is visible at 7.7~\micron, invisible at 8.7~\micron, visible again 
in the dereddened 9.6 and 10.2~\micron\ images may be due to a combination of imperfect extinction correction, intrinsic silicate emission and loss of resolution at the longest wavelengths. 

From the dereddened images, we derive the distribution of the 10~\micron\ color temperature of the underlying emission by fitting, pixel by pixel, a blackbody curve to all 6 silicate filters. The result is shown in Figure~\ref{Fig22}. The temperature across the BN/KL complex typically ranges between 180~K and 300~K, with maxima at the BN peak with $T_{col}\simeq1000$~K, and at the IRc2 position with $T_{col}\simeq420$~K. The BN temperature is certainly overestimated. Since BN is unresolved and our images are diffraction limited, the encircled energy decreases with wavelength. The dilution of flux over a larger number of pixels causes an apparent increase of the color temperature at the peak. Concerning IRc2, the warm dust appears mostly elongated in the northeastern-southwestern direction, i.e. in a direction parallel to the dark lane, and peaks $\simeq0\farcs5$ to the southeast of source~I. This effect is shown in Figure~\ref{Fig22}, where we overplot the contours of the dereddened 7.7~\micron\ image at 1.5 and 3.0 Jy~arcsec$^{-2}$ flux levels.

The bolometric luminosity can finally be estimated from the $L=\sigma T_{col}^4 \cdot 4\pi R^2$ relation, where $R$ is the radius corresponding to 1~resolution element ($0.26\arcsec \simeq 120~$AU at the distance of the Orion Nebula). We do not show its spatial distribution, which is very similar to Figure~\ref{Fig22}  with an even more prominent role of IRc2 and BN.

In Table~\ref{Tab:BN_phys}, we list the values of the three parameters we have derived---optical depth at 9.6~\micron, color temperature in the 7.7--12.4~\micron\ band and total luminosity---for the sources with photometry presented in Table~\ref{Tab:BN_sil}. The values of $\tau_{9.6}$ and $T_{col}$ are averages over a circular aperture of 1~arcsecond square area, whereas the luminosity, $L$, is integrated over the same area.  Table~\ref{Tab:BN_phys} also reports the temperatures and luminosities derived from the dereddened 10 and 20~\micron\ broad-band filters having assumed at 20~\micron\ the interstellar extinction law of \citet{Mathis90}. The color temperature between these two wavelengths (125--140~K) is typically 80~K lower than that derived from the silicate filters (200--230~K). The discrepancies up to a factor of 10 in the estimated luminosities reflect these differences.

The elementary technique we have used so far, based on a crude estimate of the extinction, still provides maps of the spatial variations
that are quite robust, since the data have been taken all at the same time and calibration errors are systematic. In the next section we shall discuss a more refined model to investigate further the characteristics of the mid-IR emission of the BN/KL complex.

\subsubsection{Silicate absorption feature}
The two component model has been successfully used in the past to model the 10~\micron\ spectrum of the main BN peak \citep{Aitken+81,Aitken+85}. Figure~\ref{Fig23} compares our broad-band photometry of BN calculated over a synthetic aperture of $5\farcs2$ diameter with the spectrophotometric data of \citet{Aitken+85} and the synthetic photometry of \citet{Gezari+98}.  The agreement between different data sets is good, considering that the broad-band values have not been color corrected and the differences in 
aperture diameter between different authors. The solid and dotted lines represent the locus of a blackbody curve at $T_{col}$ before and after reddening by an optical depth $\tau_{9.8}$ to fit our photometry. We use the Trapezium emissivity of \citet{Hanner+95} with the additional $\lambda^{-1.65}$ continuum discussed above. The best fit to our data has $T_{col}=373$~K and $\tau_{9.8}=1.37$, with a filling factor $FF=5.41\times10^{-4}$ to normalize the absolute flux levels. From the filling factor, we derive a photospheric radius $R_{BN}=27.2$~AU at 450~pc, and the resulting luminosity is $L_{BN}=4\pi R_{BN}^2\cdot\sigma T_{col}^4=600$~L$_\sun$.  For comparison, \cite{Aitken+81} derive $T_{col}=336~$K and $\tau_{9.8}=1.54$, whereas the total luminosity of BN is $L_{BN}\simeq(1-2)\times10^4$L$_\sun$ \citep{Scoville+83}).

It is known that, besides BN peak, the two component fit  does not account for the detailed shape of the 10~\micron~ spectrum of the other compact sources of the BN/KL complex \citep{Aitken+85}. The 8-13~\micron~ absorption feature appears to change across the complex, both in depth and in shape \citep{Aitken+81}. These variations reflect different physical conditions in the absorbing medium. 
\citet{Downes+81} and \citet{Werner+83} first indicated that most of the IRc sources discovered in the BN/KL complex are not self-luminous, but rather extended clumps of diffuse emission. \citet{Genzel+Stutzki89} indicated that the BN/KL complex is dominated by the radiative transport of IR radiation in a highly non-uniform medium. A further complication may arise from the fact that at shorter wavelengths scattering becomes important. \citet{Werner+83} were the first to report that at 4.8~\micron\ the BN/KL complex is predominantly a reflection nebula, but since the albedo of astronomical silicates drops by two orders of magnitudes between 4.5 and 8~\micron\ \citep{Draine85}, one can still assume that scattering is negligible at mid-IR wavelengths. In this case,  the basic two component model can be modified to account for the following effects:
\begin{enumerate}
\item
The background emission is normally optically thin. Only for a few sources, like BN and IRc2, 
the background emission may be represented by as a stellar photosphere multiplied by a proper filling factor.
\item
The presence of foreground absorbing medium (as in the two component model).
\item
A third component, optically thin, needed to account for silicate emission at 9.6~\micron.
\end{enumerate}
In a slab geometry, these three factors are described by the equation:
\begin{equation}
F_\nu = FF * B_\nu(T_b)\left[{1-\exp^{-\tau_b(\nu)}}\right]\exp^{-\tau_a(\nu)} + B_\nu(T_s)\left[{1-\exp^{-\tau_s(\nu)}}\right]
\end{equation}
Here $FF\leq 1$ and $T_b$ are the filling factor and temperature of a background medium with optical depth, $\tau_b$.  For a photosphere (e.g. BN-peak), it is $\left[{1-\exp^{-\tau_b}}\right]\simeq 1$; $\tau_a$ is the optical depth of the cold foreground absorbing material, that we assume is mostly located in the vicinity of the background sources (i.e. part of the Orion molecular cloud);  $T_e$ and $\tau_e$ are the temperature and optical depth of the foreground gas responsible for the silicate emission.  We locate this third component in the foreground of the main absorbing layer, assuming that it physically corresponds to warm dust dust heated
in the foreground cavity, i.e. contained in the optically visible Orion Nebula.

Equation~(4) contains six free parameters, and the choice of different emissivity laws allows for some extra freedom. For this exploration, we adopt the silicate emissivity law of \citet{Jaeger+94}, which provides excellent agreement with the spectra of massive young stellar objects and the Ney Allen nebula in what concerns the position of the 10~\micron\ peak and has a relatively narrow 
width that may provide the best agreement with the most recent measures of the Trapezium (see below).  Figure~\ref{Fig23} shows the best fit to the flux (averaged over 4~pixels) at the IRc2 peak. To allow the numerical routine to converge, we had to reduce the degree of complexity of our system. Thus, the dotted line represents the fit to the 7.7, 8.7, 10.2, 11.6 and 12.4 bands, i.e. all the silicate
filters except the 9.6~\micron\ one, provided by the first term at the right-hand side of Equation~(4),  having assumed $FF=1$. We obtain $T_b=710$~K, $\tau_b=0.0019$, and $\tau_a=21.6$. The dashed line represents the contribution from the second term at the right-hand side, having assumed {\sl ad-hoc} $T_s=300$~K and $\tau_s=1.5\times10^{-3}$ to approximately match the photometry
in the central part of the silicate feature. The thick solid lines represent the sum of the two contributions. Figure~\ref{Fig23} shows that
the 20~\micron\ point cannot be matched by a single blackbody temperature, due to the strength of the secondary silicate absorption peak at 18~\micron. At 10~\micron, the emissivity law of \citeauthor{Jaeger+94} is too broad to match the 8.7 and 10.2~\micron\ 
points, but the addition of the third component in emission provides a reasonable fit, including the 9.6~\micron\ point. 
A detailed look shows that the emissivity law of \citet{Jaeger+94} is still too broad to reproduce at the same time the 9.6~\micron\ peak
and the adjacent 8.7 and 10.2~\micron\ points. The finite filter band-pass, together with the presence of a range of temperatures instead of the assumption of a single 300~K blackbody, exacerbates the problem as both effects tend to reduce the observed sharpness of the feature. To illustrate the effect of the bandpass, we have convolved the sum of the two contributions with a $\Delta\lambda=1$~\micron\  rectangular bandpass. The result is presented as a thin solid line in Figure~\ref{Fig23}. Similar results are found for other sources, and in Figure~\ref{Fig23}b we show as an example the case of  IRc7, where we find $T_b=232$~K, $\tau_b=0.04$, $\tau_a=14.2$, and $\tau_s=0.0035$.

The value $\tau_{9.8}\simeq22$ we have found for IRc2 corresponds to $A_v\simeq 400$ and is compatible with the extreme value of the optical depth $\tau\gtrsim300$ at 8.0~\micron~ at the near ammonia core estimated by \citet{Greenhill+04}. A general emissivity law with a narrower peak would provide higher values of column density. On the other hand, there is no reason to assume {\sl a-priori} 
that the emissivity is the same for the three components, or in general across the region. There is evidence that, for young stellar objects and star forming regions, the 10~\micron\ silicate feature depends on the environment, due to the interplay between dust column density, cosmic ray irradiation and grain coagulation \citep{Bowey+03}. In particular, the shape of the canonical ``Trapezium emissivity'' of \citet{Forrest+75, RocheAitken84, Whittet+88}, generally assumed as a reference for molecular clouds and used to constrain the optical constants of the popular \citet{DraineLee84} model, shows a dependence on the spatial scale. Unpublished arcsecond resolution spectra of the Trapezium obtained by T.~Hayward \citep{Bowey98} show that the peak of the 10~\micron\ feature
is significantly  narrower (0.5-0.6~\micron) than that reported by \citet{Forrest+75} and \citet{Hanner+95}, these last authors with a $9\farcs4$ beam (beam size is unquoted in the \citeauthor{Forrest+75} paper). Different broadening may be due to grain size effects or to changes in the grain structure or optical properties \citep[see][for a discussion]{Jaeger+94}. \citet{Bowey98} suggests that measures made with large aperture, sampling a wide range of dust temperatures, may also cause a broadening of the 10~\micron\ silicate feature, at least in the case of Trapezium. If the optically thin emission is due to dust grains exposed to the UV radiation, the grain structure may be affected in the direction of a predominantly crystalline composition. Amorphous silicates  may be converted to crystalline minerals when annealed (heated) in vacuum at high temperature \citep{Day+Donn78}. The Trapezium grains are also exposed to strong UV radiation, but since they may have been only recently released by the disk surrounding $\theta^1$-D, they may still represent a relatively standard mixture of amorphous and crystalline silicates. On the other hand the dust grains inside the Orion nebula, and those distributed in front of the BN region in particular, could have been exposed for a longer time to the nebular UV radiation. It is intriguing to notice that the reddening law in Orion has $R_V=A_v/E(B-V)\simeq5.5$ vs.\ the standard value  $R_V=3.1$ of the interstellar medium. This anomaly could be related to the peculiar evolution of the dust grains in the harsh nebular environment.

\subsubsection{The role of IRc2 and source {\sl n}}
IRc2 has been considered for a long time as the mid-IR source more closely associated to a deeply embedded center of activity of the region, e. g. the origin of the spectacular outflow phenomena emerging from the BN-KL complex.  Evidence in this direction comes from the position of IRc2 at the center of (1) an expanding cluster of maser sources with velocity centroid nearly coincident to the LSR velocity of the hot core \citep{Genzel+81}; (2) a well defined polarization pattern \citep{Aitken+97,Chrisostomou+00}, and (3) a rich outflow phenomenology observed at near-IR wavelengths \citep{Johnes+Walker85}. Source {\sl I} of \citet{Menten+Reid95}, a compact HII region approximately 1\arcsec~ south of IRc2, has been more recently indicated as the actual center of activity of the BN/KL complex \citep[see][for a recent review]{Tan04}. Whether IRc2 is just a complex of knots heated by source {\sl I} or, as \citet{Greenhill+04} speculate, a group of individual stars associated with source {\sl I} in a young multiple system, remains unclear. An intermediate case could also be considered, with source {\sl I} and the IRc2-A peak representing deeply embedded individual sources, and the other IRc2 knots being diffuse sources heated mostly by IRc2-A peak.

Source {\sl n}, whose radio emission has been resolved into a bipolar nebula by \citet{Menten+Reid95}, has also recently gained a prominent role. It is only 3\arcsec away from IRc2, lies exactly at the center of expansion of the $H_2O$ masers of \citet{Genzel+81}, and within the narrow $2\farcs0 \times 0\farcs5$ strip where \cite{Gaume+98} locate their ``shell'' masers. Both \citet{SMB04} and \citet{Greenhill+04} have been able to resolve this source at 10~\micron~ with Keck. It appears slightly elongated in a direction perpendicular to the axis of the bipolar radio source. \citet{SMB04} suggest that source {\sl n} lies at the center of a bipolar cavity excavated by the same outflow responsible for the reach maser phenomenology. The presence of a compact HII region clearly suggest that this is a young early type star. \citet{Greenhill+04} estimate a luminosity $L\sim2000$~L$_\sun$ (mid-B type), but also notice that the recent detection of hard-X ray emission is unusual for early-type stars.

\citet{Gezari+98} have used mid-IR photometric data similar to ours to downplay the role of IRc2 vs. source {\sl n} as the dominant energy source of the region. In particular, \citet{Gezari+98} estimate for IRc2 a total luminosity $L\simeq 10^3$L$_\odot$,
approximately 2 orders of magnitude lower than the previous estimates \citep[e.g.][]{Downes+81,Wynn-Williams+84}. One may question the reliability of a technique that provides values different by orders of magnitudes in front of a rather modest gain in spatial resolution. In fact, these estimates must always be taken with great caution. First, we have seen in the previous paragraphs that different analysis techniques may lead to very different values for the optical depth at 9.8~\micron\ ($\tau_9.8=3.37$ vs. $\tau_a=21.6$), even if both use the same data set (in our particular case, both neglect the 9.6~\micron\ point). It is also known that the commonly used relation, $L=4\pi R^2\sigma T^4 \epsilon$, where $R$ and $T$ are the source radius and temperature and $\epsilon$ represents the dust emissivity, may lead to a significant underestimate of the bolometric luminosity if one assumes the IR color temperature as the blackbody temperature of the source \citep{Panagia75}.  A more subtle effect may also be present, since the same relation also assumes that the size of the underlying photosphere does not change with wavelength. In the case of IRc2, however, if the absorbing medium is clumpy or in general inhomogeneous, an unresolved source may have a filling factor that increases with wavelength as the foreground absorbing material becomes more and more transparent. The underlying source will then appear redder, and the estimated luminosity will be lower due to the strong dependence on the color temperature. In particular, a value $A_V=60^m$ corresponds to $A_{9.7}\simeq 3.4^m$ at 9.7~\micron\ and to $A_{20}=1.3^m$  at 20~\micron, using \citet{Bessel+Brett88} IR reddening law. Thus, IRc2 can be nearly optically thick at 10~\micron\ and nearly optically thin at 20~\micron. In these conditions, the mid-IR morphology is determined by a complex combination of thermal emission and variations of foreground extinction. Further multiwavelength observations will need to be coupled with 3-d radiative transfer models to disentangle each contribution.

\subsection{OMC-1 South}
The Orion-S region is characterized by the rich phenomenology typically associated with massive star formation: warm gas \citep{Ziurys+81}, large dust column densities \citep{Keene+82}, H$_2$O masers \citep{Gaume+98}, molecular outflows \citep{Schmid-Burgk+90,Rodriguez+99a}, and several optical jets and Herbig Haro objects \citep{Bally+00}. With 9 mid-IR sources in less than $1\arcmin\times1\arcmin$ field (Figure~\ref{Fig11}), this is a rather dense area and the scale of the outflows, up to a projected distance of $\simeq100\arcsec=0.21$~pc in the case of HH203/204 often makes it difficult to unambigously associate each optical outflows to a well defined driving source.  We have already mentioned the correspondence between the reddest 3.5~\micron\ sources, TPSC-1 and TPSC-2 of \citet{Lada+00}, with MAX-61 and MAX-64, respectively. A system of outflows that includes HH 529 and HH 269, and possibly HH 528, appears to depart from an ``optical outflow source'' \citep[named OOS by][]{CRO+Doi03} coincident, within the error ellipse, with both sources. Based on their proper motions, \citet{CRO+Doi03,CRO+Doi03err} estimate for the OOS coordinates
RA(2000.0)=$5^h35^m14\fs56$, DEC(2000)=$-5^{\circ}23\arcmin54\arcsec$, within $\sim 3\arcsec$ from the position of MAX-61 and MAX-64\footnote{\citet{Zapata+04} base their comparison with the present paper on a preliminary version of the manuscript. In this
final version their sources MAX-58 and MAX-60 are now listed as MAX-61 and MAX-64}. MAX-61, which is radio source 144-351 of \citet{Zapata+04}, is the most viable candidate, since our images (Figure~\ref{Fig4,5}) show two elongated filaments immediately to the east and west of MAX-61 with the same orientation of the HH~529/HH~269 outflows. These are most probably the jets emerging from the source, with the mid-IR emission due to entrained dust on their outside, along with gas. \citet{Doi+04} combine proper motions with new radial velocities of the outflows to determine that the optical jets originate from a source lying only about 0.02 pc within the PDR. Their relations between proper motion, radial velocities and the mid-IR emission would provide a self-consistent picture of this activity center, 
except that both systems of HH objects appear blue-shifted and breaking through the ionization front, a rather unusual circumstance that may indicate a more complex scenario. A curvature of the outflow toward our line of sight could result from the interaction of the jet with some stellar wind, originated e.g. from the Trapezium stars or from the nebular ionization front. This is the hypothesis of \citet{Smith+04}. It is clear that if the jet direction can be deflected by wind interaction, the identification of candidate outflow sources on the basis of proper motion and jet/counterjet alignment becomes more uncertain.  In fact, \citet{Smith+04} cast doubts about the association of HH~269 with HH~529 and rather suggest that HH~269 is the counter lobe of the HH~203/204 pair, with significant deflection with respect to both the plane of the sky and our vantage point.

Concerning the other prominent optical outflows, both \citet{Zapata+04} and \citet{Smith+04} argue that HH~202 is associated to HH~528 
rather than HH~203/204, as suggested by \citet{Rosado+01}. \citet{Zapata+04} suggest that this system originates from their radio source 143-353. At this position, we detect an extended knot of 10~\micron\ emission shown saturated in Figure~\ref{Fig4,5} and not included in our list of 177 point sources. \citet{Smith+04} suggest a different scenario, with the HH~202 jet associated to their source nr.~1 (MAX-45), but they also note that neither MAX-61 nor MAX-64 cannot be ruled out.

In what concerns the molecular outflows, the $\lambda=1.3$~mm bolometer maps of \citet{Mezger+90} show a peak of dust continuum brightness, source FIR4, located at RA(2000)=$5^h35^m13\fs3$, DEC(2000)=$-5^{\circ}24\arcmin12\farcs5$. Low velocity molecular outflows depart from the vicinity of FIR4 \citep{Ziurys+90,Schmid-Burgk+90}. A second, very young (dynamical age of $\sim 10^3$ years), very fast ($\sim 110$~\kms) and very compact ($\lesssim 0.16$~pc) molecular outflow has also been discovered by \citep{Rodriguez+99a,Rodriguez+99b}. The geometry of the high velocity gas suggests that its powering source is located $\approx20\arcsec$ north of FIR4.  We detect no IR sources within a $12\arcsec$ radius from FIR4, in particular there is no mid-IR source at the position of the 1.3~cm source 134-411 of \citet{Zapata+04}.  At the other position $20\arcsec$ north,  there are three near-IR sources of \citet{HC00}, i.e. HC~190, HC~198 and HC~199, but none of them have been detected at longer wavelengths either by us or by \citet{Lada+00}. On the other hand, sources MAX-42 and MAX-43, corresponding to TPSC-16 and TPSC-46 of \citet{Lada+00} and to the radio sources 136-359 and 136-356 of \citet{Zapata+04}, are displaced $\sim 5\arcsec-10\arcsec$ to the south-east, still in a favorable position with respect to the symmetry point of the CO outflow. Both show a remarkable $K-L\simeq 3$ color and appear as excellent candidates for the exciting source. The fainter of the two, MAX-42, is better aligned with the symmetry axis of the outflow. The best ground-based near-IR images clearly show MAX-43 is diffuse. Source MAX-47, corresponding to radio source 139-357 of \citet{Zapata+04}, also lies along the axis of the outflow. We have detected this source only at 20~\micron, whereas \citet{Smith+04}
have only a 10~\micron\ detection. Since there is no near-IR counterpart at this location, this appears to be a deeply embedded source and therefore also represents a good candidate for the source driving the molecular outflow. 

The last protostellar candidate of \citet{Lada+00}, TPSC-78, corresponding to source MAX-46, is located near a characteristic feature of the Orion Nebula, the dark arc limiting to the north the Orion-S region, $\approx 45\arcsec$ to the south-west of the Trapezium. Since the dark arc also appears in the 20~cm continuum image, it is not due to foreground extinction but rather traces a curved part of the ionization front seen edge-on, similar to the linear Orion Bar (O'Dell \& Yusef-Zadeh 2000). It is remarkable to find to the north of dark arc (i.e. on the ``wrong side'') one of the brightest ($[N]=2.74^m$) and reddest ($L-N\simeq 4.3$, $N-Q\simeq2.1$) sources of our survey. This object can still be embedded within the high density region creating the dark arc, if one assumes that the arc traces the closest part of a neutral cloud, whose surface is not exactly edge-on but rather tilted with respect to our vantage point. As we have mentioned above, \citet{Smith+04} consider this bright object as the most likely driving source for the HH~202/HH~528 outflow system.

\subsection{Point sources}
\subsubsection{Mid-IR variability}
For several 10~\micron\ sources, the comparison of the photometric results obtained in November 1998 versus December 2000 suggests the presence of mid-IR variability.  In Figure~\ref{Fig26} we plot against each other the data for the two epochs, listed in Table~\ref{Tab_stars10mic}.  Whereas the average magnitude difference between the two runs is less that 0.02~mag, the average scatter is $\simeq0.27^m$, much larger than the average measure error $\Delta[N]\simeq0.11^m$. This discrepancy is too large to be explained by measurement errors unaccounted for. In fact, for this type of comparison our photometric errors are probably overestimated,  as
they include, added in quadrature, the uncertainties on the absolute zero point. These are systematic shifts common to all magnitudes taken on the same night, that we conservatively estimated of the order of 0.05-0.07~mag at 10~\micron. Since almost all the data presented in Figure~\ref{Fig26}  were taken in two nights, November 27, 1998 and January 18, 2000, and since the systematic differences appear to be less than 0.02~mag, our zero points uncertainties may actually be too generous, hardening the evidence for variability.  

Stellar variability should be common among these sources. Typical members of the Trapezium clusters are spotted, rotating T~Tauri stars, known to be variable at visible wavelengths. According to \citet{Herbst+02}, essentially all stars in Orion Nebula bright enough to be monitored from the ground are variable, with periods having a bimodal distribution with peaks at 2 and 8 days for M$ > 0.25$M$_\odot$, and  unimodal distribution peaked near 2 days for M$\le 0.25$M$_\odot$. Future comparative studies of the optical and mid-IR variability, based on simultaneous multiwavelength photometric monitoring, may allow to understand whether the variability is due to disk activity, e.g. fluctuations  of the mass accretion rates, or to changes in the foreground extinction due to inhomogeneities of the circumstellar environment or of the circumestellar disk itself.

\subsubsection{Mid-IR excess}
To investigate the nature of the mid-IR emission of our sources, we compare in Figure~\ref{Fig27} the [K-L] vs. [N-Q] color indexes of our Trapezium stars versus a sample of classical T~Tauri stars in the Taurus-Auriga association. For the Trapezium stars, we complement our photometry with the K and L-band photometry of \citet{Muench+02},  using two different symbols to indicate the 10~\micron\ data obtained in the 1998  and 2000 observing runs.

The Taurus-Auriga sources are those listed in Table II of \citet{Strom+89}. Their K and L-band photometry is the average of the values available for each filter in the recent literature, mined using the VizieR Service\footnote{VizieR Service is a joint effort of CDS and ESA-ESRIN,  available on the internet at http://vizier.u-strasbourg.fr/viz-bin/VizieR}, and in general turns out very close to the photometry of \citet{Strom+89}. The 10 and 20~\micron\ photometry is derived from the IRAS photometry at 12 and 25~\micron\ in the following way.
The IRAS PSC2 fluxes are quoted at the effective wavelengths assuming an input energy distribution constant per $\log\nu$, i.e. $\nu F_\nu=constant$. This is also the shape of the typical ``flat'' Spectral Energy Distribution (SED) of circumstellar disks around T~Tauri stars. We therefore assume the IRAS fluxes to be color corrected and adjust their values only to account for the difference in effective wavelength between the filters. From the $\nu F_\nu$=constant condition, we derive $m(10)=m(12)+0.516$ and $m(20)=m(25)+0.694$
using the zero magnitude fluxes reported in Section~2 of this paper and in the IRAS PSC2 manual.  We have visually inspected each SED plotting, together with all the IR data available in the literature, our interpolated 10 and 20 \micron\ points. Whereas our points are always in excellent agreement with the overall shape of the SED (admittedly constrained by the IRAS points), most of the 10 and 20~\micron\ ground based data are old and suspiciously away from the best fit line.

Figure~\ref{Fig27} clearly shows that there is a systematic difference between the Orion and the Taurus-Auriga samples, as the Orion
stars have a [N-Q] excess between 3 and 5~magnitudes, whereas for Taurus-Auriga the [N-Q] index ranges between 1 and 3~magnitudes. Also, the [K-L] excesses appear on average higher for Orion, although for both samples the majority of stars have [K-L] in the range 0.7--1.5~mag. We show in Figure~\ref{Fig27} also five stars without L-band photometry, but well detected at shorter wavelengths, to which we have assigned [K-L]=0. Also for these sources the [N-Q] index appears higher than that of the stars in Taurus-Auriga. 

Extinction cannot explain these effects, either in terms of intensity or reddening.  In Figure~\ref{Fig27} we have indicated with an arrow the reddening corresponding to  $A_v=10$ magnitudes of extinction. \citet{LAH97} suggests that the mean extinction for the optically visible population of the Orion Nebula cluster is $A_V\lesssim2.5$ magnitudes, with few optical stars having $A_V\gtrsim5$ magnitudes. For the IR sample, the mean extinction is approximately 3 times higher, but only a small  fraction of sources has $A_V\gtrsim15$~magnitudes \citep{LAH+Hart98}. 

\citet{Calvet+92} have calculated the IR colors of accreting disks around classic T~Tauri stars assuming a variety of stellar and disk parameters.  Each one of their models is represented in Figure~\ref{Fig27} as a solid line,  as different orientation angles are taken into account. The \citet{Calvet+92} predictions fall in general within the region of the observed T~Tauri star colors.  In particular, the models with the highest [N-Q] $\gtrsim 2$ have in common a  higher value of $\gamma$,  the index of the power law describing the increase of the disk height with the distance from the central star. Large disk flaring angles could therefore explain the extreme [N-Q] colors of our sources. 

In a previous paper \citep{Robberto+02}, we have explored the continuum SED of circumstellar disks embedded within a HII region like Orion. We predicted that the extra energy input from the ionizing stars and from the nebular environment raises the disk temperature 
and causes an increase of the disk flaring angle. An inflated disk will also intercept a larger fraction of the flux emitted by its own central star. In the vicinity of \thC, the neutral flow produced by disk photoevaporation originates a neutral envelope bordered by a ionization front, the ``proplyd'' phenomenon \citep{OD+93,ODW94}. Together with the disk, the dusty envelope also contributes to the mid-IR emission. Our models show that there is in general a strong enhancement of the mid-IR emission due both to the higher disk temperature and the presence of the photoevaporated envelope. The UV heated dust produces a characteristic far-IR peak of emission
at 30--60~\micron. In Figure~\ref{Fig27} we show as a dotted line the locus of the colors resulting from our model, assuming  the limiting case of a disk facing both the ionizing star (for simplicity \thC\ is considered the only ionizing source) and the Earth at various
distances from \thC. The [N-Q] colors predicted by this model nicely match those of the Orion sample, supporting the hypothesis that the UV radiation is ultimately responsible for the strong enhancement of 20~\micron\ emission. 

In \citet{Robberto+02}, we have shown that the IR SEDs of proplyds depend on a number of factors,  both physical, like the distance from the ionizing star, the size of the envelope, the inner disk radius, the temperature of the central star and the dust composition, and geometrical, like the tilt angle of the disk with respect to the ionizing star and to the Earth. The projection factors in particular have a dramatic impact on the observed SED, since disks surfaces at the same distance from \thC\ can be exposed to very different UV fluxes and appear from the Earth under all possible angles, showing either the front or the back side. The 30--60~\micron\ peak stands as the main observational characteristics of  these systems. Depending on the dust temperature and grain size, IR color indexes that do not include the 20~\micron\ band, like [K-N] or [L-N],  may remain unaffected by a mid-IR peak too cold to produce strong excess at 10~\micron.  For the main group of sources detected only at 10~\micron, it is necessary to analyze in detail each individual source to constrain the disk properties. This analysis will be presented in a second paper, and will include the result of our deep 10~\micron\ survey 
of selected fields in Orion.

\section{Conclusion}
Using a new restoration method developed for this project, we have obtained the first wide-field and diffraction limited 10~\mic\ and 20~\mic\ images of the core of the Orion Nebula. The images confirm that the morphology of the Orion Nebula at mid-IR wavelengths is dominated by the BN/KL complex and by the Ney-Allen nebula. At lower signal levels, the Orion Nebula shows a filamentary structure that traces the walls of the HII region. A remarkable group of arc-like structures $\approx 1\arcmin$ is detected to the South of the Trapezium. We analyze the structure of the Ney-Allen nebula in the framework of analytic wind-wind interaction models, finding evidence for a complex kinematical status at the center of the Cluster. The B0.5V star $\theta^1$Ori-D, associated with the brightest mid-IR structure of the Trapezium region, is probably surrounded by a photoevaporated circumstellar disk. The fact that, among the main four Trapezium stars, this star is the only one without a binary companion  suggests that stellar multiplicity and massive circumstellar disks are mutually exclusive in the Trapezium.  We detect in the OMC-1 South region three point sources that can be associated to the mass loss activity detected at millimeter wavelengths. Finally, we list the position and photometry of 177 point sources, the large majority of which detected for the first time in the mid-IR. Twenty two of them lack counterparts at shorter wavelengths, and are, therefore, candidates for deeply embedded protostars. The comparison of photometric data obtained at two different epochs reveals that source variability at 10{~\mic} is present up to $\approx 1^m$ level on a time-scale $\sim 2$~yr. We compare the mid-IR color-color diagram for our sources in Orion and a similar dataset in the Taurus association, finding indications for large mid-IR excess, especially at 20~\micron, and discuss this finding in the framework of the \citet{Robberto+02} model.

We have also presented narrow band images of the BN/KL complex taken in the standard silicate filter set. The BN/KL complex is resolved into multiple and extended features. In particular, we resolve a point source in correspondence of the IRc2-A knot. We derive the extinction and temperature maps on the basis of the simple "two component" model for the IR emission, and discuss the shape of the silicate absorption feature and possible scenarios for dust evolution in the HII region.

\acknowledgments
The authors are indebted to the MPIA and UKIRT staff for their support with the MAX operations. The contribution of M.~Cohen to the MAX calibration is acknowledged. We thank Bob O'Dell for several useful comments and in particular for insights on the interpretation of the OMC-1S region. \pagebreak\ Comments from an anonymous referee also greatly contributed to improve the original manuscript. 
S.~G.~P.'s visit at STScI was supported by the STScI 2002 Summer Student Program.

\begin{figure}%4,5
\epsscale{0.85}
\plottwo{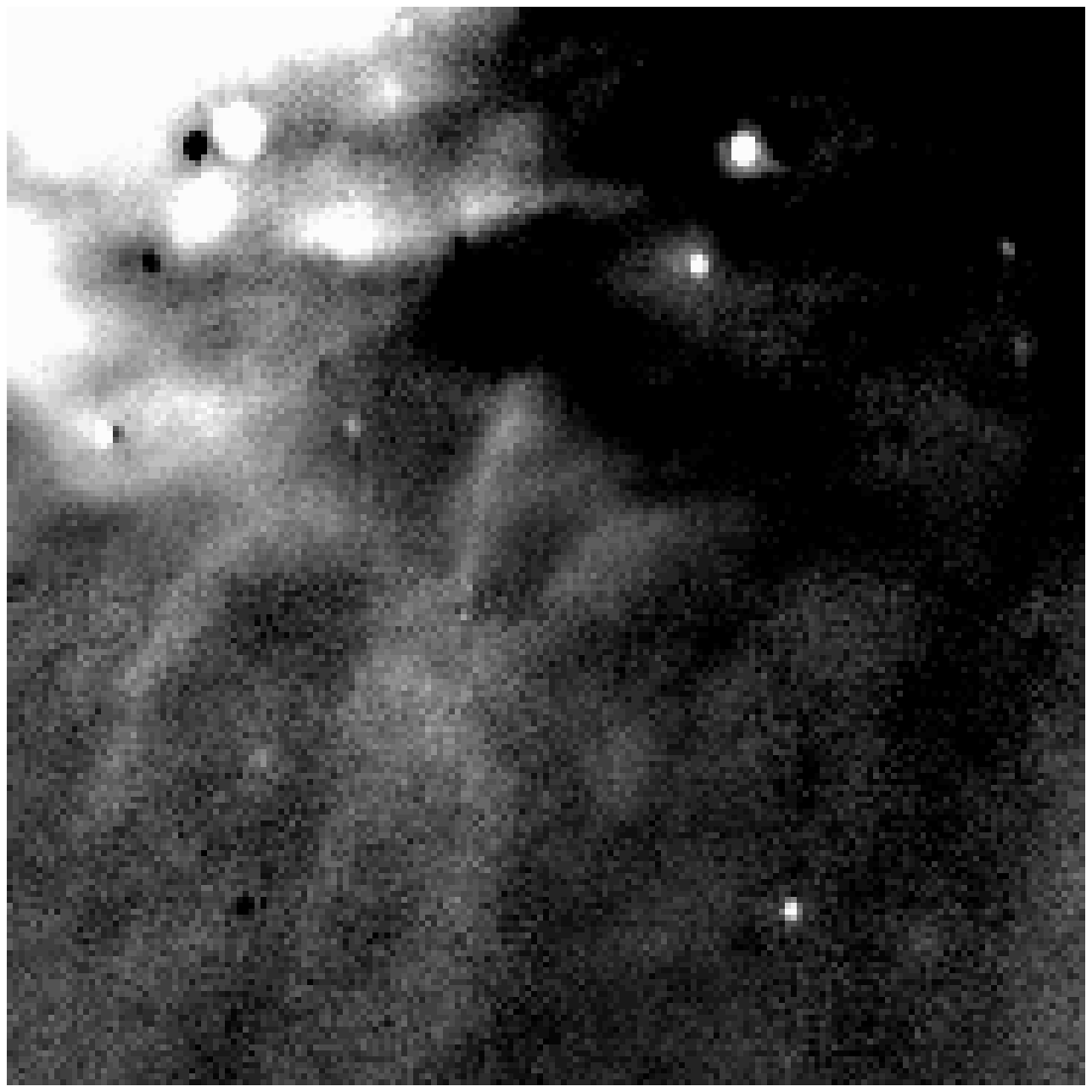}{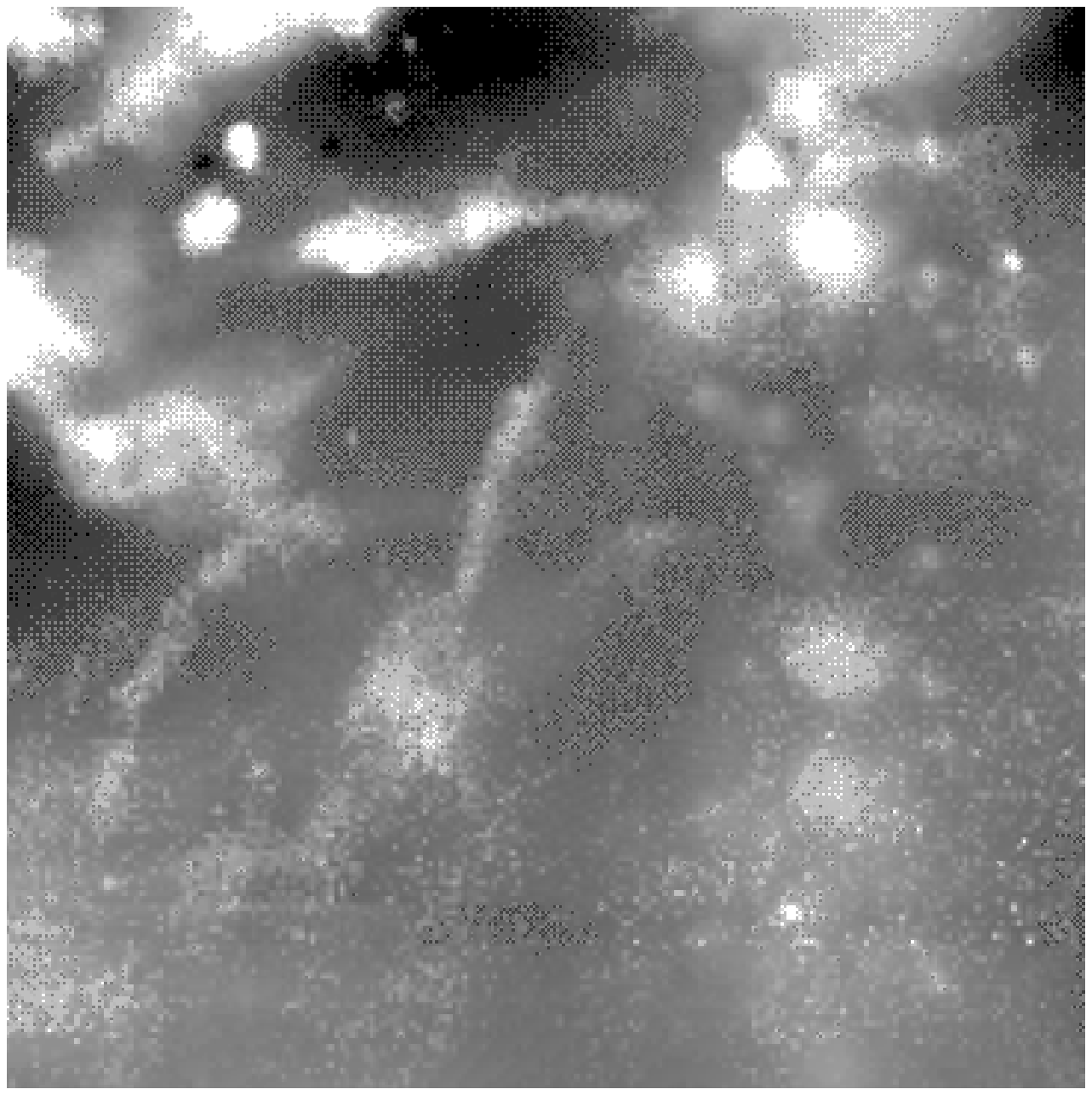}
\caption{Arcs of mid-IR emission approximately $45\arcsec $south of the Trapezium (see Figure~11 for their location). 
Left: original image before image reconstruction. Right: same field after image reconstruction and filtering. Scale is $45\arcsec\times45\arcsec$, with standard orientation. The sources in the north-western corner fall within the OMC-1 South region (see Figure~\ref{Fig11}).\label{Fig4,5}}
\end{figure}

\begin{figure}%6,7
\epsscale{.85}
\plottwo{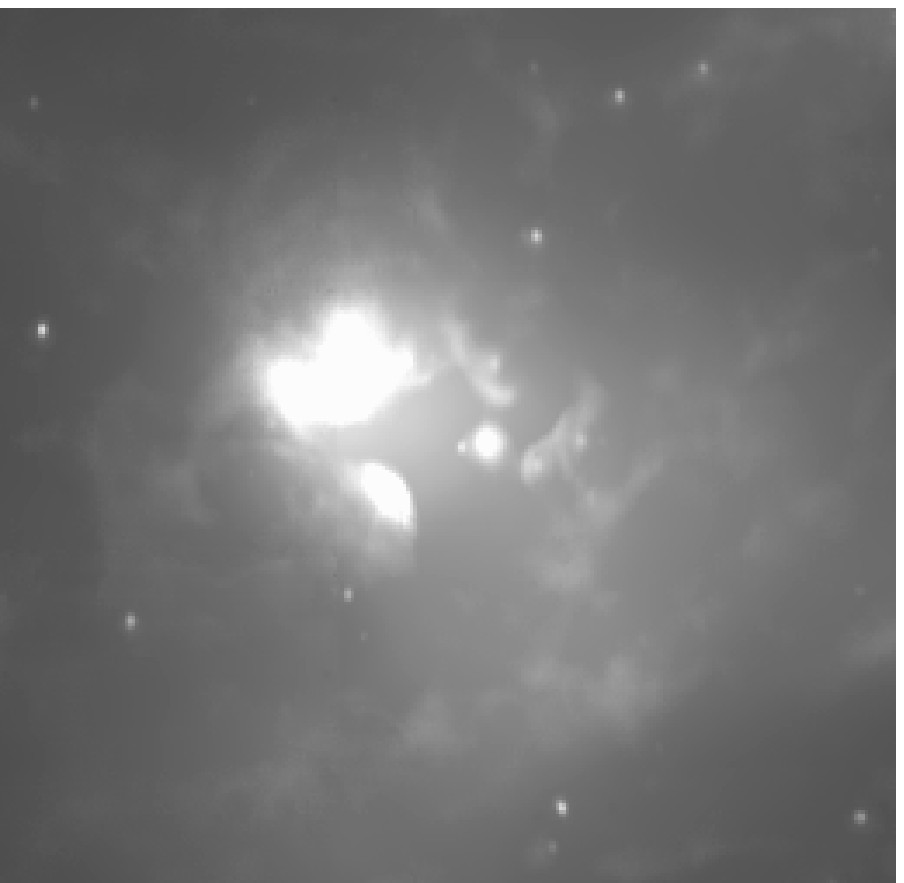}{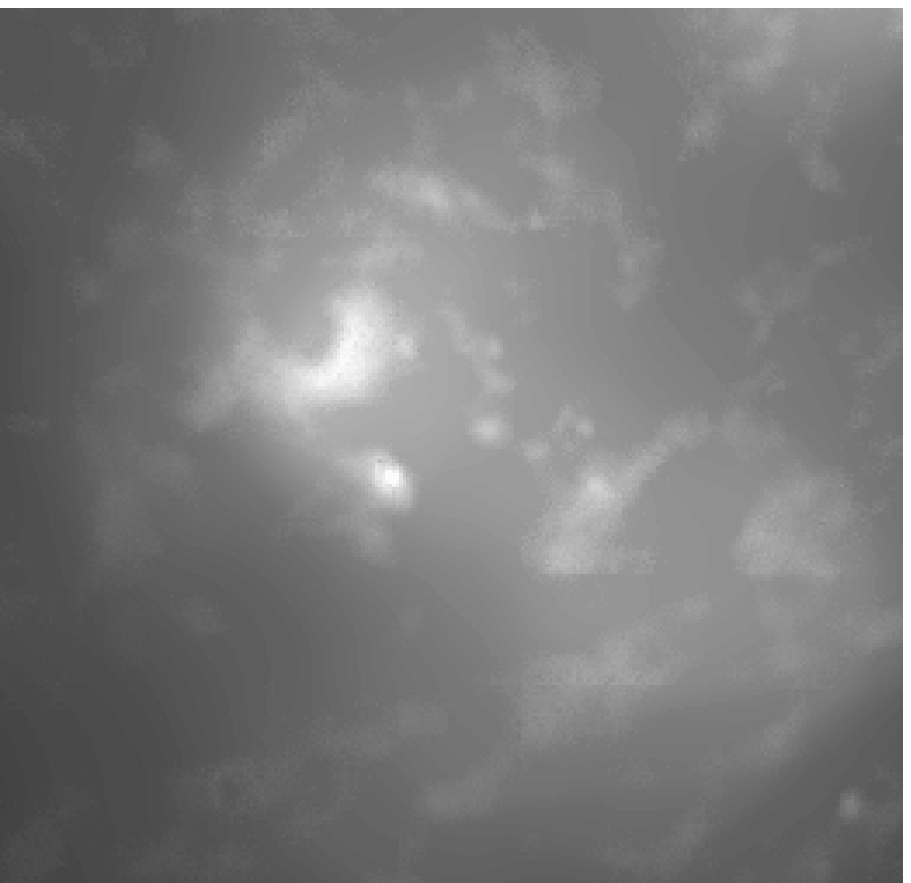}
\caption{The Trapezium region at 10~\micron{} (left) and at 20~\micron{} (right). \thC\ and SC3 are the two circular sources at the center of the 10~\mic\ image, \thC\ being the fainter one to the left. Field size is $1\arcmin \times 1\arcmin$, north is up and east is to the left.}
\label{Fig6,7}
\end{figure}

\begin{figure}%8,9
\epsscale{1.0}
\plottwo{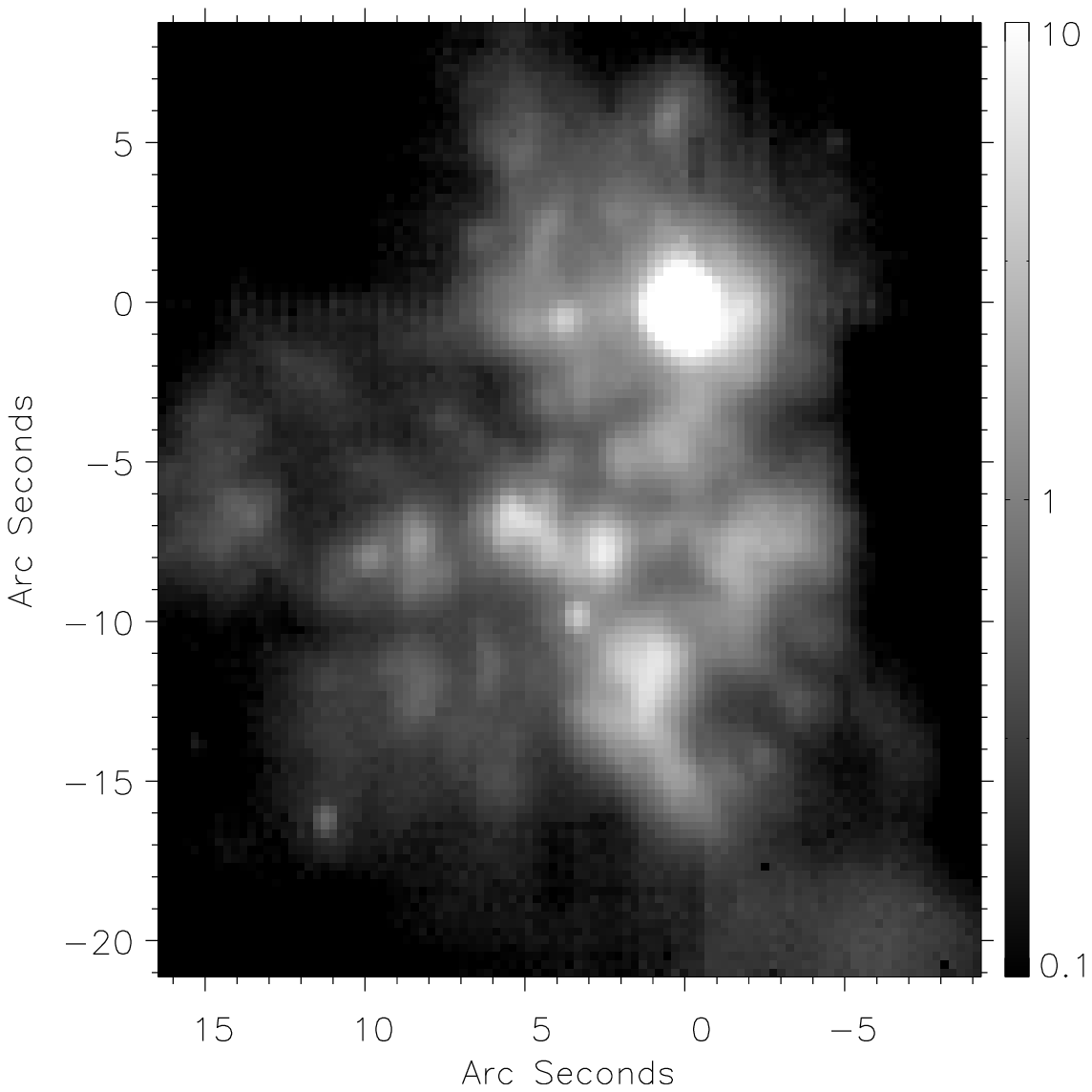}{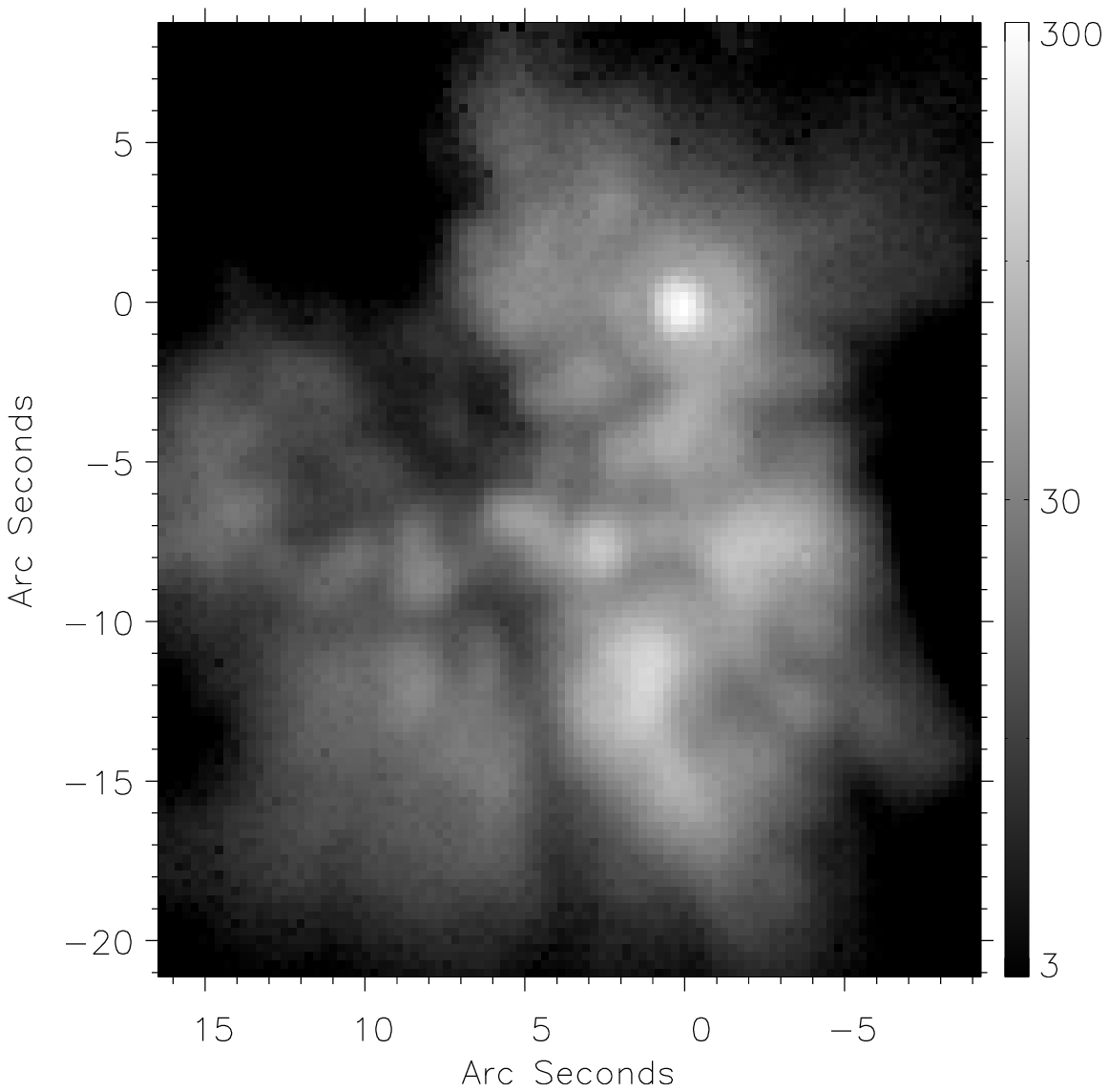}
\caption{The BN/KL complex at 10~\micron\ (left) and 20~\mic{} (right). The scale is logarithmic and ranges between 0.1 and 10 Jy/arcsec$^2$ at 10~\micron, and between 3 and 300 Jy/arcsec$^2$ at 20~\micron. 
}
\label{Fig8,9}
\end{figure}

\begin{figure}%10
\epsscale{.9}
\plotone{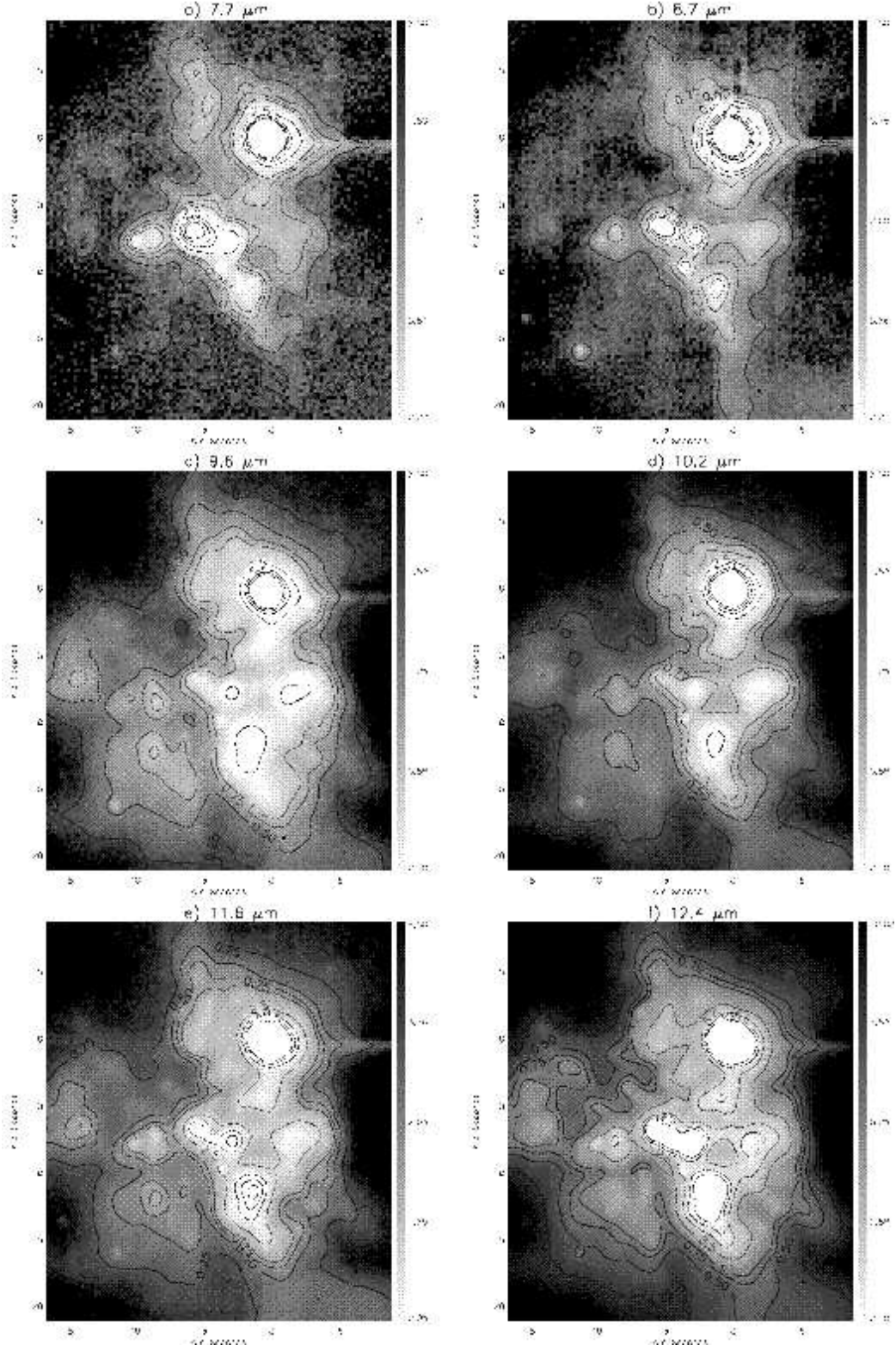}
\caption{The BN/KL complex at 7.7 and 8.7~\micron\ ({\sl top row}), 9.6 and  10.2~\micron\ ({\sl middle row}), and 11.6 and 12.4~\micron\ ({\sl bottom row}). 
All images are shown in logarithmic scale and units are Jy~arcsec$^{-2}$. To facilitate the comparison, all contours are plotted
at the same levels of 0.1, 0.25, 0.50, 0.75, 1, 2.5, 5, 7.5, and 10 Jy~arcsec$^{-2}$.
Equatorial coordinates are relative to the peak of the BN source (see Section 2.2.2).}
\label{Fig10}
\end{figure}

\begin{figure}%11,12
\epsscale{1.0}
\plottwo{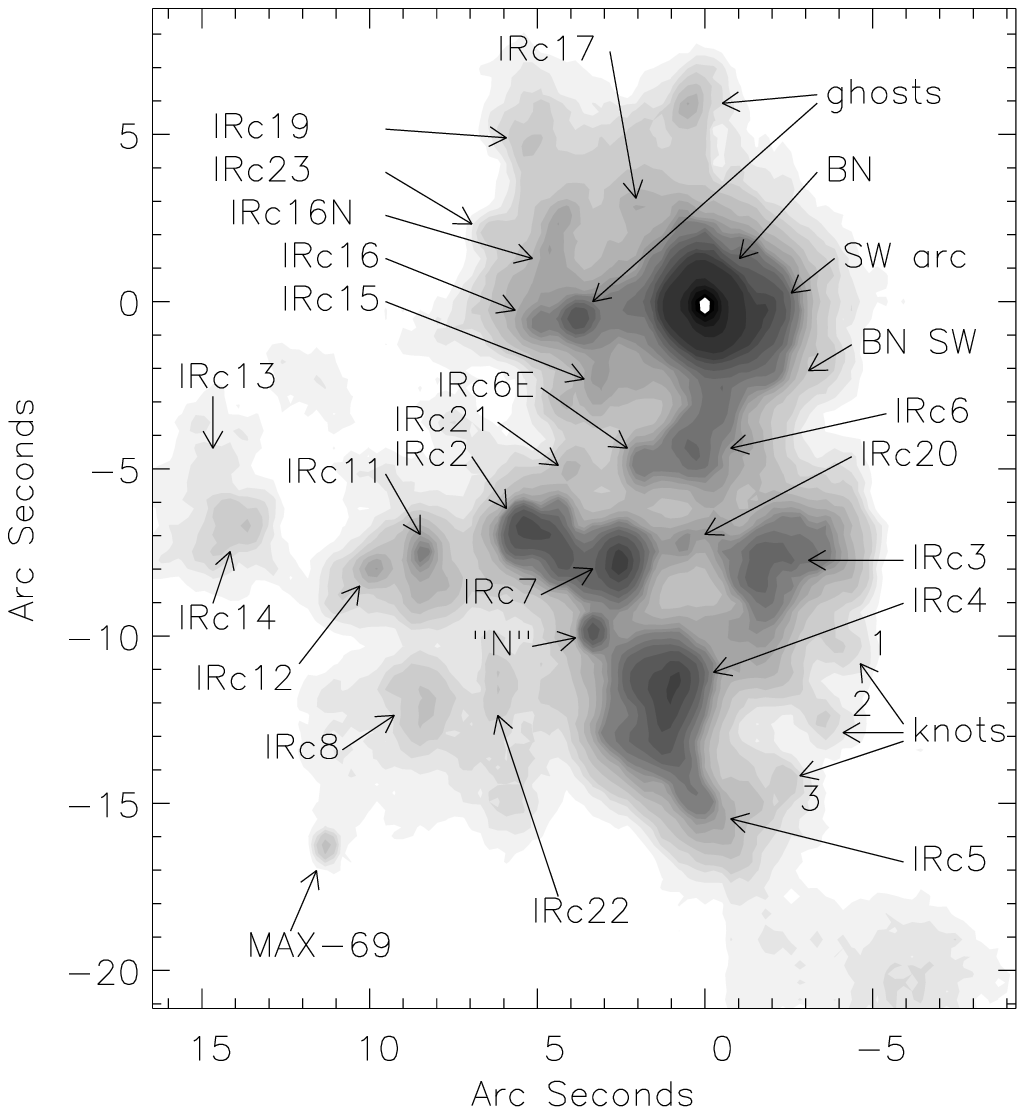}{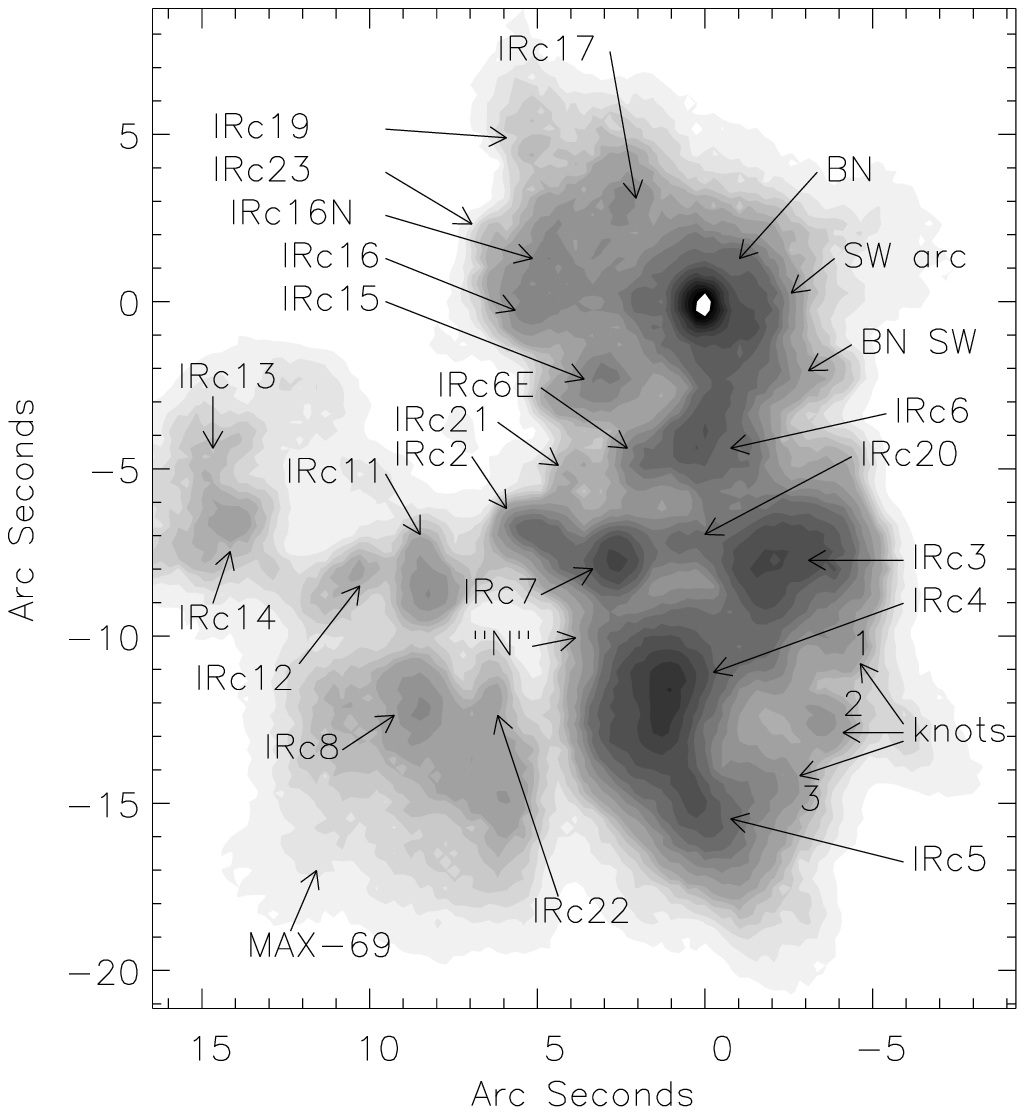}
\caption{Contour plot of the BN/KL complex at 10~\micron\ (left) and 20~\mic{} (right), with labels indicating the main features discussed in the text. Contours levels are 0.22, 0.30, 0.37, 0.45, 0.60, 0.80, 1.0, 1.2, 1.35, 1.5, 1.87, 2.25, 3.0, 4.5, 6.0, 10, 15, 30, 60, 100, 150 and 200 Jy/arcsec$^2$ at 10~\mic, and 7, 10, 12, 14, 16, 18, 20, 25, 30, 35, 40, 50, 60, 80, 100, 120, 140,1 60, 180, 200 Jy/arcsec$^2$ at 20~\mic.}
\label{Fig11,12}
\end{figure}

\begin{figure}%13
\epsscale{0.8}
\plotone{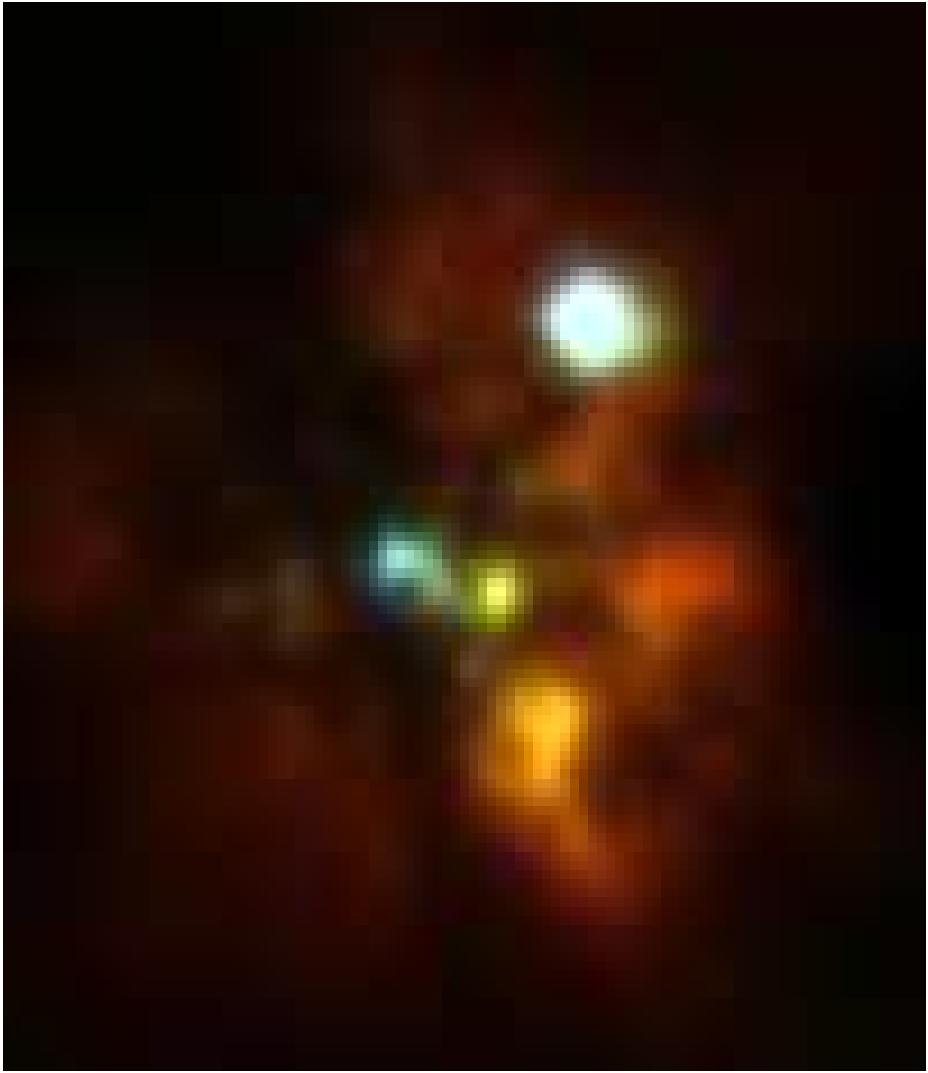}
\caption{Color composite of the BN/KL region obtained combining the 7.7~\micron\ (blue), 12.4~\micron\ (green) and 19.9~\micron\ (red) images.}
\label{Fig13.}
\end{figure}

\begin{figure}%14
\epsscale{1.0}
\plotone{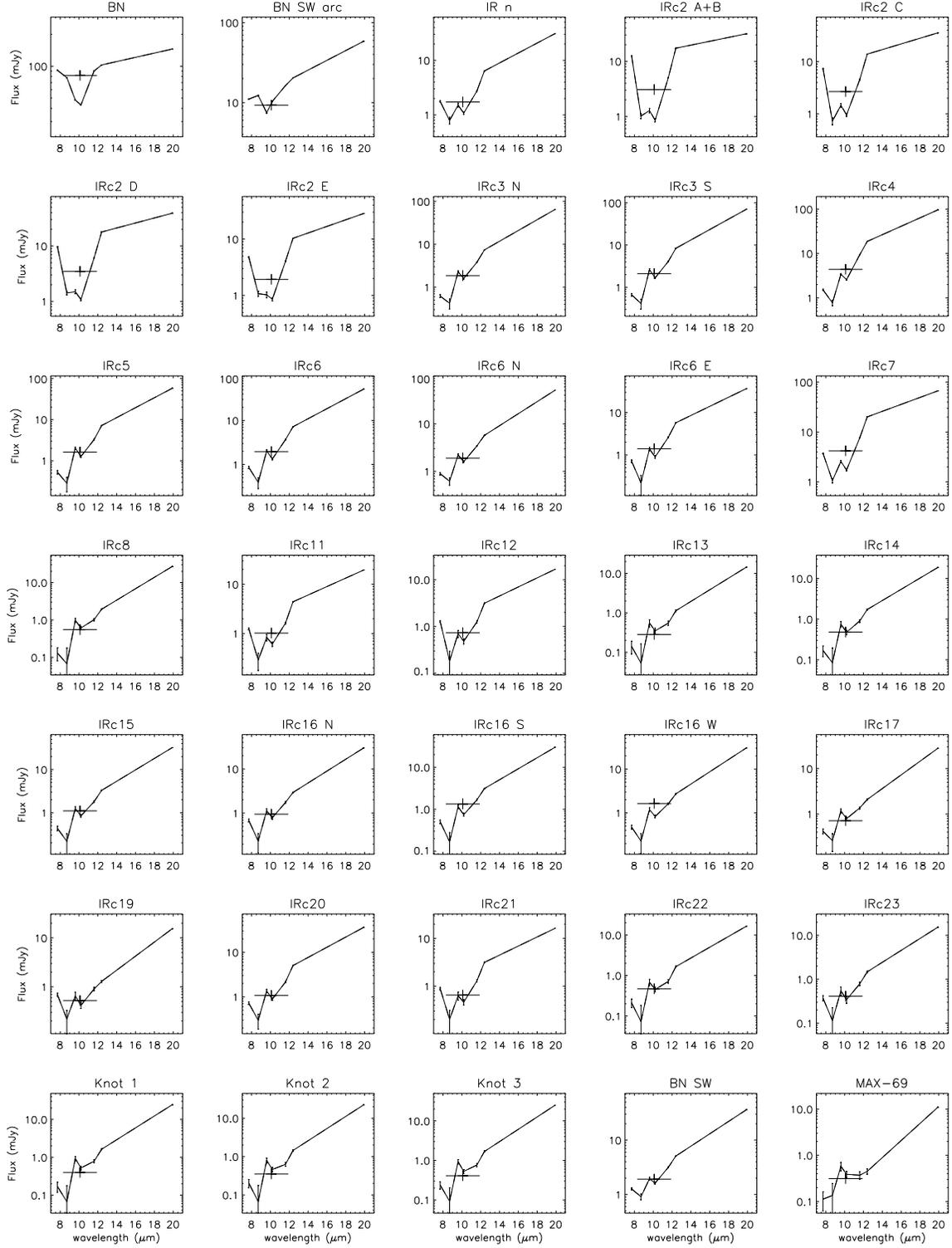}
\caption{Mid-IR spectral energy distribution for the 35 sources of Table~\ref{Tab:BN_sil}. Errors refer to the typical zero point error of our
absolute calibration. The wide horizontal line indicates the broad-band N filter.}
\label{Fig14}
\end{figure}

\begin{figure}%15
\epsscale{.6}
\plotone{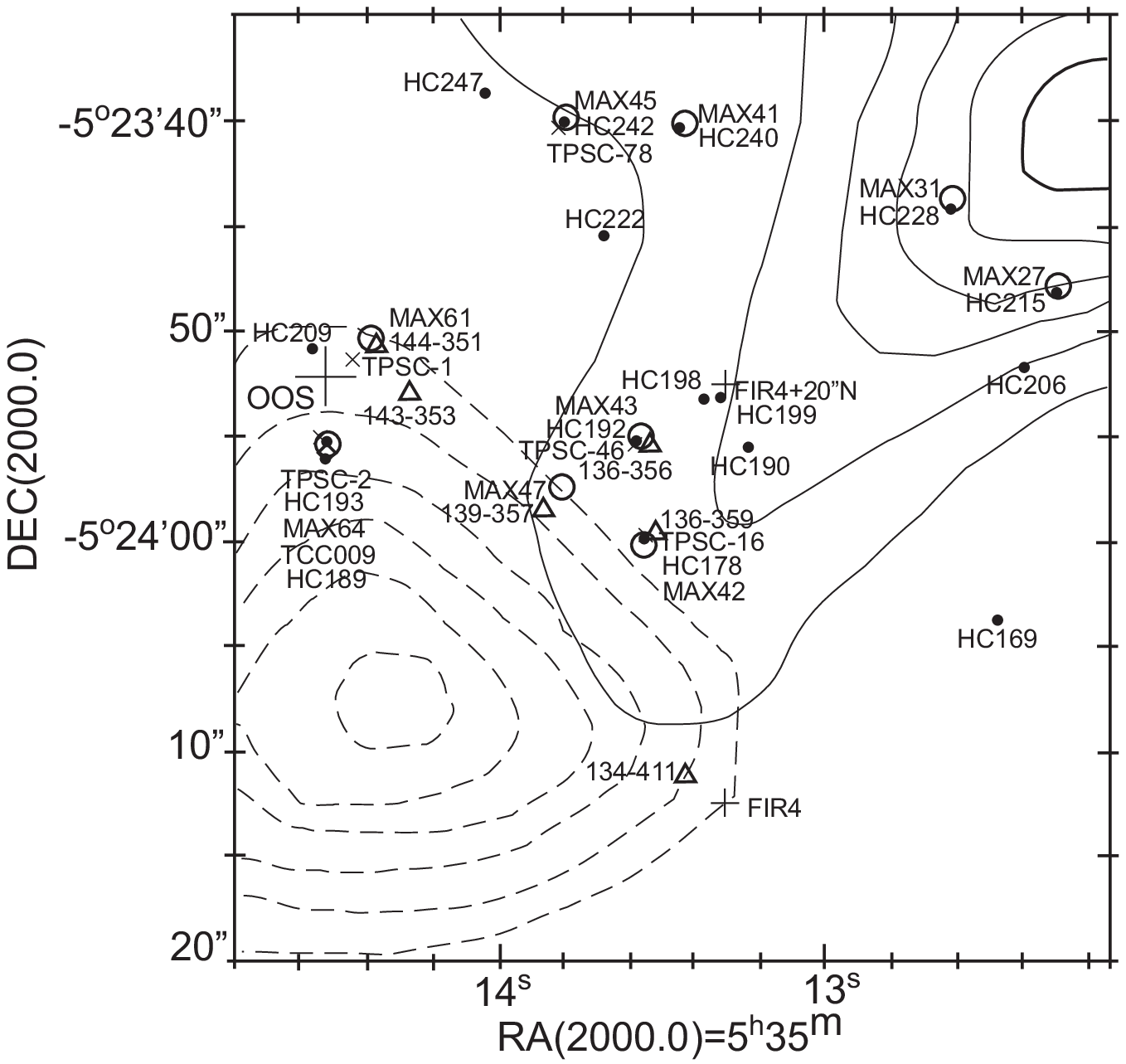}
\caption{Map of the sources located in the Orion-S region, together with the intensity contours of the CO outflow
mapped by \citet{Rodriguez+99b}. The naming conventions are: 1)~MAX sources: this paper; 2)~TPSC sources: \citet{Lada+00}; 3)~HC sources: \citet{HC00}; 134-411 ({\sl et similia}) and OOS sources: \citet{Zapata+04}; FIR~4: \citet{Rodriguez+99a}.}
\label{Fig11}
\end{figure}

\begin{figure}%16
\epsscale{.8}
\plotone{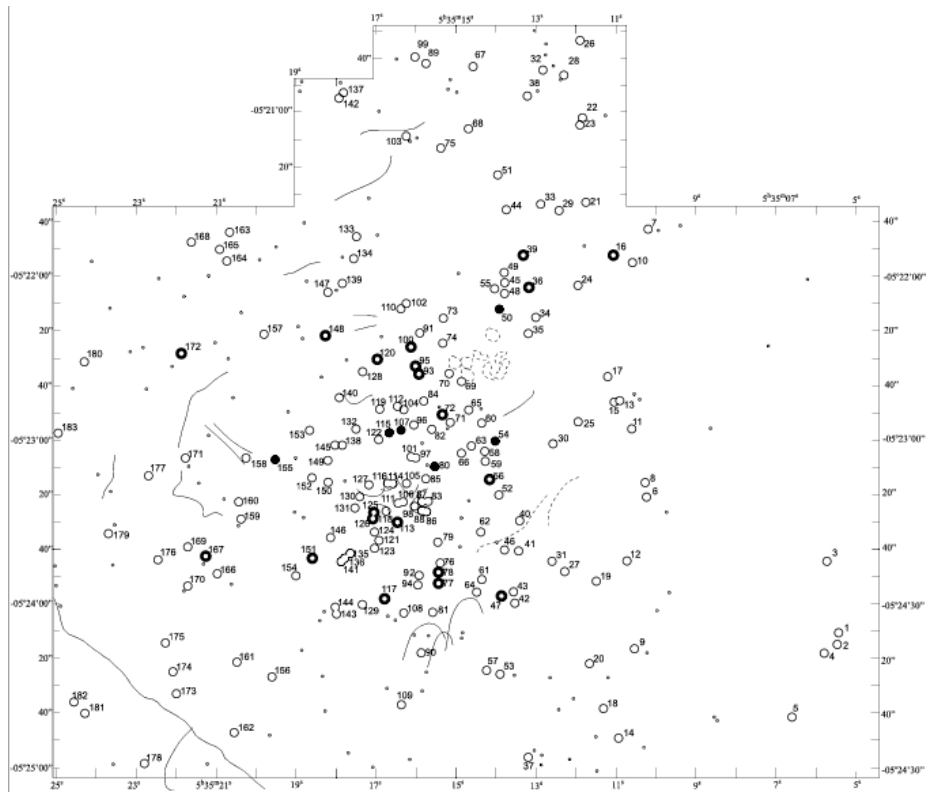}
\caption{Location of our mid-IR point sources. Thin open circles: 10~\micron\ sources with counterpart at shorter wavelengths. Thick open circles: 10~\micron\ sources detected at 10~\micron\ and not at shorter wavelengths. Filled circles: sources detected only at 20~\micron. Small circles: a subset of the near IR sources that remained undetected at 10~\micron, presented here for reference and ease of comparison with near-IR images. Thin solid lines trace the main filamentary structures and the arcs discussed in Section~4. \thC\ is source 111; $\theta^2$Ori~A is source 178.}
\label{Fig12}
\end{figure}

\begin{figure}%17
\epsscale{0.6}
\plotone{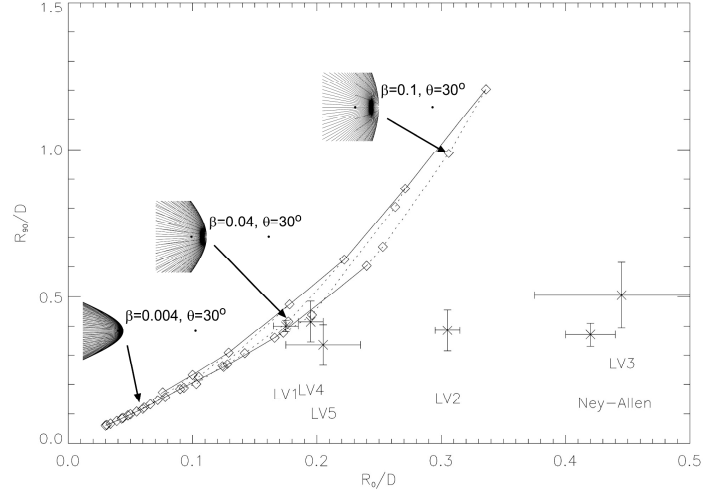}
\caption{Locus of the morphological parameters $R_0/D$ and $R_{90}/D$ describing the Trapezium arcs. Open circles 
refer from left to right to $\beta=0.001, 0.002, 0.004, 0.01, 0.02$, and $0.1$, and increase from the bottom to the top 
from $\theta_i= 0^\circ$ up to the maximum apparent tilt angle in steps of $15^\circ$. The location of the arcs 
in the vicinity of \thC\ is also shown, with the observational uncertainties.}
\label{Fig13}
\end{figure}

\begin{figure}%18
\epsscale{0.7}
\plotone{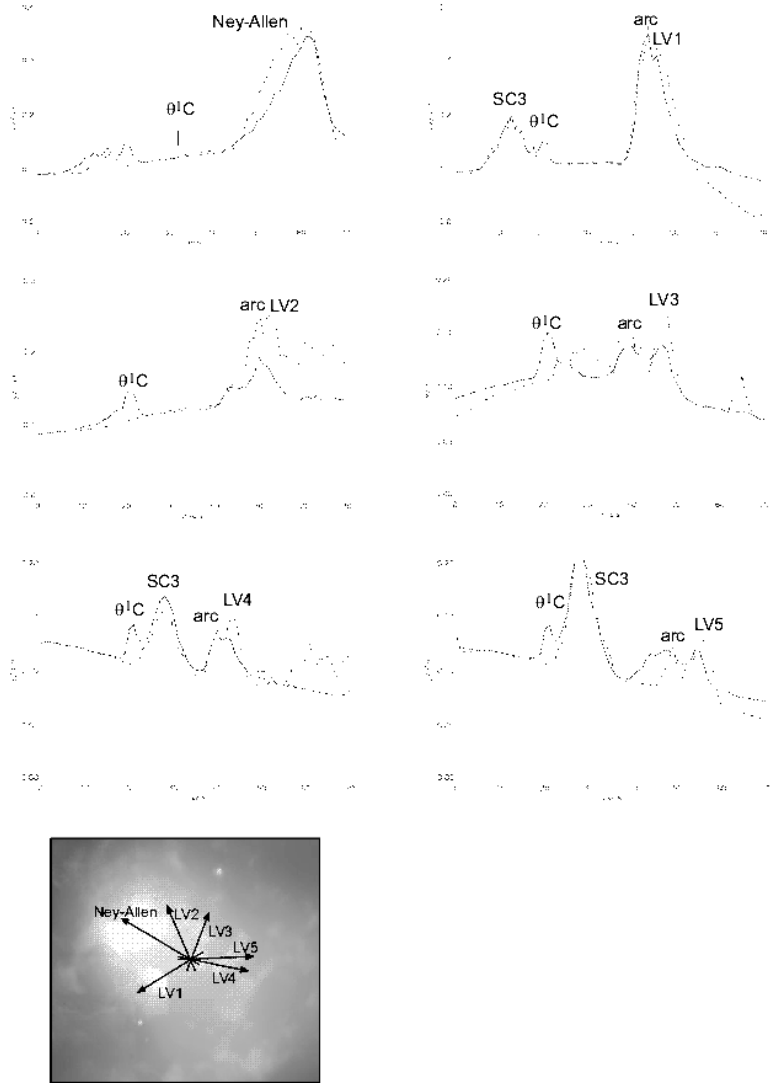}
\caption{Brightness profiles along the cuts joining \thC{} to $\theta^1$Ori~D and the LV1-5 sources. {\sl Solid line}: 10~\mic{} emission; {\sl dashed line}: 20~\mic{} emission, scaled for comparison.}
\label{Fig18}
\end{figure}

\begin{figure}%19
\epsscale{0.3}
\plotone{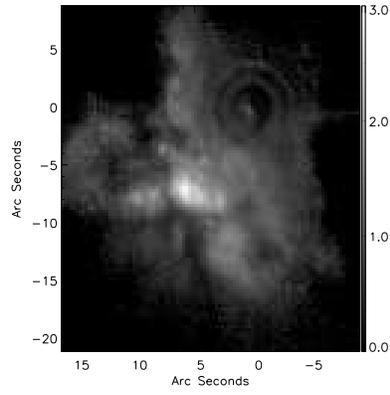}
\bigskip
\bigskip
\caption{Map of the optical depth of the BN/KL complext at 9.6~\mic.}
\label{Fig19}
\end{figure}

\begin{figure}%20
\epsscale{.8}
\plotone{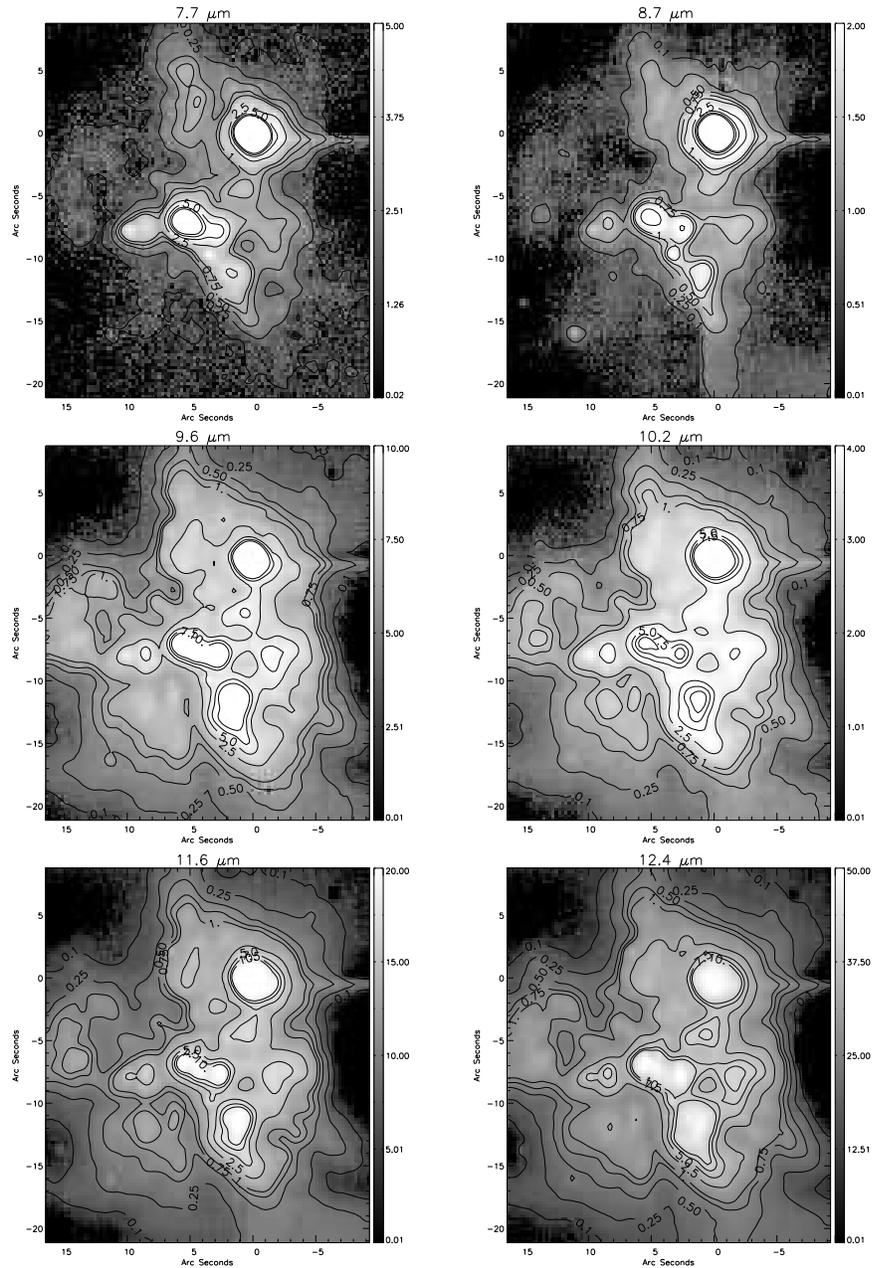}
\caption{Same as Figure~\ref{Fig10}, after reddening correction.}
\label{Fig20}
\end{figure}

\begin{figure}%21
\epsscale{0.7}
\plotone{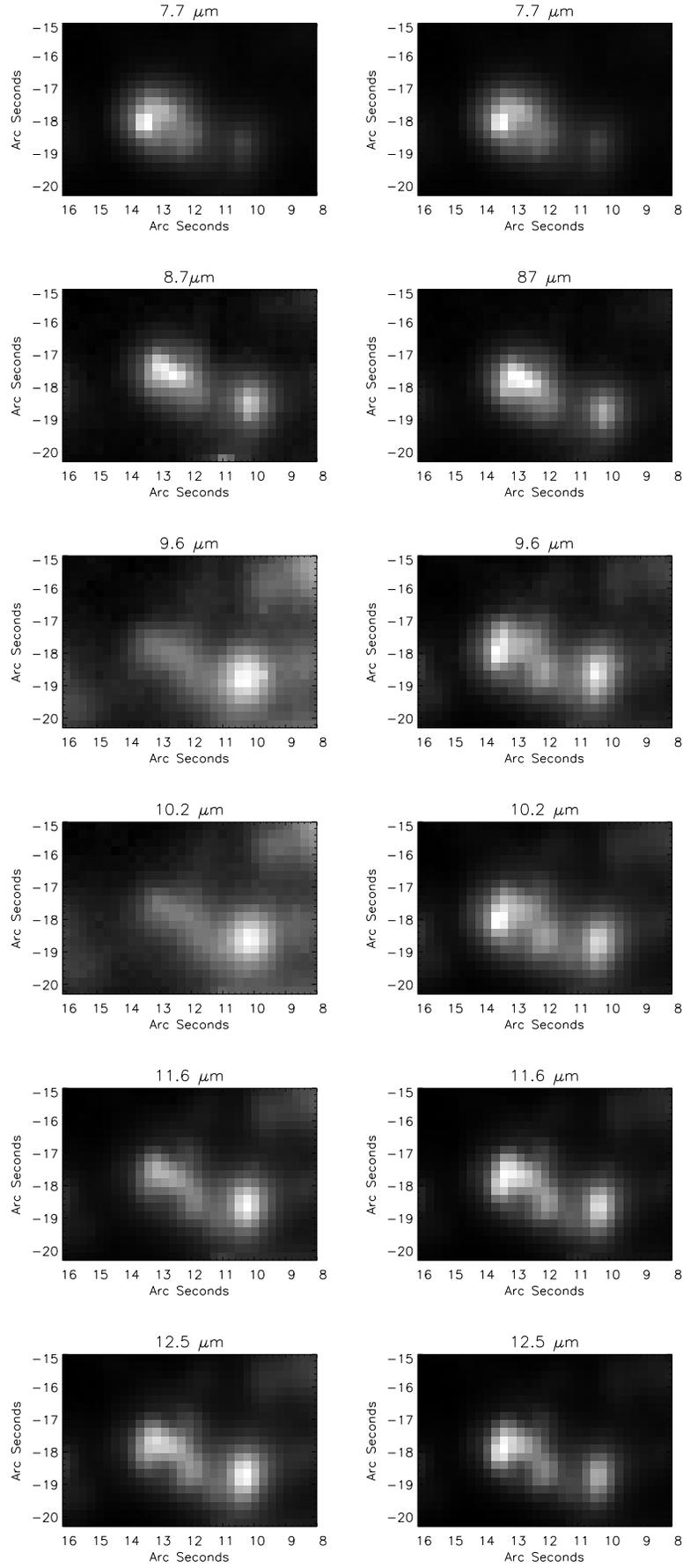}
\caption{Comparison of the morphology of the IRc2/IRc7 region before and after reddening correction for the six silicate filters.}
\label{Fig21}
\end{figure}
\clearpage

\begin{figure}%22
\epsscale{0.5}
\plotone{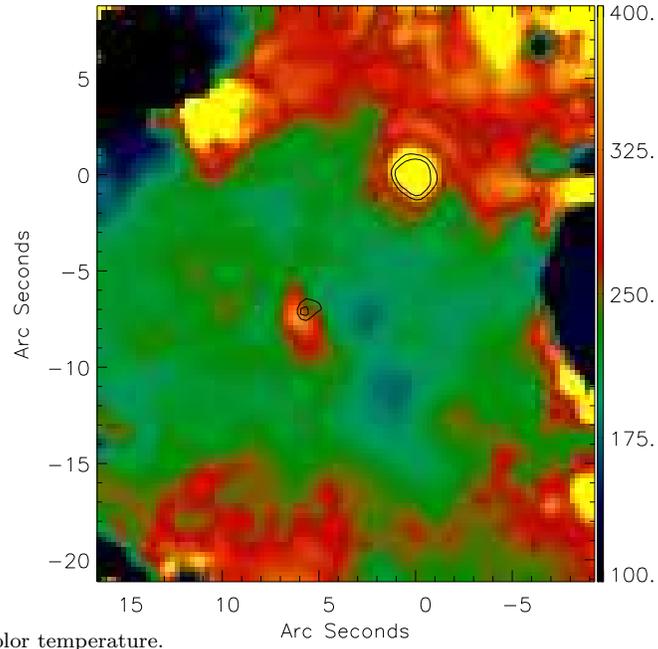}
\bigskip
\caption{Map of the 10~\micron\ color temperature.}
\label{Fig22}
\end{figure}

\begin{figure}%23
\epsscale{.7}
\plotone{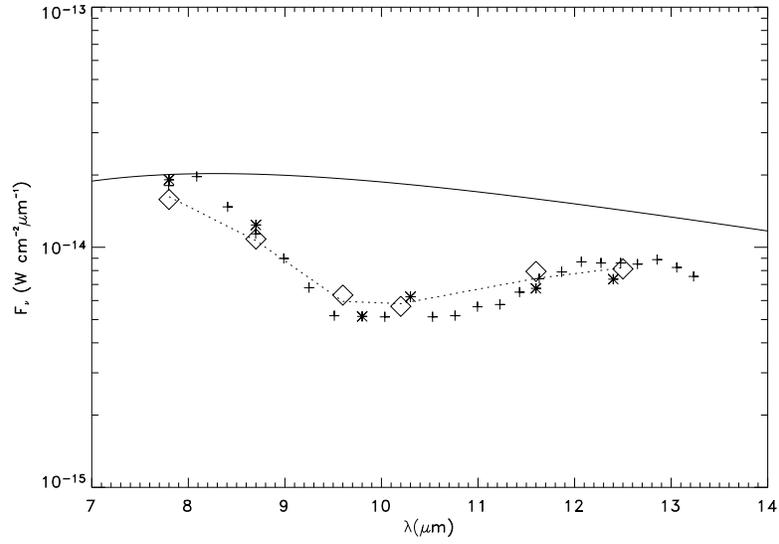}
\caption{
Silicate spectrum of BN. {\sl Diamonds}: this paper; {\sl Plus signs}: spectrophotometry of \citet{Aitken+85};
{\sl asterisks}: broad-band data of \citet{Gezari+98}. The solid line refer to a blackbody curve at $T_{col}=373$~K. The dotted line
joins the blackbody points corresponding to our effective wavelengths after extinction correction with optical depth $\tau_{9.8}$=1.37.}
\label{Fig23}
\end{figure}

\begin{figure}%24,25
\epsscale{1.0}
\plottwo{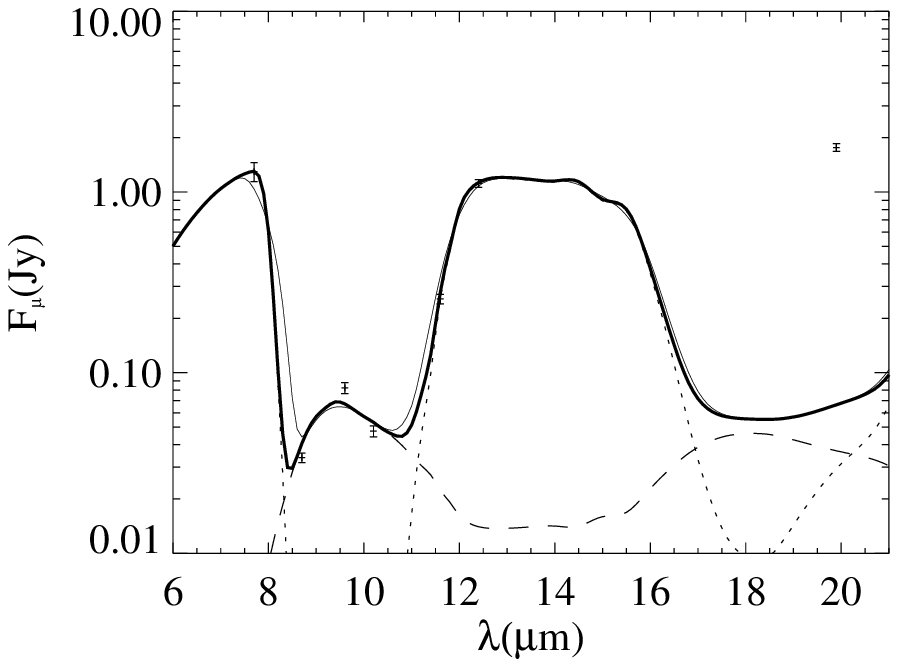}{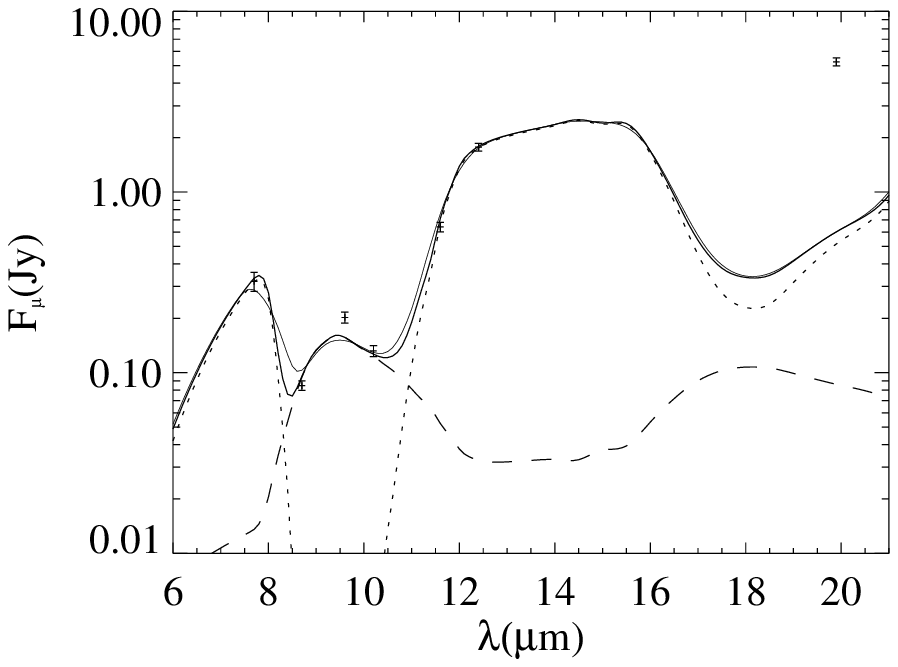}
\caption{Fit to the mid-IR photometry of IRc2 (left) and IRc7 (right) based on Equation~4. Dotted line: first term on the right side of Equation~4, relative to the bagground source and foreground extinction. Dashed line: sencond term on the right side of Equation~4, relative to the optically thin emission of 300~K dust. Thick solid line: sum of the two contributions. Thin solid line: sum of the two contribution convolved with a 1~\micron\ rectangular bandpass to reproduce the effect of finite bandwith. Error bars represent the zero point errors quoted in Section~2.2.3}
\label{Fig24,25}
\end{figure}

\begin{figure}%26
\epsscale{.8}
\plotone{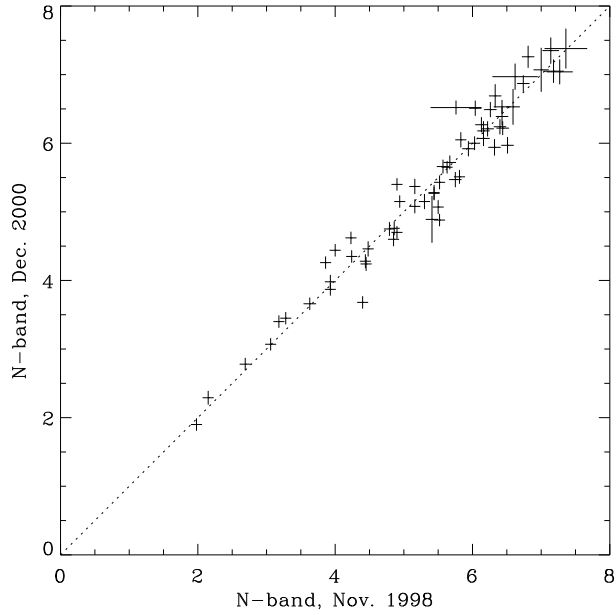}
\caption{Comparison of the MAX [N]-band photometry at the two epochs.}
\label{Fig26}
\end{figure}
\clearpage

\begin{figure}%27
\epsscale{.5}
\plotone{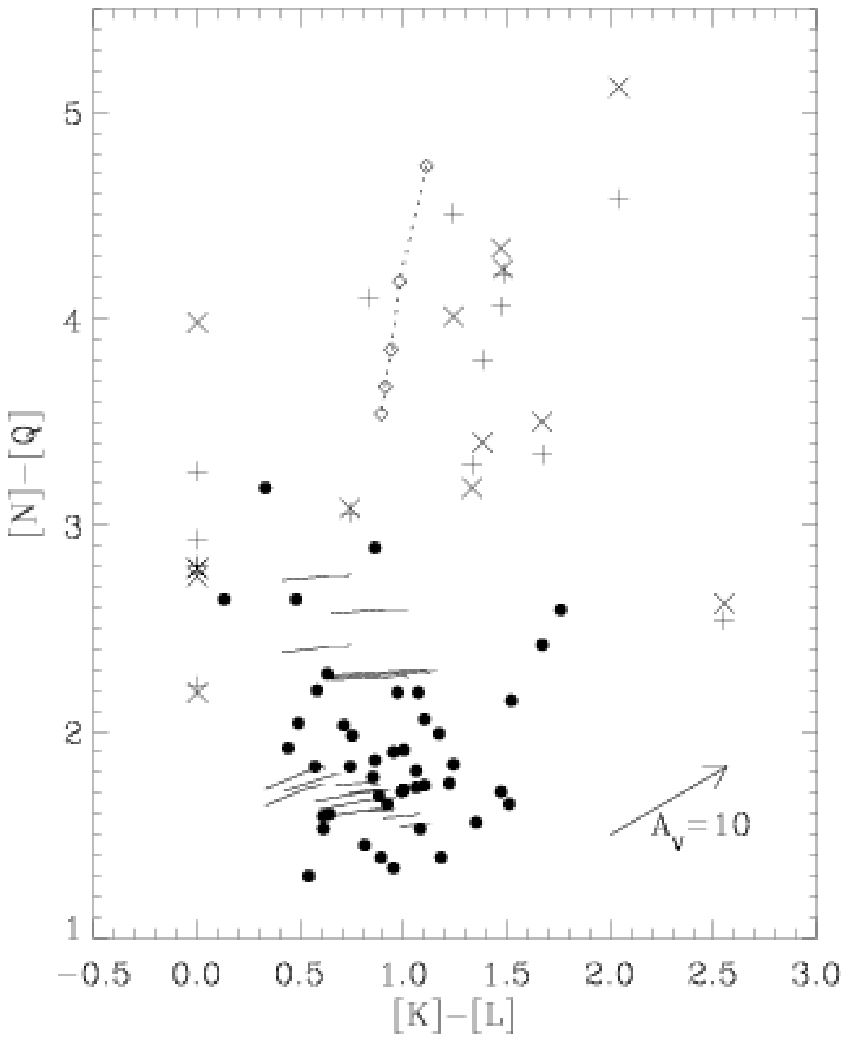}
\caption{Mid-IR color-color plot for the Orion and Taurus-Auriga sources. The K- and L-band data for Orion are from \citet{Muench+02}.
The cross ($\times$) and plus ($+$) signs refer to the MAX 10~\micron\ data taken in 1998 and 2000, respectively. Filled circles are for the Taurus dataset. Solid lines represent the colora of the \citet{Calvet+92} flaring disk models. The diamonds and the relative dotted line represent the model of \citet{Robberto+04} for a photoevaporated disk in a HII region oriented face-on with respect to both the ionizing star and the observer. The points refer to different distances from the ionizing star, from top to bottom: 30\arcsec, 60\arcsec, 90\arcsec, 120\arcsec and 150\arcsec, corresponding to 0.065, 0.13, 0.20, 0.26 and 0.33~pc, respectively, at the assumed distance $d=450$~pc of the Orion Nebula.}
\label{Fig27}
\end{figure}

\begin{deluxetable}{lcrccl}
\tablecolumns{6}
\tablewidth{0pc}
\tablecaption{Main observational parameters}
\tablehead{
 \colhead{Night} & \colhead{Filter}  &  \colhead{I.T.} &\colhead{Chop. Freq.} &\colhead{I.T./frame} &\colhead{Chop. Throw} \\
		     & 			 & \colhead{(ms)}  &\colhead{Hz}          &\colhead{ s} &\colhead{arcsec}
}
\startdata
27 November 1998 & N &	 20.2 & 4.2  & 20.07 & \phn9.80 (N-S)  \\
28 November 1998 & N &	 20.2 & 4.2  & 40.14 & 18.55 (N-S) \\
18 December 2000 & N &	 9.9 & 3.9  & 25.36 & 24.64 (E-W) \\
19 December 2000  & N &	 9.9 & 3.9  & 25.36 & 24.64 (E-W) \\
19 December 2000  & Q &	 9.9 & 3.9  & 12.68 & 24.64 (E-W) \\
\hline
18 December 2000 & silicates &20.2 & 4.12  & 12.94\rlap{4} & 24.64 (E-W) 
\enddata
\label{Tab_parameters}
\end{deluxetable}

\begin{deluxetable}{lcccccc}
\tablecolumns{7}
\tablewidth{0pc}
\tablecaption{Calibrators for narrow-band imaging}
\tablehead{
\colhead{Source}   &   \colhead{7.7~\micron} & \colhead{8.7~\micron} &\colhead{9.8~\micron} &\colhead{10.2~\micron} &\colhead{11.2~\micron} &\colhead{12.4~\micron} 
}
\startdata
HR~1040  	&  2.55    	&   	 2.55 &  	 2.55 &      2.55 &	 2.55 & 	 2.55 \\
HR~1457  	& \llap{--}2.97     &     \llap{--}2.94 &     \llap{--}3.00 &	\llap{--}3.04 & 	\llap{--}3.07 & 	\llap{--}3.07 \\
HR~1713  	&  0.06	& 	 0.06 & 	 0.06	& 	 0.06 & 	 0.06	& 	 0.06 \\
HR~4069  	& \llap{--}0.84	&  	\llap{--}0.83 &	\llap{--}0.91 &	\llap{--}0.96 &	\llap{--}1.00 &     \llap{--}1.00 \\
HR~2943  	& \llap{--}0.77	&     \llap{--}0.77 &     \llap{--}0.76	&	\llap{--}0.76 &	\llap{--}0.76 &	\llap{--}0.76 \\
HR~5340  	& \llap{--}3.11	&	\llap{--}3.10	&	\llap{--}3.14 &	\llap{--}3.17 &	\llap{--}3.18 &	\llap{--}3.18 \\	
\hline
F$_{m=0}$(Jy) & \llap{6}3.56	&	\llap{5}0.86	&	\llap{4}1.64	&	\llap{3}6.67	&	\llap{2}8.94	&	\llap{2}5.28 
\enddata
\label{Tab_SILICATES}
\end{deluxetable}

\clearpage
\LongTables
\begin{landscape}
\begin{deluxetable}{ccccccccccl}
\tablecolumns{11}
\tablewidth{0pc}
\tablecaption{Mid-IR Photometry of Sources in the Orion Nebula}
\tablehead{
\colhead{MAX}&\colhead{H97}&\colhead{HC00}&\colhead{MLLA00}&\colhead{R.A.} &
\colhead{Decl.}&\colhead{[N]}&\colhead{[N]}&\colhead{[N]}&\colhead{[Q]}&
\colhead{Note} \\
 & &  &     &(J2000.0)      &
(J2000.0)     & (Nov.~1998)      & (Dec.~2000)     & (mosaic)        &	&
}
\startdata
1   &   265 &       &   234 & $5^h35^m05\fs52$ & $-5\arcdeg24\arcmin09\farcs9$ &  			&   6.55$\pm$0.14    &       		&  & MAX source is +3.53"E,-0.28"N   \\
2   &   266 &       &   223 & $5^h35^m05\fs55$ & $-5\arcdeg24\arcmin14\farcs5$ &  			&   7.50$\pm$0.39    &       		&  & MAX source is +3.44"E,-0.19'N   \\
3   &   274 &       &   314 & $5^h35^m05\fs74$ & $-5\arcdeg23\arcmin44\farcs5$ &  			&   6.72$\pm$0.16    &       		&  & MAX source is +2.10"E,0.19"S    \\
4   &   278 &       &   161 & $5^h35^m05\fs74$ & $-5\arcdeg24\arcmin18\farcs4$ & 			&   5.90$\pm$0.11    &       		&  & MAX source is +1.55"E,0.32"S    \\
5   &   291 &   99  &   161 & $5^h35^m06\fs45$ & $-5\arcdeg24\arcmin39\farcs0$ &  			&                    &6.85$\pm$0.15	& 	\\
6   &   341 &   305 &   409 & $5^h35^m10\fs23$ & $-5\arcdeg23\arcmin21\farcs0$ & 			&   4.70$\pm$0.09    &       		&       \\
7   &       &       &       & $5^h35^m10\fs23$ & $-5\arcdeg21\arcmin43\farcs5$ & 			&   6.48$\pm$0.18    &       		&       \\
8   &       &   320 &   425 & $5^h35^m10\fs28$ & $-5\arcdeg23\arcmin15\farcs8$ &  			&   7.46$\pm$0.14    &       		&       \\
9   &   349 &   146 &   220 & $5^h35^m10\fs56$ & $-5\arcdeg24\arcmin17\farcs2$ &  			&   6.78$\pm$0.10    &       		&       \\
10  &   347 &   554 &   733 & $5^h35^m10\fs64$ & $-5\arcdeg21\arcmin56\farcs8$ &  			&   6.16$\pm$0.13    &       		&       \\
11  &   392 &   385 &   507 & $5^h35^m10\fs64$ & $-5\arcdeg22\arcmin55\farcs8$ &  			&   6.14$\pm$0.09    &       		&       \\
12  &   352 &   224 &   316 & $5^h35^m10\fs75$ & $-5\arcdeg23\arcmin44\farcs1$ &  			&   7.27$\pm$0.13    &       		&       \\
13  &   355 &   416 &   542 & $5^h35^m10\fs90$ & $-5\arcdeg22\arcmin46\farcs1$ &  			&   6.37$\pm$0.10    &       		&       \\
14  &   356 &   93  &   148 & $5^h35^m10\fs94$ & $-5\arcdeg24\arcmin48\farcs5$ &  			&   6.00$\pm$0.12    &       		&       \\
15  &       &   415 &   541 & $5^h35^m11\fs04$ & $-5\arcdeg22\arcmin46\farcs3$ &  			&   6.50$\pm$0.11    &       		&       \\
16  &       &       &       & $5^h35^m11\fs11$ & $-5\arcdeg21\arcmin53\farcs0$ &  			&   6.38$\pm$0.17    &       		&       \\
17  &       &   434 &   575 & $5^h35^m11\fs20$ & $-5\arcdeg22\arcmin37\farcs6$ &  			&   5.06$\pm$0.09    &       		&       \\
18  &       &   105 &   171 & $5^h35^m11\fs32$ & $-5\arcdeg24\arcmin38\farcs4$ &  			&   6.34$\pm$0.13    &       		&       \\
19  &       &   203 &   289 & $5^h35^m11\fs49$ & $-5\arcdeg23\arcmin51\farcs5$ &  			&   6.95$\pm$0.10    &       		&       \\
20  &   366 &   140 &   207 & $5^h35^m11\fs65$ & $-5\arcdeg24\arcmin21\farcs9$ &  			&   7.99$\pm$0.19    &       		&       \\
21  &       &   611 &   809 & $5^h35^m11\fs75$ & $-5\arcdeg21\arcmin31\farcs8$ &  			&                    &   7.23$\pm$0.3   &	\\
22  &       &   672 &   880 & $5^h35^m11\fs88$ & $-5\arcdeg20\arcmin59\farcs9$ &  			&                    &   5.70$\pm$0.2   &	\\
23  &   3058&   664 &   872 & $5^h35^m11\fs92$ & $-5\arcdeg21\arcmin02\farcs8$ &  			&                    &   6.85$\pm$0.3   &	\\
24  &       &       &   710 & $5^h35^m11\fs95$ & $-5\arcdeg22\arcmin01\farcs5$ &  			&                    &   7.17$\pm$0.4   &	\\
25  &   9009&   395 &   516 & $5^h35^m11\fs98$ & $-5\arcdeg22\arcmin53\farcs6$ &  			&   6.03$\pm$0.10    &       		&       \\
26  &  373  &       &   944 & $5^h35^m12\fs01$ & $-5\arcdeg20\arcmin33\farcs8$ &  			&                    &   6.10$\pm$0.3   &	\\
27  & 378a,b&   215 & 302A,B& $5^h35^m12\fs27$ & $-5\arcdeg23\arcmin47\farcs8$ &  			&   6.00$\pm$0.10    &       		&       \\
28  & 377a,b&   698 & 914A,B& $5^h35^m12\fs30$ & $-5\arcdeg20\arcmin45\farcs6$ &  			&                    &   5.24$\pm$0.3   &	\\
29  &       &   761 &   793 & $5^h35^m12\fs41$ & $-5\arcdeg21\arcmin36\farcs7$ &  			&   8.08$\pm$0.25    &       		&       \\
30  &   9018&   368 &   482 & $5^h35^m12\fs58$ & $-5\arcdeg23\arcmin02\farcs0$ &  			&   7.43$\pm$0.15    &       		&       \\
31  &   385 &   228 &   318 & $5^h35^m12\fs60$ & $-5\arcdeg23\arcmin43\farcs9$ &  			&   5.49$\pm$0.10    &       		&  1.39$\pm$0.20     \\
32  &   391 &   699 &   916 & $5^h35^m12\fs85$ & $-5\arcdeg20\arcmin44\farcs8$ &  			&                    &   8.97$\pm$0.3   &	\\
33  &       &606,608& 804A,B& $5^h35^m12\fs85$ & $-5\arcdeg21\arcmin33\farcs6$ &  6.33$\pm$0.13    	&   		     &       		&       \\
34  &   9029&   505 &   673 & $5^h35^m13\fs07$ & $-5\arcdeg22\arcmin16\farcs1$ &  7.36$\pm$0.31    	&   7.38$\pm$0.29    &       		&       \\
35  & 399a,b&   487 & 651A,B& $5^h35^m13\fs19$ & $-5\arcdeg22\arcmin21\farcs8$ &  7.27$\pm$0.19    	&   7.04$\pm$0.18    &       		&       \\
36  &       &       &       & $5^h35^m13\fs19$ & $-5\arcdeg22\arcmin04\farcs6$ &  7.14$\pm$0.12    	&   7.35$\pm$0.19    &       		&       \\
37  &   401 &   80  &   136 & $5^h35^m13\fs20$ & $-5\arcdeg24\arcmin55\farcs5$ &  			&   3.79$\pm$0.11    &       		&       \\
38  &   400 &   682 &   896 & $5^h35^m13\fs24$ & $-5\arcdeg20\arcmin54\farcs7$ &  6.34$\pm$0.20    	&   		     &       		&       \\
39  &       &       &       & $5^h35^m13\fs31$ & $-5\arcdeg21\arcmin53\farcs6$ &  7.55$\pm$0.26    	&   		     &       		&       \\
40  &       &   273 &   369 & $5^h35^m13\fs40$ & $-5\arcdeg23\arcmin29\farcs5$ &  6.22$\pm$0.09    	&   6.21$\pm$0.11    &       		&       \\
41  &   409 &   240 &   327 & $5^h35^m13\fs44$ & $-5\arcdeg23\arcmin40\farcs2$ &  6.44$\pm$0.09    	&   6.22$\pm$0.10    &       		&       \\
42  &       &   178 &   263 & $5^h35^m13\fs55$ & $-5\arcdeg24\arcmin00\farcs2$ &  6.74$\pm$0.09    	&   6.87$\pm$0.12    &       		&       \\
43  &       &   192 &   276 & $5^h35^m13\fs58$ & $-5\arcdeg23\arcmin55\farcs0$ &  6.40$\pm$0.09    	&   6.24$\pm$0.11    &       		&  2.9$\pm$0.3     \\
44  &       &   602 &   797 & $5^h35^m13\fs71$ & $-5\arcdeg21\arcmin35\farcs7$ &  6.43$\pm$0.09    	&   6.39$\pm$0.13    &       		&       \\
45  &   424 &   541 &   715 & $5^h35^m13\fs79$ & $-5\arcdeg22\arcmin03\farcs5$ &  			&   6.28$\pm$0.50    &       		&       \\
46  &   431 &   242 &   328 & $5^h35^m13\fs79$ & $-5\arcdeg23\arcmin40\farcs1$ &  2.69$\pm$0.08    	&   2.78$\pm$0.09    &       		&  0.68$\pm$0.19    \\
47  &       &       &       & $5^h35^m13\fs8 $ & $-5\arcdeg23\arcmin57\farcs0$ &  			&       	     &                  & 1.55$\pm$0.18    \\
48  &   423 &   703 &   703 & $5^h35^m13\fs81$ & $-5\arcdeg22\arcmin07\farcs8$ &  4.90$\pm$0.08    	&   4.70$\pm$0.09    &       		&       \\
49  &       &   548 &   726 & $5^h35^m13\fs81$ & $-5\arcdeg22\arcmin00\farcs6$ &  1.98$\pm$0.08    	&   1.90$\pm$0.09    &       		&  -0.64$\pm$0.16   \\
50  &       &       &       & $5^h35^m13\fs9 $ & $-5\arcdeg22\arcmin12\farcs0$ &  			&       	     &                  & 0.98$\pm$0.22    \\
51  &   9053&   629 &   828 & $5^h35^m13\fs96$ & $-5\arcdeg21\arcmin22\farcs8$ &  5.99$\pm$0.08    	&   		     &       		&       \\
52  &       &   314 &   413 & $5^h35^m13\fs97$ & $-5\arcdeg23\arcmin20\farcs1$ &  			&                    &   6.73$\pm$0.3   &	\\
53  &       &   134 &   199 & $5^h35^m14\fs01$ & $-5\arcdeg24\arcmin25\farcs4$ &  			&   7.47$\pm$0.21    &       		&       \\
54  &       &       &       & $5^h35^m14\fs1 $ & $-5\arcdeg23\arcmin00\farcs0$ &  			&       	     &                  & 1.65$\pm$0.21    \\
55  &       &   704 &   707 & $5^h35^m14\fs12$ & $-5\arcdeg22\arcmin05\farcs4$ &  6.78$\pm$0.50    	&   		     &       		&       \\
56  &       &       &       & $5^h35^m14\fs18$ & $-5\arcdeg23\arcmin14\farcs8$ &  			&   5.99$\pm$0.21    &       		&       \\
57  &   437 &   135 &   203 & $5^h35^m14\fs29$ & $-5\arcdeg24\arcmin24\farcs7$ &  5.57$\pm$0.09    	&   5.66$\pm$0.10    &       		&       \\
58  &       &   361 &   476 & $5^h35^m14\fs30$ & $-5\arcdeg23\arcmin04\farcs3$ &  6.04$\pm$0.09    	&   6.51$\pm$0.11    &       		&       \\
59  &   9061&   345 &   452 & $5^h35^m14\fs31$ & $-5\arcdeg23\arcmin08\farcs4$ &  5.81$\pm$0.08    	&   5.51$\pm$0.10    &       		&       \\
60  &   9062&   399 &   518 & $5^h35^m14\fs33$ & $-5\arcdeg22\arcmin53\farcs7$ &  6.16$\pm$0.09    	&   6.18$\pm$0.14    &       		&       \\
61  &       &       &   292 & $5^h35^m14\fs38$ & $-5\arcdeg23\arcmin50\farcs6$ &  4.40$\pm$0.08    	&   3.68$\pm$0.09    &       		&   0.42$\pm$0.18    \\
62  &   441 &   258 &   350 & $5^h35^m14\fs40$ & $-5\arcdeg23\arcmin33\farcs8$ &  5.83$\pm$0.08    	&   6.05$\pm$0.11    &       		&       \\
63  &   9069&   369 &   485 & $5^h35^m14\fs58$ & $-5\arcdeg23\arcmin02\farcs6$ &  			&   6.97$\pm$0.19    &       		&  & MAX source is -2.80E,+0.29"S    \\
64  &       &189,193& 273A,B& $5^h35^m14\fs54$ & $-5\arcdeg23\arcmin55\farcs7$ &  5.44$\pm$0.08    	&   5.27$\pm$0.09    &       		&   1.91$\pm$0.22    \\
65  &       &   411 &   534 & $5^h35^m14\fs63$ & $-5\arcdeg22\arcmin50\farcs2$ &  			&   6.95$\pm$0.19    &       		&  & MAX source is -2.25"E,-0.52"N   \\
66  &9073,9075& 714 & 467A,B& $5^h35^m14\fs84$ & $-5\arcdeg23\arcmin05\farcs0$ &  6.58$\pm$0.19    	&   		     &       		&  & MAX source is -3.35"E,0.24"N    \\
67  & 445a,b&       & 922A,B& $5^h35^m14\fs69$ & $-5\arcdeg20\arcmin43\farcs7$ &  			&                    &   6.60$\pm$0.3   &	\\
68  &   9071&   657 &   863 & $5^h35^m14\fs71$ & $-5\arcdeg21\arcmin05\farcs9$ &  6.32$\pm$0.10    	&   		     &       		&       \\
69  &   454 &   431 &   568 & $5^h35^m14\fs91$ & $-5\arcdeg22\arcmin39\farcs6$ &  4.79$\pm$0.08    	&   4.75$\pm$0.10    &       		&       \\
70  &       &   437 &   580 & $5^h35^m15\fs19$ & $-5\arcdeg22\arcmin37\farcs3$ &  6.59$\pm$0.10    	&   6.53$\pm$0.26    &       		&       \\
71  &   463 &   398 &   514 & $5^h35^m15\fs21$ & $-5\arcdeg22\arcmin54\farcs7$ &  4.44$\pm$0.08    	&   4.28$\pm$0.10    &       		&       \\
72  &       &       &       & $5^h35^m15\fs34$ & $-5\arcdeg22\arcmin50\farcs4$ &  7.21$\pm$0.17    	&   		     &       		&       \\
73  &   470 &   504 &   671 & $5^h35^m15\fs34$ & $-5\arcdeg22\arcmin16\farcs3$ &  4.85$\pm$0.08    	&   4.60$\pm$0.10    &       		&       \\
74  &472a,b-9096&475,476&630A,B,C& $5^h35^m15\fs35$& $-5\arcdeg22\arcmin25\farcs7$&  7.00$\pm$0.11    	&   7.07$\pm$0.32    &       		&       \\
75  &   473 &   646 &   848 & $5^h35^m15\fs40$ & $-5\arcdeg21\arcmin13\farcs6$ &  5.45$\pm$0.08    	&           	     &       		&       \\
76  &   475 &   223 &   311 & $5^h35^m15\fs43$ & $-5\arcdeg23\arcmin45\farcs0$ &  6.81$\pm$0.09    	&   7.26$\pm$0.16    &       		&       \\
77  &       &       &       & $5^h35^m15\fs49$ & $-5\arcdeg23\arcmin52\farcs3$ &  7.31$\pm$0.17    	&   		     &       		&       \\
78  &       &       &       & $5^h35^m15\fs50$ & $-5\arcdeg23\arcmin48\farcs2$ &  			&   6.21$\pm$0.11    &       		&       \\
79  &   5178&   251 &   340 & $5^h35^m15\fs52$ & $-5\arcdeg23\arcmin37\farcs5$ &  6.51$\pm$0.09    	&   5.97$\pm$0.12    &       		&   1.39$\pm$0.20    \\
80  &       &       &       & $5^h35^m15\fs6 $ & $-5\arcdeg23\arcmin10\farcs3$ &  			&       	     &                  & 1.59$\pm$0.21    \\
81  &   480 &   172 &   249 & $5^h35^m15\fs64$ & $-5\arcdeg24\arcmin03\farcs9$ &  6.26$\pm$0.09    	&   6.49$\pm$0.11    &       		&       \\
82  &   479 &   389 &   503 & $5^h35^m15\fs65$ & $-5\arcdeg22\arcmin56\farcs7$ &  3.93$\pm$0.08    	&   3.87$\pm$0.09    &       		&       \\
83  &   488w&   307 &   405 & $5^h35^m15\fs72$ & $-5\arcdeg23\arcmin22\farcs7$ &  5.75$\pm$0.09    	&   5.47$\pm$0.11    &       		&   1.41$\pm$0.24\tablenotemark{d}       \\
84  &   492 &   420 &   545 & $5^h35^m15\fs78$ & $-5\arcdeg22\arcmin46\farcs4$ &  8.33$\pm$0.82    	&   		     &       		&       \\
85  &   1865&   336 &   440 & $5^h35^m15\fs80$ & $-5\arcdeg23\arcmin14\farcs6$ &  5.43$\pm$0.25    	&   		     &       		&       \\
86  &   489 &   287 &   381 & $5^h35^m15\fs80$ & $-5\arcdeg23\arcmin26\farcs7$ &  5.44$\pm$0.08    	&   5.28$\pm$0.11    &  	        &	\\
87  &   488a&   306 &   404 & $5^h35^m15\fs86$ & $-5\arcdeg23\arcmin22\farcs5$ &  4.48$\pm$0.08    	&   4.46$\pm$0.11    &       		&   1.41$\pm$0.24\tablenotemark{d}    \\
88  &       &   291 &   385 & $5^h35^m15\fs86$ & $-5\arcdeg23\arcmin25\farcs0$ &  5.65$\pm$0.18    	&   		     &   		& & OW-158-326\tablenotemark{a}        \\
89  &   490 &       &   925 & $5^h35^m15\fs88$ & $-5\arcdeg20\arcmin39\farcs7$ &  			&                    &   8.83$\pm$0.3   &	\\
90  &   9118&   145 &   213 & $5^h35^m15\fs90$ & $-5\arcdeg24\arcmin18\farcs0$ &  			&   6.91$\pm$0.12    &       		&       \\
91  &   496 &   490 &   653 & $5^h35^m15\fs93$ & $-5\arcdeg22\arcmin21\farcs8$ &  6.26$\pm$0.09    	&   		     &       		&       \\
%88  &       &       &       & $5^h35^m15\fs95$ & $-5\arcdeg24\arcmin13\farcs9$ &  6.67$\pm$0.58    	&   		     &       		&       \\
92  &   499 &   213 &   296 & $5^h35^m15\fs95$ & $-5\arcdeg23\arcmin49\farcs9$ &  3.18$\pm$0.08    	&   3.40$\pm$0.09    &       		&   1.42$\pm$0.19    \\
93  &       &       &       & $5^h35^m15\fs97$ & $-5\arcdeg22\arcmin36\farcs1$ &  6.86$\pm$0.22    	&   		     &       		&       \\
94  &   503 &   202 &   288 & $5^h35^m16\fs01$ & $-5\arcdeg23\arcmin52\farcs8$ &  5.16$\pm$0.08    	&   5.08$\pm$0.10    &       		&       \\
95  &       &       &       & $5^h35^m16\fs05$ & $-5\arcdeg22\arcmin33\farcs4$ &  6.68$\pm$0.10    	&   		     &       		&       \\
96  &   9128&   393 &   511 & $5^h35^m16\fs07$ & $-5\arcdeg22\arcmin54\farcs4$ &  5.50$\pm$0.08    	&   5.07$\pm$0.10    &       		&       \\
97  &1863b  &   350 &   460B& $5^h35^m16\fs08$ & $-5\arcdeg23\arcmin07\farcs2$ &  3.63$\pm$0.09    	&   3.66$\pm$0.09    &       		&   1.44$\pm$0.19    \\
98  &       &   296 &   389 & $5^h35^m16\fs09$ & $-5\arcdeg23\arcmin24\farcs2$ &  3.93$\pm$0.08    	&   3.98$\pm$0.10    &       		&   1.18$\pm$0.19    \\
99  &       &  504a,b&   926 & $5^h35^m16\fs13$ & $-5\arcdeg20\arcmin38\farcs4$ &  			&  		     &   8.13$\pm$0.3   &	\\
100  &       &       &       & $5^h35^m16\fs15$ & $-5\arcdeg22\arcmin26\farcs2$ &  6.86$\pm$0.19    	&   		     &       		&   & to be confirmed  \\
101  & 1863a &   354 &   460A& $5^h35^m16\fs15$ & $-5\arcdeg23\arcmin07\farcs1$ &  5.76$\pm$0.37    	&   6.52$\pm$0.10    &       		&       \\
102 &511a,b &   522 & 689A,B& $5^h35^m16\fs28$ & $-5\arcdeg22\arcmin10\farcs5$ &  5.63$\pm$0.08    	&   5.65$\pm$0.09    &       		&       \\
103  &   9135&   652 &   857 & $5^h35^m16\fs28$ & $-5\arcdeg21\arcmin08\farcs9$ &  6.40$\pm$0.10    	&   		     &       		&       \\
104 &   513 &    412&   535 & $5^h35^m16\fs31$ & $-5\arcdeg22\arcmin49\farcs0$ &  6.62$\pm$0.33    	&   6.97$\pm$0.19    &       		&       \\
105 &   512 &   322 &   422 & $5^h35^m16\fs32$ & $-5\arcdeg23\arcmin16\farcs7$ &  4.24$\pm$0.08    	&   4.35$\pm$0.09    &       		&   1.06$\pm$0.26    \\
106 &       &       &   401B& $5^h35^m16\fs34$ & $-5\arcdeg23\arcmin22\farcs6$ &  3.06$\pm$0.08    	&   3.07$\pm$0.09    &       		&   0.27$\pm$0.17	& SC3\tablenotemark{b}    \\
107  &       &       &       & $5^h35^m16\fs4 $ & $-5\arcdeg22\arcmin56\farcs0$ &  			&       	     &                  & 1.24$\pm$0.20    \\
108 &   515 &   171 &   246 & $5^h35^m16\fs41$ & $-5\arcdeg24\arcmin04\farcs2$ &  4.94$\pm$0.08    	&   5.15$\pm$0.09    &       		&       \\
109 &       &   112 &   173 & $5^h35^m16\fs41$ & $-5\arcdeg24\arcmin36\farcs9$ &  			&   		     &   8.10$\pm$0.3   &	\\
110 &       &   514 &   682 & $5^h35^m16\fs42$ & $-5\arcdeg22\arcmin13\farcs0$ &  5.52$\pm$0.08    	&   4.88$\pm$0.09    &       		&       \\
111 &   1891&   309 &  401A & $5^h35^m16\fs46$ & $-5\arcdeg23\arcmin22\farcs7$ &  4.23$\pm$0.08    	&   4.62$\pm$0.09    &      	 	&       \\
112 &       &       &       & $5^h35^m16\fs51$ & $-5\arcdeg22\arcmin48\farcs0$ &  			&   6.79$\pm$0.65    &       		&       \\
113 &       &       &       & $5^h35^m16\fs51$ & $-5\arcdeg23\arcmin30\farcs2$ &  6.07$\pm$0.31    	&     		     &       		&  & to be confirmed  \\
114 &       &   325 &   426 & $5^h35^m16\fs52$ & $-5\arcdeg23\arcmin15\farcs5$ &  5.45$\pm$0.25    	&   		     &       		&       \\
115  &       &       &       & $5^h35^m16\fs7 $ & $-5\arcdeg22\arcmin57\farcs0$ &  			&       	     &                  & 1.12$\pm$0.21    \\
116 &   524 &   323 &   424 & $5^h35^m16\fs76$ & $-5\arcdeg23\arcmin16\farcs5$ &  3.86$\pm$0.08    	&   4.26$\pm$0.09    &       		&   0.46$\pm$0.23    \\
117 &       &       &       & $5^h35^m16\fs82$ & $-5\arcdeg23\arcmin58\farcs2$ &  			&   5.97$\pm$0.74    &       		&       \\
118 &       &   289 & 383A,B& $5^h35^m16\fs84$ & $-5\arcdeg23\arcmin26\farcs0$ &  2.15$\pm$0.08    	&   2.29$\pm$0.10    &       		&   -0.64$\pm$0.18    \\
119 &   532 &   414 &   537 & $5^h35^m16\fs98$ & $-5\arcdeg22\arcmin49\farcs1$ &  			&   6.95$\pm$0.21    &       		&       \\
120 &       &       &       & $5^h35^m16\fs98$ & $-5\arcdeg22\arcmin31\farcs0$ &  			&   6.97$\pm$0.42    &       		&       \\
121 &   534 &   252 &   341 & $5^h35^m16\fs99$ & $-5\arcdeg23\arcmin37\farcs1$ &  4.90$\pm$0.08    	&   5.40$\pm$0.09    &       		&   0.89$\pm$0.21    \\
122 &   533 &   375 &   489 & $5^h35^m16\fs99$ & $-5\arcdeg23\arcmin01\farcs3$ &  6.32$\pm$0.09    	&   5.94$\pm$0.12    &       		&       \\
123 &   537 &   244 &   330 & $5^h35^m17\fs03$ & $-5\arcdeg23\arcmin40\farcs1$ &  			&   6.65$\pm$0.19    &       		&       \\
124 &   538 &   259 &   348 & $5^h35^m17\fs06$ & $-5\arcdeg23\arcmin34\farcs0$ &  4.45$\pm$0.08    	&   4.24$\pm$0.10    &       		&       \\
125 &       &       &       & $5^h35^m17\fs11$ & $-5\arcdeg23\arcmin26\farcs6$ &  			&   5.09$\pm$0.12    &       		&   0.78$\pm$0.25  \\
126 &       &       &       & $5^h35^m17\fs11$ & $-5\arcdeg23\arcmin29\farcs3$ &  			&   5.02$\pm$0.11    &       		&  & to be confirmed  \\
127 &   1889&       & 423A,B& $5^h35^m17\fs26$ & $-5\arcdeg23\arcmin17\farcs0$ &  			&   5.97$\pm$0.46    &       		&       \\
128 &   548 &   440 &   584 & $5^h35^m17\fs37$ & $-5\arcdeg22\arcmin35\farcs6$ &  5.52$\pm$0.08    	&   5.43$\pm$0.10    &       		&       \\
129 &   549 &   180 &   261 & $5^h35^m17\fs39$ & $-5\arcdeg24\arcmin00\farcs1$ &  6.03$\pm$0.08    	&   6.00$\pm$0.10    &       		&       \\
130 &   9180&   313 &   411 & $5^h35^m17\fs46$ & $-5\arcdeg23\arcmin21\farcs1$ &  5.41$\pm$0.09    	&   4.89$\pm$0.34    &       		&       \\
131 &   554 &   295 &   387 & $5^h35^m17\fs55$ & $-5\arcdeg23\arcmin25\farcs0$ &  5.94$\pm$0.09    	&   5.92$\pm$0.11    &       		&   1.7$\pm$0.3    \\
132 & 553a,b&   388 &499.2A,B\tablenotemark{c}& $5^h35^m17\fs57$ & $-5\arcdeg22\arcmin57\farcs10$&5.67$\pm$0.09&5.72$\pm$0.10&      		&       \\
133 & 552a,b&   586 & 773A,B& $5^h35^m17\fs57$ & $-5\arcdeg21\arcmin45\farcs9$ &  5.30$\pm$0.08    	&   5.15$\pm$0.11    &       		&       \\
134 &       &   563 &   743 & $5^h35^m17\fs62$ & $-5\arcdeg21\arcmin54\farcs3$ &  6.33$\pm$0.09    	&   6.69$\pm$0.17    &       		&       \\
135 &   558 &   241 &   325A& $5^h35^m17\fs68$ & $-5\arcdeg23\arcmin41\farcs1$ &  6.44$\pm$0.10    	&   		     &       		&       \\
136 &   9195&   234 &   320 & $5^h35^m17\fs75$ & $-5\arcdeg23\arcmin42\farcs7$ &  6.88$\pm$0.11    	&   		     &       		&       \\
137 &   561 &   681 &   892 & $5^h35^m17\fs82$ & $-5\arcdeg20\arcmin54\farcs2$ &  			&   		     &   7.77$\pm$0.3   &	\\
138 &   9201&   367 &   480 & $5^h35^m17\fs90$ & $-5\arcdeg23\arcmin02\farcs0$ &  5.91$\pm$0.32    	&   		     &       		&       \\
139 &       &   542 &   714 & $5^h35^m17\fs90$ & $-5\arcdeg22\arcmin03\farcs4$ &  6.16$\pm$0.09    	&   6.07$\pm$0.11    &      		&       \\
140 &   567 &   425 &   549 & $5^h35^m17\fs92$ & $-5\arcdeg22\arcmin44\farcs6$ &  6.13$\pm$0.09    	&   6.27$\pm$0.11    &       		&       \\
141 &   9196&   230 &   317 & $5^h35^m17\fs93$ & $-5\arcdeg23\arcmin44\farcs5$ &  7.46$\pm$0.34    	&   		     &       		&       \\
142 &       &   679 &   890 & $5^h35^m17\fs96$ & $-5\arcdeg20\arcmin55\farcs8$ &  			&   		     &   6.34$\pm$0.3   &	\\
143 &   9210&   174 &   248 & $5^h35^m18\fs04$ & $-5\arcdeg24\arcmin03\farcs0$ &  4.86$\pm$0.08    	&   4.76$\pm$0.10    &       		&       \\
144 &   575 &   177 &   256 & $5^h35^m18\fs08$ & $-5\arcdeg24\arcmin01\farcs2$ &  6.43$\pm$0.12    	&   6.53$\pm$0.10    &       		&       \\
145 & MS-107&   372 &   486 & $5^h35^m18\fs11$ & $-5\arcdeg23\arcmin02\farcs8$ &  6.71$\pm$0.23    	&   		     &       		&       \\
146 &   581 &   253 &   344 & $5^h35^m18\fs21$ & $-5\arcdeg23\arcmin36\farcs0$ &  4.00$\pm$0.08    	&   4.44$\pm$0.09    &       		&       \\
147 &       &   533 &   705 & $5^h35^m18\fs24$ & $-5\arcdeg22\arcmin05\farcs7$ &  7.18$\pm$0.15    	&   7.05$\pm$0.17    &       		&       \\
148 &       &       &       & $5^h35^m18\fs31$ & $-5\arcdeg22\arcmin22\farcs6$ &  7.38$\pm$0.16    	&   		     &       		&       \\
149 &   9220&   348 &   455 & $5^h35^m18\fs36$ & $-5\arcdeg23\arcmin06\farcs7$ &  6.71$\pm$0.17    	&   		     &       		&       \\
150 &   582 &   331 &   431 & $5^h35^m18\fs37$ & $-5\arcdeg23\arcmin15\farcs7$ &  6.93$\pm$0.20    	&   	             &       		&       \\
151 &       &       &       & $5^h35^m18\fs65$ & $-5\arcdeg23\arcmin43\farcs1$ &  			&   7.85$\pm$0.17    &       		&       \\
152 &   596 &   337 &   441 & $5^h35^m18\fs65$ & $-5\arcdeg23\arcmin13\farcs9$ &  3.28$\pm$0.08    	&   3.45$\pm$0.09    &       		&       \\
153 & 598a,b&   713 &499.1A,B\tablenotemark{c}& $5^h35^m18\fs70$ & $-5\arcdeg22\arcmin56\farcs9$&5.16$\pm$0.08&5.37$\pm$0.11&       		&       \\
154 &   607 &   214 &   298 & $5^h35^m19\fs04$ & $-5\arcdeg23\arcmin49\farcs6$ &  			&   3.91$\pm$0.09    &       		&       \\
155 &       &   351 &       & $5^h35^m19\fs6 $ & $-5\arcdeg23\arcmin07\farcs0$ &  			&       	     &                  & 2.8 $\pm$0.25    \\
156 &   622 &   133 &   193 & $5^h35^m19\fs67$ & $-5\arcdeg24\arcmin26\farcs5$ &  			&   5.71$\pm$0.10    &       		&       \\
157 &   624 &   492 &   648 & $5^h35^m19\fs84$ & $-5\arcdeg22\arcmin21\farcs9$ &  			&   6.32$\pm$0.10    &       		&       \\
158 &   3075&   346 &   450 & $5^h35^m20\fs31$ & $-5\arcdeg23\arcmin06\farcs9$ &  			&   		     &   7.57$\pm$0.3   &   	\\
159 & 648a,b&277,279& 366A,B& $5^h35^m20\fs45$ & $-5\arcdeg23\arcmin29\farcs9$ &  			&   6.21$\pm$0.10    &       		&       \\
160 &   9266&   308 &   398 & $5^h35^m20\fs48$ & $-5\arcdeg23\arcmin22\farcs9$ &  			&   7.60$\pm$0.18    &       		&       \\
161 &   652 &   141 &   208 & $5^h35^m20\fs56$ & $-5\arcdeg24\arcmin20\farcs9$ &  			&   7.37$\pm$0.14    &       		&       \\
162 &   658 &   97  &   154 & $5^h35^m20\fs62$ & $-5\arcdeg24\arcmin46\farcs1$ &  			&   5.78$\pm$0.17    &       		&       \\
163 &   660 &   590 &   777 & $5^h35^m20\fs74$ & $-5\arcdeg21\arcmin45\farcs2$ &  			&   6.23$\pm$0.13    &       		&       \\
164 &   661 &   560 &   737 & $5^h35^m20\fs80$ & $-5\arcdeg21\arcmin56\farcs1$ &  			&   6.64$\pm$0.14    &       		&       \\
165 &   665 &   574 &   753 & $5^h35^m20\fs95$ & $-5\arcdeg21\arcmin51\farcs5$ &  			&   6.02$\pm$0.12    &       		&       \\
166 &   669 &   217 &   300 & $5^h35^m21\fs02$ & $-5\arcdeg23\arcmin49\farcs2$ &  			&   7.51$\pm$0.18    &       		&       \\
167 &       &       &       & $5^h35^m21\fs33$ & $-5\arcdeg23\arcmin42\farcs7$ &  			&   6.86$\pm$0.13    &       		&       \\
168 & 687a,b&585,255& 769A,B& $5^h35^m21\fs67$ & $-5\arcdeg21\arcmin48\farcs1$ &  			&   5.86$\pm$0.13    &       		&       \\
169 &   694 &   249 &   332 & $5^h35^m21\fs78$ & $-5\arcdeg23\arcmin39\farcs5$ &  			&   7.12$\pm$0.12    &       		&       \\
170 &   698 &   204 &   282 & $5^h35^m21\fs79$ & $-5\arcdeg23\arcmin53\farcs8$ &  			&   7.57$\pm$0.21    &       		&       \\
171 &   9294&   358 &   465 & $5^h35^m21\fs84$ & $-5\arcdeg23\arcmin06\farcs9$ &  			&   		     &   8.01$\pm$0.3   &	\\
172 &       &       &       & $5^h35^m21\fs94$ & $-5\arcdeg22\arcmin28\farcs9$ &  			&   7.28$\pm$0.42    &       		&       \\
173 &   5179&   121 &   182 & $5^h35^m22\fs10$ & $-5\arcdeg24\arcmin32\farcs9$ &  			&   6.48$\pm$0.10    &       		&       \\
174 &   707 &   139 &   201 & $5^h35^m22\fs20$ & $-5\arcdeg24\arcmin25\farcs0$ &  			&   6.71$\pm$0.11    &       		&       \\
175 &   710 &   155 &   224 & $5^h35^m22\fs32$ & $-5\arcdeg24\arcmin14\farcs5$ &  			&   6.66$\pm$0.11    &       		&       \\
176 &   721 &   233 &   319 & $5^h35^m22\fs54$ & $-5\arcdeg23\arcmin43\farcs8$ &  			&   		     &   7.90$\pm$0.3   &	\\
177 &   723 &   339 &   443 & $5^h35^m22\fs80$ & $-5\arcdeg23\arcmin13\farcs8$ &  			&   5.82$\pm$0.09    &       		&       \\
178 &   1993&   82  & 129A,B& $5^h35^m22\fs90$ & $-5\arcdeg24\arcmin57\farcs5$ &  			&   4.97$\pm$0.18    &       		&       \\
179 &   744 &   262 &   347 & $5^h35^m23\fs78$ & $-5\arcdeg23\arcmin34\farcs6$ &  			&   		     &   6.28$\pm$0.3   &	\\
180 &   9317&   460 &   601 & $5^h35^m24\fs41$ & $-5\arcdeg22\arcmin31\farcs9$ &  			&   		     &   6.41$\pm$0.3   &	\\
181 &   756 &   110 &   164 & $5^h35^m24\fs44$ & $-5\arcdeg24\arcmin41\farcs0$ &  			&   5.50$\pm$0.21    &       		&       \\
182 &   762 &   119 &   176 & $5^h35^m24\fs70$ & $-5\arcdeg24\arcmin36\farcs0$ &  			&   6.31$\pm$0.31    &       		&       \\
183 & 766a,b&   387 & 494A,B& $5^h35^m25\fs07$ & $-5\arcdeg22\arcmin59\farcs4$ &  			&       	     &   6.01$\pm$0.3  
\enddata
\label{Tab_stars10mic}
\tablenotetext{b}{From \citet{ODW94}}
\tablenotetext{b}{From \citet{Hayward+94}}
\tablenotetext{c}{The identification number 499 is used 
the list of Muench et al. (2002) for two different stars (both are binaries).}
\tablenotetext{d}{Source nr.\ 84 and 87 are unresolved in the Q-band}
\end{deluxetable}
\clearpage
\end{landscape}

\begin{deluxetable}{crrrrrrrrrr}
\tablecolumns{11}
\tablewidth{0pc}
\tablecaption{Peak Surface Brightness of the BN/KL complex}
\tablehead{
\colhead{} & \colhead{} & \colhead{} & \multicolumn{6}{c}{Narrow-band} & \multicolumn{2}{c}{Broad-band} \\
\cline{4-9} \cline{10-11} \\
\colhead{Source Name} & \colhead{Offset R.A.}  & \colhead{Offset Dec.}  & \colhead{7.7~\micron}  & \colhead{8.7~\micron} & 
\colhead{9.8~\micron} & \colhead{10.2~\micron} & \colhead{11.2~\micron} & \colhead{12.4~\micron} & \colhead{N-band} & \colhead{Q-band} } 
\startdata
        BN &        0.0 &        0.0 &     122.30 &     106.11 &      62.59 &      57.21 &     126.80 &     143.19 &     119.58 &     171.34 \\
 BN SW arc &       --1.4 &       --0.6 &       9.82 &      11.35 &       7.22 &      10.31 &      16.29 &      20.78 &       8.65 &      59.76 \\
      IR n &        3.3 &       --9.8 &       2.11 &       1.06 &       1.62 &       1.22 &       3.23 &       7.14 &       2.12 &      31.53 \\
  IRc2 A+B &        5.7 &       --7.0 &      15.21 &       1.07 &       1.38 &       0.89 &       5.62 &      20.24 &       3.52 &      34.07 \\
    IRc2 C &        4.6 &       --7.8 &       7.98 &       0.73 &       1.50 &       0.95 &       4.67 &      14.94 &       2.77 &      36.67 \\
    IRc2 D &        4.8 &       --7.0 &      10.37 &       1.58 &       1.59 &       1.14 &       6.67 &      19.64 &       3.85 &      41.70 \\
    IRc2 E &        4.5 &       --6.3 &       4.55 &       1.08 &       1.02 &       0.86 &       3.96 &      10.17 &       1.86 &      28.45 \\
    IRc3 N &       --1.6 &       --7.2 &       0.62 &       0.44 &       2.32 &       1.64 &       4.02 &       7.75 &       1.94 &      66.69 \\
    IRc3 S &       --1.4 &       --7.9 &       0.69 &       0.43 &       2.71 &       1.75 &       4.32 &       8.81 &       2.23 &      74.34 \\
      IRc4 &        1.1 &      --11.1 &       1.63 &       0.83 &       3.58 &       2.72 &      10.13 &      20.15 &       4.76 &     101.36 \\
      IRc5 &        0.5 &      --14.5 &       0.56 &       0.30 &       2.08 &       1.30 &       3.45 &       7.56 &       1.73 &      59.99 \\
      IRc6 &        0.4 &       --4.5 &       0.91 &       0.42 &       2.13 &       1.40 &       3.92 &       7.74 &       2.11 &      54.24 \\
    IRc6 N &       --0.2 &       --3.2 &       0.89 &       0.61 &       2.30 &       1.64 &       3.58 &       5.97 &       1.99 &      52.92 \\
    IRc6 E &        1.8 &       --4.8 &       0.75 &       0.24 &       1.43 &       0.96 &       2.81 &       6.13 &       1.52 &      38.26 \\
      IRc7 &        2.8 &       --7.8 &       4.29 &       1.21 &       2.84 &       1.89 &       8.86 &      23.88 &       4.93 &      73.93 \\
      IRc8 &        8.4 &      --12.2 &       0.13 &       0.07 &       1.02 &       0.63 &       1.07 &       2.04 &       0.58 &      28.24 \\
     IRc11 &        8.5 &       --7.4 &       1.38 &       0.32 &       0.89 &       0.64 &       1.90 &       5.28 &       1.19 &      20.76 \\
     IRc12 &        9.9 &       --7.9 &       1.58 &       0.22 &       0.72 &       0.51 &       1.45 &       3.64 &       0.85 &      16.94 \\
     IRc13 &       14.1 &       --4.8 &       0.14 &       0.06 &       0.56 &       0.35 &       0.55 &       1.14 &       0.29 &      14.37 \\
     IRc14 &       13.8 &       --6.7 &       0.18 &       0.09 &       0.78 &       0.49 &       0.96 &       1.87 &       0.52 &      19.42 \\
     IRc15 &        3.1 &       --2.3 &       0.45 &       0.23 &       1.30 &       0.87 &       1.91 &       3.46 &       1.15 &      33.98 \\
   IRc16 N &        4.9 &        1.3 &       0.71 &       0.24 &       1.17 &       0.81 &       1.83 &       3.09 &       0.99 &      30.66 \\
   IRc16 S &        5.0 &       --0.3 &       0.51 &       0.18 &       1.17 &       0.77 &       1.77 &       3.26 &       1.37 &      31.12 \\
   IRc16 W &        4.0 &        0.2 &       0.47 &       0.23 &       1.20 &       0.83 &       1.61 &       2.73 &       1.58 &      30.73 \\
     IRc17 &        3.0 &        2.3 &       0.42 &       0.26 &       1.17 &       0.77 &       1.35 &       2.08 &       0.69 &      28.38 \\
     IRc19 &        5.2 &        4.7 &       0.74 &       0.24 &       0.68 &       0.44 &       0.97 &       1.37 &       0.56 &      16.06 \\
     IRc20 &        0.7 &       --6.9 &       0.74 &       0.31 &       1.36 &       0.90 &       2.22 &       5.17 &       1.10 &      37.28 \\
     IRc21 &        4.4 &       --5.0 &       0.84 &       0.19 &       0.64 &       0.46 &       1.29 &       3.21 &       0.67 &      16.82 \\
     IRc22 &        6.4 &      --11.3 &       0.23 &       0.08 &       0.72 &       0.46 &       0.76 &       1.83 &       0.50 &      17.30 \\
     IRc23 &        6.3 &        2.2 &       0.38 &       0.12 &       0.56 &       0.35 &       0.84 &       1.59 &       0.45 &      15.81 \\
    Knot 1 &       --4.2 &      --10.4 &       0.17 &       0.07 &       0.96 &       0.53 &       0.82 &       1.68 &       0.42 &      25.13 \\
    Knot 2 &       --3.9 &      --12.7 &       0.21 &       0.08 &       0.82 &       0.48 &       0.66 &       1.53 &       0.38 &      23.44 \\
    Knot 3 &       --2.6 &      --14.6 &       0.25 &       0.10 &       0.95 &       0.54 &       0.81 &       1.80 &       0.45 &      25.70 \\
     BN SW &       --2.0 &       --2.0 &       1.21 &       0.88 &       1.90 &       1.54 &       3.12 &       5.05 &       1.90 &      36.66 \\
    MAX--69 &       11.4 &      --16.6 &       0.13 &       0.15 &       0.65 &       0.42 &       0.40 &       0.48 &       0.38 &      11.17 
\enddata
\label{Tab:BN_sil}
\end{deluxetable}
\clearpage

\begin{deluxetable}{cccccc}
\tablecolumns{11}
\tablewidth{0pc}
\tablecaption{Physical parameters of the IRc sources in the BN/KL complex\label{Tab:BN_phys}}
\tablehead{
\colhead{} & \colhead{} & \multicolumn{2}{c}{7.7--12.4~\micron} & \multicolumn{2}{c}{10--20~\micron} \\
\cline{3-4} \cline{5-6} %\\
\colhead{Source Name} & \colhead{$\tau_{9.6}$}  & \colhead{T$_{col}$}  &  \colhead{L}	& \colhead{T$_{col}$}     &  \colhead{L}\\
   & 				& \colhead{(K)} 	    	  &  \colhead{(L$_\odot$)}	& \colhead{(K)} 	    	  &  \colhead{(L$_\odot$)}} 
\startdata
        BN &       0.51 &       490  &  9.21$\times10^{5}$ &       296  &  7.98$\times10^{4}$\\
 BN SW arc &       0.33 &       288  &  8.48$\times10^{4}$ &       176  &  8.39$\times10^{3}$\\
      IR n &       1.30 &       227  &  2.41$\times10^{4}$ &       149  &  4.37$\times10^{3}$\\
  IRc2 A+B &       2.91 &       308  &  1.14$\times10^{5}$ &       164  &  6.27$\times10^{3}$\\
    IRc2 C &       2.53 &       258  &  4.49$\times10^{4}$ &       153  &  4.70$\times10^{3}$\\
    IRc2 D &       2.36 &       219  &  2.11$\times10^{4}$ &       160  &  5.68$\times10^{3}$\\
    IRc2 E &       1.94 &       208  &  1.63$\times10^{4}$ &       148  &  4.06$\times10^{3}$\\
    IRc3 N &       1.11 &       204  &  1.46$\times10^{4}$ &       127  &  2.27$\times10^{3}$\\
    IRc3 S &       1.18 &       209  &  1.61$\times10^{4}$ &       128  &  2.35$\times10^{3}$\\
      IRc4 &       1.53 &       184  &  9.55$\times10^{3}$ &       138  &  3.12$\times10^{3}$\\
      IRc5 &       1.32 &       206  &  1.53$\times10^{4}$ &       126  &  2.15$\times10^{3}$\\
      IRc6 &       1.30 &       210  &  1.62$\times10^{4}$ &       133  &  2.69$\times10^{3}$\\
    IRc6 N &       0.87 &       223  &  2.09$\times10^{4}$ &       133  &  2.72$\times10^{3}$\\
    IRc6 E &       1.48 &       206  &  1.52$\times10^{4}$ &       134  &  2.75$\times10^{3}$\\
      IRc7 &       2.02 &       198  &  1.14$\times10^{4}$ &       148  &  4.17$\times10^{3}$\\
      IRc8 &       0.79 &       225  &  2.16$\times10^{4}$ &       119  &  1.69$\times10^{3}$\\
     IRc11 &       1.73 &       215  &  1.87$\times10^{4}$ &       144  &  3.69$\times10^{3}$\\
     IRc12 &       1.79 &       258  &  3.93$\times10^{4}$ &       140  &  3.34$\times10^{3}$\\
     IRc13 &       0.84 &       233  &  2.49$\times10^{4}$ &       118  &  1.66$\times10^{3}$\\
     IRc14 &       0.95 &       223  &  2.11$\times10^{4}$ &       124  &  2.03$\times10^{3}$\\
     IRc15 &       1.00 &       222  &  2.06$\times10^{4}$ &       130  &  2.41$\times10^{3}$\\
   IRc16 N &       1.06 &       238  &  2.70$\times10^{4}$ &       128  &  2.32$\times10^{3}$\\
   IRc16 S &       1.11 &       227  &  2.25$\times10^{4}$ &       136  &  2.97$\times10^{3}$\\
   IRc16 W &       0.86 &       233  &  2.47$\times10^{4}$ &       140  &  3.38$\times10^{3}$\\
     IRc17 &       0.65 &       247  &  3.16$\times10^{4}$ &       122  &  1.89$\times10^{3}$\\
     IRc19 &       1.02 &       303  &  7.30$\times10^{4}$ &       131  &  2.48$\times10^{3}$\\
     IRc20 &       1.33 &       214  &  1.75$\times10^{4}$ &       126  &  2.18$\times10^{3}$\\
     IRc21 &       1.63 &       221  &  2.03$\times10^{4}$ &       134  &  2.74$\times10^{3}$\\
     IRc22 &       1.05 &       229  &  2.36$\times10^{4}$ &       126  &  2.15$\times10^{3}$\\
     IRc23 &       1.21 &       239  &  2.80$\times10^{4}$ &       126  &  2.12$\times10^{3}$\\
    Knot 1 &       0.81 &       240  &  2.83$\times10^{4}$ &       115  &  1.48$\times10^{3}$\\
    Knot 2 &       0.84 &       246  &  3.12$\times10^{4}$ &       114  &  1.43$\times10^{3}$\\
    Knot 3 &       0.85 &       245  &  3.06$\times10^{4}$ &       115  &  1.50$\times10^{3}$\\
     BN SW &       0.77 &       234  &  2.53$\times10^{4}$ &       140  &  3.30$\times10^{3}$\\
    MAX-69 &        ---  &       297  &  6.79$\times10^{4}$ &       128  &  2.36$\times10^{3}$
\enddata
\end{deluxetable}

\typeout{LaTeX Warning: Label(s) may have changed. Rerun}

\end{document}